\documentclass[aps,prd,twocolumn,showpacs,notitlepage,eqsecnum,
superscriptaddress]{revtex4-1}

\usepackage[centertags]{amsmath}
\usepackage{amssymb}
\usepackage{latexsym}
\usepackage{enumerate}
\usepackage{graphicx}
\usepackage{mathrsfs}
\usepackage{hyperref}
\usepackage{stmaryrd}
\usepackage{import}
\usepackage{tensor}
\usepackage{color}
\usepackage{bm}
\usepackage{multirow}
\usepackage{breakurl}
\usepackage{float}
\usepackage{placeins}

\newcommand{\bi}{\begin{itemize}}
\newcommand{\ei}{\end{itemize}}
\newcommand{\be}{\begin{equation}}
\newcommand{\ee}{\end{equation}}

\renewcommand{\l}{\left(}
\renewcommand{\r}{\right)}
\renewcommand{\a}{\alpha}
\renewcommand{\b}{\beta}
\newcommand{\g}{\gamma}
\newcommand{\G}{\Gamma}
\renewcommand{\d}{\delta}
\newcommand{\D}{\Delta}
\newcommand{\e}{\epsilon}
\newcommand{\ve}{\varepsilon}
\newcommand{\ka}{\kappa}
\newcommand{\La}{\Lambda}
\newcommand{\la}{\lambda}
\renewcommand{\O}{\Omega}
\renewcommand{\o}{\omega}
\renewcommand{\th}{\theta}

\newcommand{\q}{\quad}
\newcommand{\qq}{\qquad}

\newcommand{\vp}{\varphi}

\newcommand{\pa}{\partial}

\allowdisplaybreaks

\begin{document}

\title{Analytic post-Newtonian expansion of the energy and angular momentum radiated to infinity by eccentric-orbit
non-spinning extreme-mass-ratio inspirals to 19PN}

\author{Christopher Munna}
\affiliation{Department of Physics and Astronomy, University of North 
Carolina, Chapel Hill, North Carolina 27599, USA}

\begin{abstract}
We develop new high-order results for the post-Newtonian (PN) expansions of the energy and angular momentum 
fluxes at infinity for eccentric-orbit extreme-mass-ratio inspirals (EMRIs) on a Schwarzschild background.  The series 
are derived through direct expansion of the MST solutions within the RWZ formalism for first-order black hole 
perturbation theory (BHPT).  By utilizing factorization and a few computational simplifications, we are able to 
compute the fluxes to 19PN, with each PN term calculated as a power series in (Darwin) eccentricity to $e^{10}$.
This compares favorably with the numeric fitting approach used in previous work.  We also compute PN terms to 
$e^{20}$ through 10PN.  Then, we analyze the convergence properties of the composite energy flux expansion by 
checking against numeric data for several orbits, both for the full flux and also for the individual 220 mode, 
with various resummation schemes tried for each.  The match between the high-order series and numerical 
calculations is generally strong, maintaining relative error better than $10^{-5}$ except when $p$ (the semi-latus 
rectum) is small and $e$ is large.  However, the full-flux expansion demonstrates superior fidelity (particularly at high 
$e$), as it is able to incorporate additional information from PN theory.  For the orbit $(p=10, e=1/2)$, the full flux 
achieves a best error near $10^{-5}$, while the 220 mode exhibits error worse than $1\%$.  Finally, we describe a 
procedure for transforming these expansions to the harmonic gauge of PN theory by analyzing Schwarzschild 
geodesic motion in harmonic coordinates.  This will facilitate future comparisons between BHPT and PN theory.
\end{abstract}


\maketitle

\section{Introduction}
\label{sec:intro}

With the launch of the LISA mission rapidly approaching, advancements in our understanding of generic 
extreme-mass-ratio inspirals (EMRIs), a major source of gravitational waves in LISA's frequency band, are of 
paramount importance \cite{BaraETC18,AmarETC07,BerrETC19,LISA}.  Theoretical models of these systems 
permit the creation of waveform templates, which can then be used to isolate gravitational-wave signals from LISA's 
data stream.  If sufficiently accurate, these templates can also be used to estimate various characteristics of the 
bodies.

To aid this effort, the last few decades have seen significant research in the field of black hole perturbation theory 
(BHPT), in which the Einstein field equations are expanded under the assumption that one of the two masses
is much larger than the other, or $\mu/M \ll 1$.  In this approximation the zeroth-order system is given by the 
spacetime of the large central black hole (Schwarzschild or Kerr), with the smaller body following a geodesic on 
that background.  The geodesic motion of the smaller body sources the first-order perturbation, which then interacts
with the body itself, causing a self-force or radiation reaction, driving the two bodies together.  See 
\cite{BaraPoun18} for a review.  By design this scheme is naturally suited to the description of EMRIs; however, the 
expansion process is quite complex.  The first-order perturbation is effectively understood on both 
Schwarzschild \cite{ReggWhee57, Zeri70, MartPois05, HoppEvan10, HoppEvan13} and Kerr \cite{Teuk73, 
Chrz75, KegeCohe79, Vand18}, but neither is yet implemented computationally in a manner sufficient for LISA.  
Worse, the second-order solution is not yet known theoretically, though significant progress is being made 
\cite{Poun12a, PounMill14, MillWardPoun16, Poun17, PounETC20}.  Thus, much work remains before LISA data analysis can begin.

There is another approximation scheme which applies for binary systems with slowly moving (or equivalently, widely 
separated) bodies.  This is the classic post-Newtonian (PN) expansion, which utilizes the small parameter $v/c \ll 1$ 
\cite{Blan14}.  The PN approximation is accurate early in the lifetime of any inspiral but generally breaks down before 
the point of merger.  Thus, this expansion method must be supplemented with additional information in order to 
capture the inspiral's full behavior.  For the comparable-mass binaries observed by LIGO, the late-stage orbit is 
typically described using full numerical relativity (NR) simulations.  Information from PN and NR (as well as BHPT) 
can be joined using a compact interpolation scheme known as the effective-one-body (EOB) formalism 
\cite{BuonDamo99, Damo01, Damo10}.  The combination of the three (PN, NR, EOB) has led to a large library of 
waveform templates for LIGO, which have allowed for its unprecedented success in detecting and characterizing 
mergers (see, e.g. \cite{AbboETC16a, AbboETC16b, AbboETC16c, LVC1606.01210, LVC1706.01812, 
LVC1811.12907}).  PN waveforms in particular were essential to the detection of the first binary neutron star merger 
\cite{LVC171016}, which heralded in a new era of multi-messenger astronomy \cite{LVC1710.05833, 
LVC1710.05834}.

Once the mass ratio deviates from unity, models from BHPT become necessary for the construction of accurate 
waveform templates.  This fact has already been validated through LIGO: Detections of 1/3.6 mass ratio 
and 1/8.8 mass ratio binaries \cite{LVC2004.08342, LVC2006.12611} utilized EOB waveforms partially calibrated 
using BHPT \cite{BoheETC17, CoteETC18, OssoETC20}.  For the EMRIs that will be observable by LISA, BHPT 
will almost surely serve as the central analytic framework in the construction of waveform templates.  However, due 
to the theoretical and computational complexity of BHPT methods, some combination with the other 
approximation schemes may be required to achieve rapid, accurate simulations across the possible parameter 
space.  To that end, there has recently been surge in work at the intersection of BHPT and PN theory, with frequent 
application to EOB models \cite{Detw08, SagoBaraDetw08, BaraSago09, BlanETC09, BlanETC10, Fuji12a,Fuji12b, 
ShahFrieWhit14, Shah14, DolaETC14b, JohnMcDaShahWhit15, BiniDamoGera15, KavaOtteWard15, AkcaETC15, 
SagoFuji15, ForsEvanHopp16,HoppKavaOtte16,KavaOtteWard16,BiniDamoGera16a, BiniDamoGera16b, 
BiniDamoGera16c, NagaShah16, SagoFujiNaka16, KavaETC17, BiniETC18, BiniETC18b, BiniGera18a, 
BiniGera18b, BiniGera18c, BiniDamoGera18, BiniGera19a, BiniGera19b, BiniGera19c, NagaETC19, MunnEvan19a,
MunnEvan19b, AntoETC20, BiniDamoGera20a, BiniDamoGera20b}.  As evidenced in those papers, combining the 
methodologies in this manner often yields progress in the separate theories that would be more difficult to derive 
otherwise.

The present work advances this effort by combining BHPT with PN theory to determine high-order PN series for 
the (orbit-averaged) energy and angular momentum radiated out to infinity by eccentric, non-spinning EMRIs.  
The expansions are pursued via the formalism of Mano, Suzuki, Takasugi (MST), which solves the 
Regge-Wheeler-Zerilli (RWZ) equations of first-order BHPT using infinite summations of analytic functions  
\cite{ReggWhee57, Zeri70, MartPois05, ManoSuzuTaka96a}.  In particular, the MST solutions to the homogeneous 
RW equation contain small PN parameters with which expansions can be made rapidly using algebraic computing 
software like \textsc{Mathematica} \cite{BiniDamo14a,BiniDamo14b,BiniDamo14c,KavaOtteWard15, 
HoppKavaOtte16}.  These series can then be joined with PN expansion of the source motion to 
compute analytic series for the normalization constants and fluxes.  The resulting flux representations can be rapidly 
evaluated to produce numeric values along and across orbits.  Related methods have already been successfully 
applied to extract high-order series for various orbital quantities in both the conservative and 
dissipative sectors, on both Schwarzschild and Kerr backgrounds \cite{Fuji12a, Fuji12b, Fuji15, BiniDamoGera15, 
AkcaETC15, HoppKavaOtte16, KavaOtteWard16, BiniDamoGera16a, BiniDamoGera16b, BiniDamoGera16c, 
KavaETC17, BiniETC18, BiniETC18b, BiniGera18a, BiniGera18b, BiniGera18c, BiniDamoGera18, BiniGera19a, 
BiniGera19b, BiniGera19c, AntoETC20, BiniDamoGera20a, BiniDamoGera20b}. 

The fluxes serve as the largest contribution to the inspiral's orbital phase and therefore require high accuracy in the 
construction of waveforms \cite{HindFlan08, OsbuWarbEvan16}, suggesting the utility of high-order expansions.  In 
2012 Fujita applied similar techniques to derive the circular-orbit limit of the EMRI energy flux to 22PN \cite{Fuji12b}. 
In the eccentric case each PN term must also be expanded as a Taylor series in Darwin eccentricity $e$ to complete 
the source integration, compounding the complexity by an order of magnitude.  Nevertheless, because LISA is 
expected to be sensitive to binaries with moderate or high eccentricity (unlike LIGO, which primarily observes 
circular-orbit binaries), expansions which reach high orders in both $v/c$ and $e$ may be needed 
\cite{BaraCutl07, HopmAlex05, AmarETC07}.

To that end, four other papers have also made recent progress on the eccentric EMRI flux expansions, though with 
different techniques \cite{ForsEvanHopp16, MunnETC20, MunnEvan19a, MunnEvan19b}.  Specifically, the two 
\cite{ForsEvanHopp16, MunnETC20} utilized thousands of numeric flux calculations to perform numeric fits to 
the forms of the two PN expansions.  These numeric fits were then partially converted to analytic form using an 
integer relation algorithm \cite{FergBailArno99}, resulting in series to 9PN and varying orders in eccentricity 
(frequently, $e^{30}$).  The other two \cite{MunnEvan19a, MunnEvan19b} combined separate discoveries from 
BHPT and PN theory to derive certain logarithmic contributions to the fluxes to arbitrary order in eccentricity.    

The analytic expansion procedure of the present work confers several advantages over the fitting methods of 
\cite{ForsEvanHopp16, MunnETC20}, allowing us to extend those results to much higher PN order.  In total, we
compute each of the two fluxes to 10PN through $e^{20}$ and to 19PN through $e^{10}$, with the latter 
PN order nearing the state of the art for circular orbits \cite{Fuji12b}.  Like all PN series, these flux representations 
produce numeric values that are accurate early in the EMRI's lifetime; however, it has also more recently been found 
that sufficiently high-order expansions have the potential to match numerical calculations near the point of merger.  
Indeed, \cite{Fuji12b} found that the 22PN expansion of the circular-orbit fluxes was sufficient to track the inspiral's 
evolution all the way to the separatrix.  Other works have demonstrated convergence for similar PN series to the 
light ring of the system \cite{JohnMcDaShahWhit15}.  We therefore use the present results to assess whether the 
same convergence properties continue to hold for eccentric orbits, both for an individual mode as well as for the full 
flux.  We evaluate different orbits and vary the PN order, and we also try a few resummation schemes 
like those mentioned in \cite{JohnMcDa14} to compare fidelity.  

We find that for the 220 mode flux, the expansion for $p=10$ can be made to maintain relative error better than 
$10^{-6}$ for $e \lesssim 1/4$ by factoring out the separatrix $p-6-2 e$.  However, at larger $e$ the series breaks 
down, barely reaching an error of $1\%$ at $e=1/2$ in the best case (though typically far worse).  Wider orbits exhibit 
better fidelity.  The full-flux expansion proves superior to the factorized mode flux, due to the fact that the former 
can incorporate derivations from PN theory and also utilize eccentricity singular factors.  The best match of the
full flux at $(p=10, e=1/2)$ is roughly $10^{-5}$.  Because of the large size of the expressions, it will not be useful 
to supply them here; however, the full series will be posted on the black hole perturbation toolkit \cite{BHPTK18} for 
easy retrieval.  This effort serves as a necessary intermediate step on the path to fully generic expansions on a 
Kerr background.  

It is of note that BHPT-PN series reproduce the small-mass-ratio limit of the full PN theory \cite{Blan14}, but 
in terms of parameters suited to BHPT coordinate systems --- like the Darwin eccentricity $e$, semi-latus 
rectum $p$, and relativistic anomaly $\chi$ in Schwarzschild coordinates.  Direct comparisons to derivations within
the full PN framework require transforming back to the more standard PN representation involving quasi-Keplerian 
(QK) parameters like the time eccentricity $e_t$ or true anomaly $V$ in (modified) harmonic or ADM coordinates.  
This was generally said to be possible only to the highest known order of the full PN equations of motion.  The 
equations of motion have been recently found to 4PN \cite{MarcETC18}, though the quasi-Keplerian parameters 
have only been explicitly derived to 3PN \cite{ArunETC08a, ArunETC08b, MemmGopaScha04}.  This 
allows for the confirmation of mutual agreement between the two theories in low-order cases.  

However, the litany of high-order BHPT-PN expansions, as well as of newer techniques permitting the extraction of 
PN terms in non-sequential order (often by combining BHPT and PN theory \cite{BlanETC09, BlanETC10, 
LetiBlanWhit12, DamoNagaBern13, ForsEvanHopp16, BiniDamoGera20a, BiniDamoGera20a}), has made it 
desirable to be able to convert to and from QK harmonic parameters to higher order.  This can be done 
by analyzing the harmonic-coordinate solution of the Schwarzschild metric.  This solution will serve as the 
small-mass-ratio limit of full PN harmonic coordinates.  We will show that the simple relationship between 
Schwarzschild harmonic coordinates and basic Schwarzschild coordinates can be used to form a QK 
representation of Schwarzschild geodesic motion to arbitrary order.  This representation will give definition to the QK 
parameters to arbitrary PN order at lowest order in the mass ratio, allowing them to be connected to the BHPT 
expansion parameters to arbitrary order as well.

The structure of this paper is as follows.  In Sec.~\ref{sec:RWZ}, we broadly review the RWZ formalism for
generic bound orbits on a Schwarzschild background and discuss the significance of high-order flux series.  
Then, in Sec.~\ref{sec:MSTexp}, we describe the analytic expansion of the MST homogeneous solutions, following 
the methods of \cite{KavaOtteWard15}.  In Sec.~\ref{sec:InhomExps} we proceed to the inhomogeneous problem, 
applying the methods of \cite{HoppKavaOtte16}.  Sec.~\ref{sec:fluxExps} describes the flux results, which are then 
compared to numerical calculations to test the accuracy and convergence of the series.  Sec.~\ref{sec:QKSchw} 
then describes the process by which these and other BHPT-PN expansions can be converted to a QK representation 
with harmonic-gauge parameters.  We finish in Sec.~\ref{sec:conclusions} with conclusions and outlook.

Unless otherwise noted, we set $c = G = 1$ and use metric signature $(-+++)$ and
sign conventions of Misner, Thorne, and Wheeler \cite{MisnThorWhee73}.  Our notation for the RWZ formalism 
follows that found in \cite{ForsEvanHopp16,MunnETC20,HoppKavaOtte16}, which in part 
derives from notational changes for tensor spherical harmonics and
perturbation amplitudes made by Martel and Poisson \cite{MartPois05}.  For 
the MST formalism, we largely make use of the discussion and notation found 
in the review by Sasaki and Tagoshi \cite{SasaTago03}.

\section{Review of the RWZ formalism and first-order fluxes}
\label{sec:RWZ}

\subsection{The RWZ master equations}
\label{sec:masterEq}

We begin by outlining the RWZ formalism used to compute the fluxes within first-order BHPT for a 
point mass $\mu$ in eccentric orbit around a Schwarzschild black hole of mass $M$ in the equatorial plane.  
We use Schwarzschild coordinates $x^{\mu} = (t,r,\theta, \varphi )$ with the line element
\be
\label{eqn:SchwLE}
ds^2 = -f dt^2 + f^{-1} dr^2 + r^2 \left( d\theta^2 + \sin^2\theta \, d\varphi^2 \right)
\ee
for $f = (1 - 2M/r)$.  The Schwarzschild metric is thus the zeroth-order piece of the full metric $g_{\mu \nu}$.  To 
obtain the first-order portion, the RWZ approach decomposes the linearized field equations in RW gauge over tensor 
spherical harmonics with indices $lm$.  The angular components decouple, resulting in two sets of partial differential 
equations (PDEs) for the $t$- and $r$-dependent spherical harmonic amplitudes, one for odd parity and the other for 
even parity.  It is found that each set can be encoded to a single (mode-dependent) PDE for a corresponding master 
function.  The first-order metric perturbations can then be recovered from these master functions.  The reader
should refer to \cite{MartPois05, HoppEvan13} for a complete description.  

The odd-parity PDE, known as the time domain (TD) RW equation, is given by
\begin{widetext}
\be
\label{eqn:TDRW}
\left[-\frac{\pa^2}{\pa t^2} + f \frac{\pa}{\pa r} f \frac{\pa}{\pa r} + f \left(\frac{l(l+1)}{r^2} - 
\frac{6M}{r^3} \right) \right]  \Psi^o_{lm} = S_{lm}^o,
\ee
where $\Psi^o_{lm}$ is the odd-parity master function.  The source term results from the tensor 
spherical harmonic decomposition of the smaller body's (point particle) stress energy tensor.  It can be written as 
\be
S_{lm}^o(t) = G_{lm}^o(t) \, \d [r - r_p(t)] + F_{lm}^o(t) \, \delta'[r - r_p(t)] ,
\ee
for functions $G_{lm}^o(t)$ and $F_{lm}^o(t)$ that will be given explicitly in Sec.~\ref{sec:orbits} below.  

Similarly, the even-parity Zerilli equation is given by
\be
\label{eqn:TDZ}
\left\{-\frac{\pa^2}{\pa t^2} + f \frac{\pa}{\pa r} f \frac{\pa}{\pa r} 
+  \frac{f}{\La^2} \biggl[ \ka^2 \biggl(\frac{\ka+2}{r^2} + \frac{6M}{r^3} \biggr) 
+ \frac{36M^2}{r^4} \biggl(\ka + \frac{2M}{r} \biggr) \biggr] \right\}  \Psi^e_{lm} = S_{lm}^e.
\ee
where $\ka = (l-1)(l+2), \La = \ka + 6M/r$, and
\be
S_{lm}^e(t) = G_{lm}^e(t) \, \d [r - r_p(t)] + F_{lm}^e(t) \, \delta'[r - r_p(t)] .
\ee

We now exploit the biperiodicity of the source motion and make a transformation to the frequency domain (FD).  
In this way the odd-parity equation becomes
\be
\label{eqn:FDRW}
\left[f \frac{\pa}{\pa r} f \frac{\pa}{\pa r} + \o^2 + f \left(\frac{l(l+1)}{r^2} - 
\frac{6M}{r^3} \right) \right]  X_{lmn}(r) = Z_{lmn} (r),
\ee
\end{widetext}
where $\o = \o_{mn} = m \O_\vp + n \O_r$ and 
\begin{align}
X_{lmn}(r) &= \frac{1}{T_r} \int_0^{2 \pi} \Psi_{lm}^o e^{i \o t} dt, \notag \\
Z_{lmn}(r) &= \frac{1}{T_r} \int_0^{2 \pi} S_{lm}^o e^{i \o t} dt.
\end{align}
For simplicity $X_{lmn}(r)$ and $Z_{lmn}(r)$ are written without a superscript denoting odd-parity, as we shall work 
almost exclusively in the odd-parity sector.  The TD solutions are reconstructed in the usual way:
\begin{align}
\Psi_{lm}^o &= \sum_{n = -\infty}^\infty X_{lmn}(r) e^{- i \o t}, \notag \\
S_{lm}^o &= \sum_{n = -\infty}^\infty Z_{lmn}(r) e^{- i \o t}.
\end{align}

The homogeneous solutions to \eqref{eqn:FDRW} can be derived analytically using the MST method 
\cite{ManoSuzuTaka96a}.  This leads to the pair of functions $X_{lmn}^+ = X_{lmn}^{\rm up}$, with proper behavior 
for $r > r_p$, and $X_{lmn}^- = X_{lmn}^{\rm in}$, with proper behavior for $2M < r < r_p$.  The subscript $p$ 
represents the location of the smaller body.  Explicit representations for $X_{lmn}^\pm$ will be given in 
Sec.~\ref{sec:MSTexp}.  
The corresponding even-parity homogeneous solutions can be found directly from $X_{lmn}^+$ and $X_{lmn}^-$ 
using the Detweiler-Chandrasekar transformation \cite{Chan75, ChanDetw75, Chan83, Bern07}. 

\subsection{Particular solution to the Regge-Wheeler equation}
\label{sec:InhSoln}

A suitable set of particular solutions for this problem can be found using the method of extended homogeneous 
solutions (EHS) \cite{BaraOriSago08,HoppEvan10}.  Explicitly, the result is
\begin{align}
\label{eqn:XEHS}
X_{lmn} &= C^+_{lmn} X^+_{lmn}(r) \Theta[r - r_p(t)] \notag \\ &
\hspace{5em} + C^-_{lmn} X^-_{lmn}(r) \Theta[r_p(t) - r],
\end{align}
where $\Theta$ is the Heaviside function and $C^{\pm}_{lmn}$ are constants.  Even though \eqref{eqn:XEHS} is not 
a valid solution to \eqref{eqn:FDRW}, the time-domain solution found by taking
\be
\Psi_{lm}^o = \sum_{n = -\infty}^{\infty} X_{lmn} e^{-i \o t}
\ee
is in fact a valid solution to its time-domain counterpart, as can be shown by direct evaluation \cite{HoppEvan10}.  
The proper normalization constants are given by
\begin{align}
C^{\pm}_{lmn} &= \frac{1}{W_{lmn} T_r} \int_0^{2 \pi}  \bigg[ \frac{1}{f_p}G_{lm}^o(t) X^{\mp}_{lmn} \, +  \notag \\&
\left( \frac{2 M}{r_p^2 f_p^2}  X^{\mp}_{lmn} - \frac{1}{f_p}
 \frac{d X^{\mp}_{lmn}}{dr} \right) 
  F_{lm}^o(t) \bigg] e^{i \o t} dt,
\label{eqn:ClmnIntT}
\end{align}
where $T_r$ is the period of radial libration and $W_{lmn}$ is the Wronskian,
\be
W_{lmn} = f \frac{d X^+_{lmn}}{dr} X^-_{lmn} - f \frac{d X^-_{lmn}}{dr} X^+_{lmn}.
\ee
Note that the corresponding even-parity expression is identical, but with the even-parity source terms $G_{lm}^e$ 
and $F_{lm}^e$ and homogeneous functions $X^{\mp,e}_{lmn}$. Interestingly, it can be shown through direct 
evaluation that the Wronskian $W_{lmn}$ is parity-independent.  Note also that the various functions are evaluated 
at the location of the smaller body, meaning that it is necessary to possess compact expressions for the zeroth-order 
motion of body.  This is simply the geodesic motion of a test mass on a Schwarzschild background.

\subsection{Bound orbits on a Schwarzschild background and the corresponding source terms}
\label{sec:orbits}

At zeroth order the motion is geodesic in the static background.  The geodesic equations can be integrated to
yield the four-velocity as
\be
\label{eqn:four_velocity}
u^\a = \l \frac{{\mathcal{E}}}{f_{p}}, u^r, 0, \frac{{\mathcal{L}}}{r_p^2} \r 
\ee
for energy $\mathcal{E}$ and angular momentum $\mathcal{L}$.
The radial motion is found using the constraint on the four-velocity $u^\a u_\a = -1$, or
\be
(u^r)^2 =  \mathcal{E}^2  - f_p  \l 1 + \frac{{\mathcal{L}}^2}{r_p^2} \r  .
\ee
Bound orbits have ${\mathcal{E}} < 1$ and ${\mathcal{L}} > 2 \sqrt{3} M$.  

In Sec.~\ref{sec:InhomExps} the motion will be reparameterized using geometric features of the orbit to simplify the 
process of PN expansion.  For now, the four-velocity \eqref{eqn:four_velocity} can be used to derive compact forms 
for the source terms.  The process is straightforward though cumbersome, involving combinations of components of 
the stress energy tensor integrated over tensor spherical harmonics.  The odd-parity source terms are found to be
\begin{widetext}
\begin{align}
F_{o}^{lm}(t) &= \frac{32 \pi \mu \mathcal{L} f_p^3 (r_p^2 + \mathcal{L}^2)}{(l-1)l(l+1)(l+2) 
\mathcal{E}^2 r_p^3} X_{\vp}^{*lm} ,   \notag \\
G_{o}^{lm}(t) &= \frac{32 \pi \mu \mathcal{L} f_p}{(l-1)l(l+1)(l+2) \mathcal{E}^2 r_p^5} \Big[ 
\mathcal{L} \mathcal{E} r_p^2 \dot{r}_p (-i m) - f_p \left( 5Mr_p^2 + 7 M \mathcal{L}^2 
+ (2 \mathcal{E}^2 - 1)r_p^3 - 2 \mathcal{L}^2 r_p \right)  \Big]  X^{*lm}_{\vp}.
\label{eqn:OPFGsimp}
\end{align}
Note that these expressions were published in \cite{HoppKavaOtte16}.  The even-parity terms follow as
\begin{align}
F_{e}^{lm}(t) &=  \frac{32 \pi \mu f_p^3 (r_p^2 + \mathcal{L}^2)}{l(l+1) (r_p(l-1)(l+2) + 6M)
 \mathcal{E} r_p} Y_{lm}^* ,  \notag \\
G_{e}^{lm}(t) &=\frac{16 \pi \mu f_p}{(l-1)l(l+1)(l+2) r_p^3 ((l-1)(l+2)r_p + 6M)^2 \mathcal{E}}
\Big[2 f_p^2 (l-1) (2+l) \mathcal{L}^2 r_p (6 M+(l-1) (2+l) r_p)   \notag \\ &
-f_p \mathcal{L} (6 M+(l-1) (2+l) r_p)
\left(\mathcal{L} \left(l+l^2-2 m^2\right) (6 M+(l-1) (2+l) r_p)+4 i (l-1) (2+l) m r_p^2 u^r\right)  \notag \\ &
+(l-1) (2+l) r_p^2 (\mathcal{E}^2 \left(-60 M^2-12 (l-1) (2+l) M r_p-(l-1) l (1+l) (2+l) r_p^2\right)  \notag \\ &
+\left(12 M^2+12 l (1+l) M  r_p+(l-1) l (1+l) (2+l) r_p^2\right) (u^r)^2)\Big] Y_{lm}^*.
\label{eqn:EPFGsimp}
\end{align}
The definitions for the scalar spherical harmonic $Y_{lm}$ and vector spherical harmonic $X_\vp^{lm}$ are given
in \cite{MartPois05}.  Both are evaluated at the location of the smaller body.

\subsection{The energy and angular momentum radiated to infinity}
\label{sec:fluxFormulas}

The RWZ method leads to elegant expressions for the energy and angular momentum radiated out to 
infinity.  These are described in \cite{MartPois05} by analyzing the $r \rightarrow \infty$ limit of the metric 
perturbations.  Explicitly, they are given by
\begin{align}
\label{eqn:EfluxPsi}
\bigg\langle\frac{dE}{dt}\bigg\rangle^\infty &= 
\frac{1}{64 \pi} \sum_{lm} (l+2)(l+1)(l)(l-1) \bigg\langle | \dot{\Psi}_{lm}^{e}(t,r=\infty) |^2
+| \dot{\Psi}_{lm}^{o}(t,r=\infty) |^2 \bigg\rangle,  \\
\bigg\langle\frac{dL}{dt}\bigg\rangle^\infty &=
\frac{1}{64 \pi} \sum_{lm} (l+2)(l+1)(l)(l-1) (-i m) \bigg\langle  \Psi_{lm}^{*e}
\dot{\Psi}_{lm}^{e} + \Psi_{lm}^{*o} \dot{\Psi}_{lm}^{o}  \bigg\rangle,
\label{eqn:LfluxPsi}
\end{align}
\end{widetext}
The horizon fluxes are similar, except with evaluation at $r = 2M$.  

These can be simplified further by rewriting the $\Psi$ functions in terms of their EHS Fourier sums.  
These will involve factors like $C^+_{lmn} X^+_{lmn} $ evaluated in the appropriate limits.  
However, it is easier to instead work with normalized homogeneous functions $\hat{X}^+_{lmn}$ which are 
constructed to approach unity at infinity.  With these normalized functions, the flux 
summations simply become
\begin{align}
\label{eqn:EfluxC}
\bigg\langle\frac{dE}{dt}\bigg\rangle^{\infty} &= \frac{1}{64 \pi} \sum_{lmn} (l+2)(l+1)(l)(l-1) \o^2 
| C^+_{lmn} |^2,  \\
\bigg\langle\frac{dL}{dt}\bigg\rangle^{\infty} &= \frac{1}{64 \pi} \sum_{lmn} (l+2)(l+1)(l)(l-1) m \o | C^+_{lmn} |^2,
\label{eqn:LfluxC}
\end{align}
where the $C_{lmn}^+$ are now constructed by integrating $\hat{X}^-_{lmn}$ in \eqref{eqn:ClmnIntT}.

\subsection{Significance of the EMRI fluxes and of their PN expansions}
\label{sec:significance}

The two fluxes \eqref{eqn:EfluxC} and \eqref{eqn:LfluxC}, along with the additional pair at the larger black hole's 
horizon, are critical 
to our understanding and description of EMRI radiation.  It has been shown through use of multiple 
timescale analysis that these first-order, orbit-averaged fluxes form the dominant contribution to the system's 
cumulative phase \cite{HindFlan08, FlanHind12}.  This phase must be known to within a fraction of a radian over the
inspiral's lifetime for the successful detection and characterization of EMRIs by LISA \cite{BaraETC18}.

Within the multi-scale framework, the exclusive use of the fluxes to model EMRIs is known as the adiabatic 
approximation.  Such an approach has been used to simulate EMRIs and generate coarse but efficient waveforms 
across various realms of parameter space \cite{DrasFlanHugh05, HughETC05, DrasHugh06, FujiHikiTago09, 
Fuji12b}.  However, it is known that adiabatic waveforms will be insufficient for EMRI parameter estimation with 
LISA, which will require knowledge of all contributions through post-1 adiabatic order, including resonance effects, 
the oscillatory first-order self-force, and the second-order fluxes \cite{HindFlan08, FlanHind12}.  

Nevertheless, as the leading contribution, the first-order fluxes must be known to significantly higher accuracy than 
the other quantities, all of which appear in the cumulative phase at higher order in the mass ratio.  As a result, there 
has been a great deal of past work analyzing the total energy and angular momentum radiated by EMRIs.  The 
primary approach to flux determination has historically been direct numerical calculation (e.g., \cite{DrasHugh06, 
TaraETC13}).  Large swaths of numeric flux computations can be combined with suitable interpolation schemes to
obtain accurate expressions across parameter space, which can then be used to simulate the inspiral 
\cite{OsbuWarbEvan16}.  

Low-order analytic PN expansions of the flux formulas \eqref{eqn:EfluxC} and \eqref{eqn:LfluxC} using the MST 
solutions have also been known quite some time \cite{TanaTagoSasa96}.  These have been useful for verifying 
numerical calculations, informing PN theory, and generating rapid inspiral simulations.  However, because they lose 
accuracy near the point of merger, it was generally thought that their utility in the generation of full 
waveform templates would be limited.

More recently, there have been a number of discoveries on the accuracy of high-order BHPT-PN expansions which 
have led to increased confidence in their relevance in the strong-field regime.  In particular, it was shown directly 
in the last decade that high-PN-order expansions in many cases converge (albeit slowly) all the way to the orbit's 
separatrix, if not its light ring.  This was first demonstrated for the energy flux at infinity in 2012 for the case 
of circular orbits about a Schwarzschild background, \cite{Fuji12a, Fuji12b}.  There, the author found that 22PN 
expressions were able to match numerical adiabatic simulations all the way to the innermost stable circular orbit 
(ISCO).  Similar results have been obtained for high-order expansions of certain conservative-sector quantities, 
again for the case of circular orbits on a Schwarzschild background \cite{JohnMcDaShahWhit15, KavaOtteWard15}.  

It has also been found that the use of factorization schemes or resummation methods can improve the convergence 
of these expansions even further \cite{JohnMcDa14, Fuji15, NagaShah16, NagaETC19}.  For instance, as proposed 
in \cite{IsoyETC13b}, the simple process of re-expanding the logarithm of the $lm$ (circular-orbit) mode fluxes, 
evaluating numerically, and then exponentiating the result often improves agreement with numerical calculations at 
the ISCO \cite{JohnMcDa14}.  Resummed high-order BHPT-PN expansions are especially well suited to the 
development of EOB waveforms, which have thus far been highly successful at simulating binaries across large 
regions of parameter space \cite{DamoNaga07, DamoIyerNaga09, PanETC11, PanETC11, MessNaga17, 
BoheETC17, MessNaga17, CoteETC18, MessMaldNaga18, NagaETC19, NagaETC20, OssoETC20, ChiaNaga20}.

Still, the convergence properties of the expansions and overall utility of this approach in more intricate EMRIs 
remains an open question of study, though progress has been made.  The energy flux series for circular, 
equatorial orbits on a Kerr background was derived to 11PN in \cite{Fuji15}, and this was found to agree with 
numerical calculations up to velocities of about $0.4$ for both prograde and retrograde orbits \cite{SagoFujiNaka16, 
FujiSagoNaka18}.  In the case of fully generic orbits on a Kerr background, series are presently published only to 
4PN and $e^6$, and the strong-field behavior is unknown \cite{SagoFuji15}.  Results that incorporate secondary spin 
have also been published in \cite{AkcaETC20} with some analysis of strong-field convergence.  

The present work focuses on high-order flux expansions for eccentric orbits on a Schwarzschild background, 
offering a necessary intermediate step on the path to fully generic orbits.  The convergence properties of the 
expansions can be analyzed, both with and without the use of basic resummation methods, 
allowing for comparison to the circular-orbit results of \cite{Fuji12b, Fuji15, SagoFujiNaka16, FujiSagoNaka18}.
We will find that the convergence worsens with increasing $e$, but there will still be large regions of parameter 
space where the high-order expansions are useful.

\section{Analytic expansion of the MST homogeneous solutions}
\label{sec:MSTexp}

The previous section offered a broad overview of the formalism that will be used to compute the two fluxes at infinity 
for eccentric-orbit inspirals.  Now we move to the specific implementation used to construct our high-order PN 
expansions.  This section will detail the process of expanding the odd-parity MST homogeneous solutions, 
generally following the methods of \cite{KavaOtteWard15}.  The next section will then cover the inhomogeneous 
integral and source motion.  

To begin, the odd-parity infinity- and horizon-side MST homogeneous solutions can be written as
\cite{ManoSuzuTaka96a, SasaTago03, KavaOtteWard15}:
\begin{widetext}
\begin{align}
\label{eqn:XupMST}
X^{+}_{lmn} &= e^{iz} z^{\nu+1} \left(1- \frac{\e}{z}\right)^{-i \e} \sum_{j=-\infty}^{\infty} a_j (-2 i z)^{j} 
\frac{ \G(j + \nu + 1 - i \e) \G(j + \nu - 1 - i \e) }{\G(j + \nu + 3 + i \e) \G(j + \nu + 1 + i \e) } \times \notag \\
& \hspace{25em} U(j + \nu + 1 - i \e, 2 j + 2 \nu + 2, -2 i z),  \\
X^{-}_{lmn} &= e^{-iz} \left(\frac{z}{\e} - 1\right)^{-i \e} \left(\frac{\e}{z}\right)^{i \e + 1}  
\sum_{j= -\infty}^{\infty} a_j \frac{\G(j + \nu - 1 - i\e) \G(-j - \nu - 2 - i \e)}{\G(1 - 2i\e)} \times \notag \\
& \hspace{18em} {}_2F_1(j + \nu - 1 - i\e, -j - \nu - 2 - i\e; 1 - 2 i \e; 1 - z/\e).
\label{eqn:XinMST}
\end{align}
In these expressions, $\nu=\nu(l,\e)$ is the renormalized angular momentum, a special parameter chosen to make
the summations converge (see \cite{ManoSuzuTaka96a, SasaTago03}), and $a_j=a_j(l,\e)$ are 
$\nu$-dependent series coefficients.  $U(a,b,\zeta)$ is the irregular confluent hypergeometric function,
and ${}_2F_1(a,b,c,\zeta)$ is the hypergeometric function.  Finally, $\e = 2 M \o \eta^3$ and $z = r \o \eta$, with 
$\eta = 1/c$, serve as the expansion parameters.  (In this section, factors of $c$ are restored to track PN order.)

Everything contained within $X^{+}_{lmn}$ and $X^{-}_{lmn}$ depends upon $\e$ and $z$ and thus on $\eta$.  
By definition $\eta^2$ corresponds to 1PN order.  Therefore, both $X^{+}_{lmn}$ and $X^{-}_{lmn}$ can be directly 
expanded in $\eta$ analytically, and this is achieved here to high order using the algebraic computing software 
\textsc{Mathematica}.  Note briefly that the MST solutions presented above are slightly different from those given in 
\cite{ManoSuzuTaka96a, SasaTago03, KavaOtteWard15}.  Here, we have preemptively canceled a few 
$z$-independent factors that do not contribute to the radiation.

\subsection{Expansion of $\nu$ and $a_j$}
\label{sec:expNuaj}

The PN expansion procedure is best begun with renormalized angular momentum $\nu$ and series coefficients 
$a_j$, which identically appear in both $X^{+}_{lmn}$ and $X^{-}_{lmn}$.  These terms are computed via the 
resolution of a continued fraction equation, defined to make the sum converge as $j \rightarrow \pm \infty$.  This 
equation is given by \cite{SasaTago03}
\be \alpha_j a_{j+1} + \beta_j a_j + \gamma_j a_{j-1} = 0, \ee
with
\begin{gather}
\alpha_j = - \frac{ i \e (-1 - i \e + j + \nu) (-1 + i \e + j + \nu) (1 - i \e + j + \nu)}{(1 + j + \nu) (3 + 2 j + 2 \nu)}, \notag \\
\beta_j = 2 \e^2 - l (l + 1) + \frac{\e^2 (\e^2 + 4)}{(j + \nu) (1 + j + \nu)} + (j + \nu) (1 + j + \nu), \notag \\
\gamma_j = \frac{ i \e (i \e + j + \nu) (2 - i \e + j + \nu) (2 + i \e + j + \nu)}{(j + \nu) (-1 + 2 j + 2 \nu)}. 
\end{gather}

This is solved to some desired power of $\e$ by setting 
\be \alpha_j R_{j+1} + \beta_j + \gamma_j L_{j-1} = 0, \ee
where $R$ and $L$ are the continued fractions:
\begin{gather}
R_{j+1} = \frac{a_{j+1}}{a_j} = - \frac{\g_{j+1}}{\b_{j+1} -} \frac{\a_{j+1} \g_{j+2}}{\b_{j+2} -}
 \frac{\a_{j+2} \g_{j+3}}{\b_{j+3} - \cdots} , \qq
L_{j-1} = \frac{a_{j-1}}{a_j} = - \frac{\a_{j-1}}{\b_{j-1} -} \frac{\g_{j-1} \a_{j-2}}{\b_{j-2} -}
 \frac{\g_{j-2} \a_{j-3}}{\b_{j+3} - \cdots}.
\end{gather}
First, $\nu$ is found to some given order in $\e$ by fixing $j$ and truncating the fractions at the needed 
depth.  An ansatz of $\nu = \nu_0 + \nu_2 \e^2 + \nu_4 \e^4 \cdots$ can be substituted, and the 
resulting equation can be solved order by order to extract each $\nu_i$ \cite{CasaOtte15}.  Then, the series 
coefficients $a_j$ can be iteratively built up using $a_{j+1} = R_{j+1} a_j$ and $a_{j-1} = L_{j-1} a_j$.

As an example, $\nu$ and some of the series coefficients for $l=2$ can be found as \cite{CasaOtte15}
\be
\nu = 2-\frac{107 \e^2}{210}-\frac{1695233 \e^4}{9261000}
-\frac{76720109901233 \e^6}{480698687700000}
-\frac{71638806585865707261481 \e^8}{389235629236738284000000} + \mathcal{O}\left(\e^{10}\right)
\ee
\begin{align}
a_{-4} &= -\frac{7 i  \e^5}{856}-\frac{53 \e^6}{6420}+\mathcal{O}\left(\e^7\right), \notag \\
a_{-3} &= -\frac{7 \e^4}{1926}+\frac{211 i \e^5}{28890}-\frac{3985481 \e^6}{370947600}
+\mathcal{O}\left(\e^7\right),  \notag \\
a_{-2} &= \frac{11 \e^4}{12840}-\frac{11 i \e^5}{8560}+\frac{18652901 \e^6}{15147027000}
+\mathcal{O}\left(\e^7\right),\notag \\
a_{-1} &= -\frac{i \e^3}{20}-\frac{\e^4}{40}-\frac{4920329 i \e^5}{94374000}
-\frac{3061237 \e^6}{94374000}+\mathcal{O}\left(\e^7\right),  \notag \\
a_0 &= 1, \notag \\
a_1 &= -\frac{5 i \e}{6}+\frac{5 \e^2}{18}-\frac{12029 i \e^3}{52920}+\frac{19519 \e^4}{158760}
-\frac{4807626493 i \e^5}{25671492000}+\frac{2573708771 \e^6}{25671492000}
+\mathcal{O}\left(\e^7\right), \notag \\
a_2 &= -\frac{15 \e^2}{49}-\frac{5 i \e^3}{28} -\frac{730781 \e^4}{6338640}-\frac{2691 i \e^5}{24640}
-\frac{921715511273 \e^6}{8882970096000}+\mathcal{O}\left(\e^7\right),  \notag \\
a_3 &= \frac{5 i \e^3}{72}-\frac{47 \e^4}{864}+\frac{1379137 i \e^5}{49533120}-\frac{58088509 \e^6}{1485993600}
+\mathcal{O}\left(\e^7\right), \notag \\
a_4 &= \frac{10 \e^4}{891}+\frac{19 i \e^5}{1782}+\frac{8914057 \e^6}{2074675680}
+\mathcal{O}\left(\e^7\right) .
\end{align}
As we can see, the results for negative $j$ are not quite regular.  This is due to the fact that $L_{j}$ 
experiences cancelation in its denominator for certain values of $j<0$.  Thus, the corresponding
$L_{j}$ either gains or loses additional powers of $\e$.  Fortunately, this behavior can be precisely determined, 
and the starting orders are listed in \cite{KavaOtteWard15}.

\subsection{Expansion of the infinity-side homogeneous solution $X^{+}_{lmn}$}
\label{sec:expXup}

The remaining factors in the homogeneous solutions have additional subtleties that complicate their expansions.  
A complete description of the process was provided in \cite{KavaOtteWard15}; therefore, our treatment here will be
brief.  We start with the odd-parity solution $X^{+}_{lmn}$ in \eqref{eqn:XupMST}.  This function
is most easily expanded in a few separate pieces, which can then be combined with $\nu$ and the $a_j$ to 
produce the full solution.

\subsubsection{The initial prefactor, $C_{\rm up}$}

First, the expansion of the prefactor is straightforward.  We slightly modify the expansion variables and write
\be
C_{\rm up} = e^{iz} z^{\nu+1} \left(1- \frac{\e}{z}\right)^{-i \e} = e^{i\bar{z} \eta} (\bar{z} \eta)^{\nu+1} 
\left(1- \frac{\bar{\e}}{\bar{z}} \eta^2 \right)^{-i \bar{\e}\eta^3},
\ee
where we have defined $\bar{z} = r \o$ and $\bar{\e} = 2 M \o$.  Thus, $z = \bar{z} \eta, \e = \bar{\e} \eta^3$, which 
allows for straightforward expansion in $\eta$.  This substitution can be utilized throughout the procedure to more 
easily track powers of $\eta$.  Note that the factor of $z^{\nu+1}$ ensures that $C_{\rm up}$ will begin at order 
$\mathcal{O}(\eta^{l+1})$.  As an example, this can be found for $l=2$ to give
\begin{align}
C_{\rm up}^{l=2} =&  \bar{z}^3 \eta^3 + i \bar{z}^4 \eta^4 - \frac{\bar{z}^5}{2} \eta^5 - \frac{1}{6} i \bar{z}^6 \eta^6
+\frac{\bar{z}^7}{24} \eta^7  + \left(i \bar{\e}^2 \bar{z}^2+\frac{i \bar{z}^8}{120}\right) \eta^8  \notag \\& \hspace{17em}
-  \left(\bar{\e}^2 \bar{z}^3 + \frac{\bar{z}^9}{720}
+\frac{107}{210} \bar{\e}^2 \bar{z}^3 \log (\bar{z} \eta) \right) \eta^9 + \mathcal{O}(\eta^{10})
\end{align}

\subsubsection{Manipulation of the hypergeometric function}

The primary remaining complication is the irregular confluent hypergeometric function $U$, which must be recast 
using hypergeometric identities into a form more suitable for PN expansion.  One useful choice is
\be
U(a,b,\zeta) = \frac{\G(1-b)}{\G(a-b+1)} M(a,b,\zeta) + \frac{\G(b-1)}{\G(a)} \zeta^{1-b} M(a-b+1,2-b, \zeta)
\ee
for Kummer hypergeometric function $M(a,b,\zeta) = {}_1F_1(a,b,\zeta)$ \cite{KavaOtteWard15}.  

Taking the two instances of $M$ separately, and including the other factors in the summation for $X_{lmn}^+$, 
the first portion can be written as
\begin{align}
U_1^{lj} \equiv (-2 i z)^{j} \frac{ \G(j + \nu - 1 - i \e) \G(j + \nu + 1 - i \e) 
\G(-2j - 2\nu -1)}{\G(j + \nu + 3 + i \e) \G(j + \nu + 1 + i \e) \G(-j-\nu- i \e)}
M(j + \nu + 1 - i \e, 2j + 2 \nu + 2, -2 i z).
\end{align}
The function $U_1^{lj}$ exhibits PN irregularities in both the $\G$ prefactors and the function $M$. In the product of
$\G$ functions, factors of $\e$ are lost whenever a term in the numerator has an argument 
$\le 0$, and they are gained whenever a term in the denominator has an argument $\le 0$. 
Once this is accounted for, the $\G$ product can be properly 
expanded to any order in $\e$, though the basic execution in \textsc{Mathematica} can be slow. 

For $M(j + \nu + 1 - i \e, 2j + 2 \nu + 2, -2 i z)$, irregular behavior behavior occurs when $j + l < 0$.
This can be observed in the hypergeometric series:
\be
M(a,b,z) = \sum_{k = 0}^{\infty} \frac{(a)_k}{(b)_k} \frac{z^k}{k!}, \qq \qq 
(a)_k = \frac{\G(a + k)}{\G(a)} = (a) (a+1) (a+2) \cdots (a+k-1),
\ee
where $(a)_k$ is the Pochhammer symbol.  Thus, when $j + l = -1,$ the PN series for 
$M(j + \nu + 1 - i \e, 2j + 2 \nu + 2, -2 i z)$ starts at $\mathcal{O}(1/\eta^2)$. 

The second piece, given by
\begin{align}
U_2^{lj} = (-2 i z)^{(-j - 2\nu - 1)} \frac{ \G(j + \nu - 1 - i \e)\G(2j + 2\nu + 1)}{\G(j + \nu + 3 + i \e) \G(j + \nu + 1 + i \e)}
M(-j - \nu - i \e, - 2j - 2 \nu, -2 i z),
\end{align}
is handled similarly.

\subsubsection{The full $X_{lmn}^+$ for $l=2$}

With the components expanded, we can now proceed to the computation of $X^{+}_{lmn}$.
To that end, Table \ref{tab:XupOrds} establishes the leading PN orders of $X^{+}_{lmn}$ for each $l$ and $j$.
An equivalent table is given in \cite{KavaOtteWard15}.
\begin{table*}[t]
\caption{Leading powers of $\eta$ in $C_{\rm up} a_j U_1^{lj}$ and $C_{\rm up} a_j U_2^{lj}$ as 
functions of $l$ and $j$.}
\label{tab:XupOrds}
\begin{center}
\begin{tabular}{|| c | c | c | c | c | c ||}
\hline\hline
  &  $j \le -2l-1$  & $-2l \le j \le -l - 3$ & $-l -2 \le j \le -l-1$ & $-l \le j \le -l+1$ & $j \ge -l +2$ \\
\hline
$C_{\rm up} a_j U_1^{lj}$ & $2|j| + l-2$ & $2|j|+l+4$ & $3|j| - j - l - 3$  & $3|j|+ j +l + 1$ & $3|j| + j + l-2$\\
$C_{\rm up} a_j U_2^{lj}$  &  $4|j| - l - 6$  & $4|j| - l$ & $3|j|- j - l - 3$ & $3|j| + j + l + 1$ & $3|j|-j-l$ \\
\hline \hline
\end{tabular}
\end{center}
\end{table*}
In this way it can be determined how many $j$ values must retained for a given $l$ to reach any desired order.  
For example, in order to calculate 
$X^{+}_{2mn}$ to, say, $\eta^4$, we must include $0 \le j \le 1$ for $U_1^{2j}$ and $0 \le j \le 3$ for $U_2^{2j}$.  A
low-order expansion for the normalized version of $X^{+}_{2mn}$ will be shown in Sec.~\ref{sec:normXFuncs}.

\subsection{Expansion of the horizon solution $X^{-}_{lmn}$}
\label{sec:expXin}

\subsubsection{Separating and expanding the hypergeometric function}

The prefactor $C_{\rm in} = e^{-iz} \left(\frac{z}{\e} - 1\right)^{-i \e} \left(\frac{\e}{z}\right)^{i \e + 1}$ is expanded 
similarly to its infinity-side counterpart.   The hypergeometric function ${}_2F_1(a,b,c,\zeta)$, meanwhile, can be 
separated into a form more amenable to the present expansion \cite{KavaOtteWard15}:
\begin{align}
{}_2F_1(a,b,c,\zeta) &= \frac{\G(c) \G(b-a)}{\G(b) \G(c-a)} (1-\zeta)^{-a} \, 
{}_2F_1\left(a,c-b,a-b+1, \frac{1}{1-\zeta}\right)   \notag \\ & \hspace{12em}
+ \frac{\G(c) \G(a-b)}{\G(a) \G(c-b)} (1-\zeta)^{-b} \, {}_2F_1\left(c-a,\, b, \, b-a+1, \frac{1}{1-\zeta}\right).
\end{align}
The first appearance ${}_2F_1$ can be combined with remaining factors in the summand to produce the function
\begin{align}
F_1^{lj} =  \frac{ \G(j + \nu - 1 - i \e) \G(-2 j - 2 \nu -1)}{\G(-j - \nu - i \e + 2)} \left( \frac{\e}{z} \right)^{j+\nu-1- i \e}
{}_2F_1(j + \nu - 1 - i \e, j + \nu + 3 - i \e, 2 j + 2 \nu + 2, \e/z).
\end{align}

Once again, irregularities in leading PN order exist in the $\G$ functions and in 
${}_2F_1(j + \nu - 1 - i \e, j + \nu + 3 - i \e, 2 j + 2 \nu + 2, \e/z)$ itself.  
For ${}_2F_1(j + \nu - 1 - i \e, j + \nu + 3 - i \e, 2 j + 2 \nu + 2, \e/z)$, irregular 
behavior occurs when the various arguments are non-positive, as can be observed from the hypergeometric series:
\be
_2F_1(a,b;c;z) = \sum_{k = 0}^{\infty} \frac{(a)_k (b)_k}{(c)_k} \frac{z^k}{k!}.
\ee
The series will start at $\mathcal{O}(\eta^{-4})$ for $j = -l -1$ and at $\mathcal{O}(1)$ otherwise.

The second appearance of ${}_2F_1$ is combined with its multiplicative factors to yield a second function:
\begin{align}
F_2^{lj} = \frac{ \G(-j - \nu - 2 - i \e) \G(2 j + 2 \nu +1)}{\G(j + \nu - i \e + 3)} \l\frac{\e}{z}\r^{-j-\nu-2- i \e}
{}_2F_1(-j - \nu + 2 - i \e,-j - \nu - 2 - i \e, -2 j - 2 \nu , \e/z).
\end{align}
The hypergeometric function here has leading behavior of $\mathcal{O}(\eta^{-4})$ for 
$j = -l$ and of $\mathcal{O}(1)$ otherwise.  

\subsubsection{The full horizon-side homogeneous solution}

The computation of $X^{-}_{lmn}$ follows from these component pieces.  The combined leading behavior is
given in Table \ref{tab:XinOrds}.
\begin{table*}[t]
\caption{Leading powers of $\eta$ in $C_{\rm in} a_j F_1^{lj}$ and $C_{\rm in} a_j F_2^{lj}$ as 
functions of $l$ and $j$}
\label{tab:XinOrds}
\begin{center}
\begin{tabular}{|| c | c | c | c | c | c | c ||}
\hline\hline
 &  $j \le -2l-1$  & $-2l \le j \le -l - 3$ & $j = -l -2$ & $-l -1 \le j \le -l$ & $j = -l+1$ & $j \ge -l +2$ \\
\hline
$C_{\rm in} a_j F_1^{lj}$ & $|j| + 2l - 6$ & $|j|+ 2l$ & $3 |j| - 4$ & $3 l-3$ & $3|j|-1$ & 
\begin{tabular}{@{}c@{}}  $3|j| + 2 j$  \vspace{-.7em} \\   $+ 2l - 3$  \end{tabular} \\
$C_{\rm in} a_j F_2^{lj}$  &  $5|j| -2l - 8$  & $5|j| -2l -2$ & $3 |j| - 4$ & $3 l-3$ & $3|j|-1$ & 
\begin{tabular}{@{}c@{}}  $3|j| - 2j$  \vspace{-.7em} \\   $- 2l - 5$  \end{tabular} \\
\hline \hline
\end{tabular}
\end{center}
\end{table*}
Note that this corrects a few small mistakes in Table~III of \cite{KavaOtteWard15}.  Thus, calculation of $X_{2mn}^-$ 
to, say, $1/\eta$ requires no $j$ for $F_1^{2j}$ and $0 \le n \le 8$ for $F_2^{2j}$.   An
expansion for a normalized version of $X^{-}_{2mn}$ will be given in Sec.~\ref{sec:normXFuncs} below.

\subsection{The normalized functions, $\hat{X}^{+}_{lmn}$ and $\hat{X}^{-}_{lmn}$}
\label{sec:normXFuncs}

The functions $X^{\pm}_{lmn}$ will each have some amplitude at infinity or the horizon
$X^{\pm}_{lmn} \sim A^{\pm}_{lmn} \, e^{\pm i w r_{*}}, \,\, r_* \rightarrow \pm \infty$, where 
$r_{*} = r + 2M \ln(r/2M-1)$ is the tortoise coordinate.  As mentioned in Sec.~\ref{sec:RWZ}, it is advantageous in 
the computation of the fluxes to normalize these functions so that we have 
$\hat{X}^{\pm}_{lmn} \sim e^{\pm i w r_{*}}$ as $r \rightarrow \infty$ or $r \rightarrow 2M$.  
This is done by dividing off the initial amplitudes $A^{\pm}_{lmn}$, found by analyzing the appropriate limits.

Explicitly, the function $X^{+}_{lmn}$ can be normalized by taking the limit $r \rightarrow \infty$ or, equivalently,  
$z \rightarrow \infty$.  Noting that $U(a,b,z)$ limits to $z^{-a}$ as $z \rightarrow \infty$, we find that 
the desired amplitude is given by
\be
A^{+}_{lmn} = (\e)^{i \e} (-2 i)^{-\nu-1+i \e} \sum_{j=-\infty} a_j 
\frac{ \G(j + \nu - 1 - i \e) \G(j + \nu + 1 - i \e) }{\G(j + \nu + 3 + i \e) \G(j + \nu + 1 + i \e) } 
=  (\e)^{i \e} (-2 i)^{-\nu-1+i \e} A_{\rm up}^{\rm sum}.
\ee
$\hat{X}^{+}_{lmn}$ follows by dividing off this amplitude.  Absorbing the amplitude into the prefactor,
we find
\begin{align}
\hat{C}_{\rm up} &= \frac{C_{\rm up}}{A^{+}_{lmn}}
= \frac{e^{iz} (-2 i z)^{\nu+1}}{A_{\rm up}^{\rm sum}} (-2 i \e)^{-i \e} \left(1- \frac{\e}{z}\right)^{-i \e} . \notag
\end{align}
For $l=2$ the expansion for the full normalized homogeneous solution begins
\begin{align}
\label{eqn:Xup2mn}
\hat{X}^{+}_{2mn} &= -\frac{3}{\bar{z}^2 \eta^2}  -  \left(\frac{1}{2}+\frac{5 \bar{\e}}{2 \bar{z}^3}\right)
+  \left(-\frac{5 i \bar{\e}}{\bar{z}^2}+\frac{3 i \bar{\e} \gamma }{\bar{z}^2}+\frac{3 \bar{\e} \pi }{2 \bar{z}^2}
+\frac{3 i \bar{\e} }{\bar{z}^2} \log \left(2 \bar{\e} \eta^3\right) \right)\eta  
+\left(-\frac{15 \bar{\e}^2}{7 \bar{z}^4}-\frac{7 \bar{\e}}{4 \bar{z}}-\frac{\bar{z}^2}{8}\right)\eta^2 \notag \\ &
+\left(-\frac{5 i \bar{\e}}{6} + \frac{i \bar{\e} \gamma }{2}+\frac{\bar{\e} \pi }{4}-\frac{25 i \bar{\e}^2}{6 \bar{z}^3}
+\frac{5 i \bar{\e}^2 \gamma }{2 \bar{z}^3}+\frac{5 \bar{\e}^2 \pi }{4 \bar{z}^3}-\frac{i \bar{z}^3}{15}
+\frac{1}{2} i \bar{\e} \log \left(2 \bar{\e} \eta^3\right)
+\frac{5 i \bar{\e}^2}{2 \bar{z}^3} \log \left(2 \bar{\e} \eta ^3\right)  \right)\eta^3   \notag \\&
+ \bigg(-\frac{15 \bar{\e}^3}{8 \bar{z}^5}+\frac{3757 \bar{\e}^2}{420 \bar{z}^2}    
-\frac{457 \bar{\e}^2 \gamma }{70 \bar{z}^2}+\frac{3 \bar{\e}^2 \gamma^2}{2
\bar{z}^2}+\frac{457 i \bar{\e}^2 \pi }{140 \bar{z}^2}-\frac{3 i \bar{\e}^2 \gamma  \pi }{2 \bar{z}^2} 
-\frac{5 \bar{\e}^2 \pi ^2}{8 \bar{z}^2}-\frac{7 \bar{\e} \bar{z}}{16}+\frac{\bar{z}^4}{48}
-\frac{5 \bar{\e}^2}{\bar{z}^2} \log \left(2 \bar{\e} \eta^3\right)  \notag \\&
+\frac{3 \bar{\e}^2 \gamma }{\bar{z}^2}  \log \left(2 \bar{\e} \eta^3\right) 
-\frac{3 i \bar{\e}^2 \pi }{2 \bar{z}^2} \log \left(2 \bar{\e} \eta^3\right)  +  \frac{3 \bar{\e}^2 }{2 \bar{z}^2}\log^2
\left(2 \bar{\e} \eta^3\right) - \frac{107 \bar{\e}^2}{70 \bar{z}^2} \log (2 \bar{z} \eta) \bigg) \eta^4 
+ \mathcal{O}(\eta^5).
\end{align}
\\

The function $\hat{X}^{-}_{lmn}$, meanwhile, is normalized by taking the limit $r \rightarrow 2M$, which implies 
$z \rightarrow \e$.  Because $_2F_1(a,b,c,1-r/2M)$ limits to $1$ as $r \rightarrow 2M$ for 
any $(a, b, c)$, we find the amplitude
\begin{align}
A^{-}_{lmn} =  \sum_{n=-\infty} a_j  \frac{ \G(n+\nu-1- i\e) \G(-n - \nu - 2 - i \e) }{\G(1-2 i \e) } .
\end{align}
$\hat{X}^{-}_{lmn}$ follows by dividing off this amplitude.  The series for $l=2$ is found to be
\begin{align}
\hat{X}_{2mn}^{-} &= \frac{\bar{z}^3}{\bar{\e}^3 \eta^6}-\frac{\bar{z}^5}{14 \bar{\e}^3 \eta^4}
+\frac{13 i \bar{z}^3}{12 \bar{\e}^2 \eta ^3}+\left(-\frac{13 \bar{z}^4}{42\bar{\e}^2}
+\frac{\bar{z}^7}{504 \bar{\e}^3}\right)\frac{1}{\eta ^2}-\frac{13 i \bar{z}^5}{168 \bar{\e}^2 \eta } \notag \\ &
+ \left(-\frac{95 \bar{z}^3}{48 \bar{\e}}-\frac{\pi ^2 \bar{z}^3}{6 \bar{\e}}+\frac{\bar{z}^6}{54 \bar{\e}^2}-\frac{\bar{z}^9}{33264 \bar{\e}^3}
+\frac{107 \bar{z}^3 }{210 \bar{\e}} \log\left(\frac{\bar{\e}}{\bar{z}} \eta^2 \right)  \right) +
\left(-\frac{169 i \bar{z}^4}{504 \bar{\e}}+\frac{13 i \bar{z}^7}{6048 \bar{\e}^2} \right) \eta   \notag \\ &
+ \left( \frac{319 \bar{z}^2}{420}+\frac{85429 \bar{z}^5}{493920 \bar{\e}}+\frac{\pi ^2 \bar{z}^5}{84 \bar{\e}}
-\frac{53 \bar{z}^8}{118800 \bar{\e}^2} +  \frac{\bar{z}^{11}}{3459456 \bar{\e}^3}
-\frac{107 \bar{z}^5}{2940 \bar{\e}} \log \left(\frac{\bar{\e}}{\bar{z}} \eta^2 \right) \right) \eta^2 + \mathcal{O}(\eta^3).
\end{align}
\end{widetext}

\subsection{Optimizing expansions through $\G$ function identities and factorization}
\label{sec:optExps}

\subsubsection{Rewriting $\G$ functions using Pochhammer symbols}

The procedure detailed above is sufficient to produce PN series; however, the expressions are too 
computationally expensive as written, primarily due to the complexity of the $\G$ functions, which are difficult
to expand when the arguments are arbitrary.  Fortunately, it is possible to reformulate the $\G$ functions slightly 
to construct series in a much more efficient manner.  This is done by first repeatedly applying the standard 
identity $z \G(z) = \G(z+1)$ to put all such functions into the form $\G(1 + g(\e))$ for some small function $g(\e)$ and
then, because the resulting $\G(1 + g(\e))$ expressions can be pulled out of the summations over $j$, 
finding opportunities to cancel or simplify these factors.

Explicitly, we write
\begin{align}
\G(k + g(\e)) &= 
\G(1 + g(\e)) \left( \frac{\G(k + g(\e))}{\G(1 + g(\e))} \right)  \notag \\
& = \G(1 + g(\e)) (1 + g(\e))_{k-1},
\end{align}
where $k$ is some integer and $(a)_n$ is the Pochammer symbol.  In this context, Pochhammer symbol takes one 
of two values, depending on the value of $k$:
\begin{align}
(1 + g(\e))_{k-1} &= \prod_{i=1}^{k-1} (i + g(\e))  \qq   (k \ge 1),  \notag \\
(1 + g(\e))_{k-1} &= \prod_{i=k}^{0} \left( \frac{1}{i + g(\e)} \right)   \hspace{1.1em} (k < 1).
\end{align}
In each case, this yields a purely rational series in $\e$, one which can be rapidly expanded in \textsc{Mathematica}.
Doing this for each $\G$ function in $\hat{X}^{+}_{lmn}$ and $\hat{X}^{-}_{lmn}$ creates significant cancelations 
of $\G$ functions in $A^{\pm}_{lmn}$ with those in $U_1^{lj}, U_2^{lj}, F_1^{lj}, F_2^{lj}$, drastically reducing the
computational cost.  In what follows, we will call the functions that remain after such cancelations
$\bar{U}_1^{lj}, \bar{U}_2^{lj}, \bar{F}_1^{lj}, \bar{F}_2^{lj}$, respectively.

\subsubsection{Factorization}

We can simplify $\hat{X}^{\pm}_{lmn}$ further by preemptively factoring out certain complicated $z$-independent 
terms.  In certain cases these factors will eventually cancel through division by the Wronskian 
\cite{KavaOtteWard15}, but the rest of the time, we will simply multiply these factors back in at the end, after 
$|C^{\pm}_{lmn}|^2$ is constructed for the fluxes.  This serves to accelerate the integral for the normalization 
coefficients $C^{\pm}_{lmn}$ (the rate-limiting step in the expansion of the fluxes) by an order of magnitude.  

To give an immediate example, the expression for $\hat{X}_{lmn}^+$ contains the $z$-independent factor
\be
(-2 i \e)^{-i \e} = \exp\left[-i \e \log(-2 i \e) \right].
\ee
This piece expands into a sequence of logarithms that greatly increases the expression length and computational
cost.  Therefore, this factor is removed from the outset.

More subtly, we can simplify the summations in $X_{lmn}^-$ and $X_{lmn}^+$ by analyzing more closely the leading
behavior of the functions $\bar{U}_1^{lj}, \bar{U}_2^{lj}, \bar{F}_1^{lj}, \bar{F}_2^{lj}$.  
On the horizon side, multiplying in the $(\e/z)^{i\e+1}$ factor from $C_{\rm in}$, we have 
\begin{align}
\left(\frac{\e}{z}\right)^{i\e+1}  \bar{F}_1^{lj} &\propto \left(\frac{\e}{z}\right)^{j+\nu}, \notag \\
\left(\frac{\e}{z}\right)^{i\e+1}  \bar{F}_2^{lj} &\propto \left(\frac{\e}{z}\right)^{-j-\nu-1}.
\end{align}
Note that the latter expression controls the leading behavior $((z/\e)^{l+1}$ for $j=0$), while the former holds an 
additional factor of $(\e/z)^{2 l+1} \propto \eta^{4 l + 2}$.  Therefore, when attempting to reach a given PN order, 
expansions of $\bar{F}_2^{lj}$ must be computed for more $j$ and to higher relative order than for those of 
$\bar{F}_1^{lj}$.

Similarly, in $X_{lmn}^+$ $\bar{U}_2^{lj}$ has an extra factor of $(z^{-l})$ over $\bar{U}_1^{lj}$, though the difference 
there is more modest.  Because the $\bar{F}_1^{lj}$ and $\bar{U}_1^{lj}$ calculations are simpler and fewer in 
number, we can reduce the total computations by ``moving" all the $j$-independent $\G(1 + g(\e))$ functions from 
$\bar{F}_2^{lj}$ and $\bar{U}_2^{lj}$ to $\bar{F}_1^{lj}$ and $\bar{U}_1^{lj}$ via division.  When necessary, these 
factors will be multiplied back in at the end \cite{JohnMcDa14}.

Finally, it is possible to identify one additional simplifying factor: the lowest appearance of each eulerlog-like function
\cite{DamoIyerNaga09, JohnMcDa14, JohnMcDaShahWhit15, MunnETC20, MunnEvan19a}. 
These functions are produced by the leading-order behavior within $\hat{U}_2$ and $\hat{F}_2$ \cite{JohnMcDa14}. 
Upon evaluation at the location of the particle (see Sec.~\ref{sec:InhomExps} below), this adduces the factors
\begin{align}
X^+_{lmn} &: (-i \e p)^{-\D \nu}, \notag \\
X^-_{lmn}  &: \left(\frac{2}{p}\right)^{-\D \nu},
\end{align}
where $\D \nu = \nu - l$.

Once all such quantities are canceled or factored out of the homogeneous solutions, the resulting expansions
are multiple orders of magnitude simpler and faster to execute.  However, when constructed in this manner, the 
$X^{\pm}_{lmn}$ functions are no longer normalized, so we no longer mark them with hats.  The missing factors
will eventually be resupplied in the final construction of the fluxes.  They are given explicitly in 
Sec.~\ref{sec:InhomExps}.

\subsection{The functions $\hat{X}_{\rm up}^{\rm e}$ and $\hat{X}_{\rm in}^{\rm e}$ (even parity)}

The even-parity functions can be found by using the Detweiler-Chandrasekar transformation 
\cite{Chan75, ChanDetw75, Chan83, Bern07}:
\begin{align}
\label{eqn:XevenDC}
X_{\pm}^{\rm even} &= \left(\frac{4}{\la_l \pm 6 i \e}\right) \bigg[ \frac{3 \e}{2} \left(1 - \frac{\e}{z} \right) 
\frac{d X_{\rm \pm}^{\rm odd}}{d z} +   \notag \\ &  \left( \frac{1}{4}\la_l + 
 \frac{9 \e^2 \left(1 - \frac{\e}{z} \right)}{2(l-1)(l+2)z^2 + 6 z \e } \right) X_{\rm \pm}^{\rm odd} \bigg],
\end{align}
where $\la_l = (l-1)l(l+1)(l+2)$.  This transformation is constructed such that whenever the
odd-parity functions are normalized, the even-parity ones will be as well.  This can be checked directly by
taking the appropriate limits.  Thus, the bulk of the expansion procedure remains unchanged in the even-parity case.

\section{Analytic expansion of the normalization constants}
\label{sec:InhomExps}

\subsection{Obtaining PN series for the geodesic motion of the smaller body}
\label{sec:geoExps}

The prescription above allows for expansion of the homogeneous solutions to effectively arbitrary PN order.  These
can be used to construct the normalization constants $C^{\pm}_{lmn}$ defined in Sec.~\ref{sec:InhSoln}.
To do so, we follow and refine the methods of \cite{HoppKavaOtte16}.  The process requires that $X^{\pm}_{lmn}$ 
be evaluated at the location of the particle as it follows a generic, bound geodesic on the Schwarzschild background. 
In order to maintain a consistent PN description of the system, we must PN expand this motion, something that can 
be done to arbitrary order.  

The basic framework for Schwarzschild geodesic motion was described in Sec.~\ref{sec:orbits}.  An alternative 
description of the geodesic orbit known as the Darwin parameterization is much more useful for PN expansions.  
The Darwin parameterization recasts $\mathcal{E}$ and $\mathcal{L}$ in terms of the geometric quantities $p$, the 
(dimensionless) semi-latus rectum, and $e$, the eccentricity \cite{Darw59, CutlKennPois94, BaraSago10}.  These 
are related by 
\be
\label{eqn:defeandp}
{\mathcal{E}}^2 = \frac{(p-2)^2-4e^2}{p(p-3-e^2)},
\qq  
{\mathcal{L}}^2 = \frac{p^2 M^2}{p-3-e^2}.
\ee
Bound orbits now satisfy $p > 6 + 2 e$, with the boundary $p = 6 + 2 e$ representing the separatrix 
\cite{CutlKennPois94}.  

It is of note that $1/p$ is a 1PN quantity, meaning PN series can be equivalently constructed by expanding in terms 
of $1/p$.  We will thus expand the coordinate position of the particle in $1/p$ and eventually use this to expand the 
homogeneous solutions (evaluated at $r_p$) in $1/p$ as well.  This formulation is well suited for the expansions of 
the fluxes, and it will also be used in Sec.~\ref{sec:QKSchw} to derive the relationship between certain BHPT-PN 
expansions in Schwarzschild coordinates and more standard PN expansions in modified harmonic coordinates.  For 
the fluxes series in $e$ will also be made at each order in $1/p$ to make the normalization constants integrable.

The Darwin parameterization also shifts the curve parameter from proper time $\tau$ to 
the relativistic anomaly $\chi$, putting the radial position into the form 
\be
r_p \l \chi \r = \frac{pM}{1+ e \cos \chi} .
\ee
One radial libration makes a change $\Delta\chi = 2\pi$.  The remaining coordinates can be found as functions of
$\chi$ through a set of ordinary differential equations:
\begin{align}
\label{eqn:tPhiODEs}
\frac{dt_p}{d \chi} &= \frac{r_p \l \chi \r^2}{M (p - 2 - 2 e \cos \chi)}
 \left[\frac{(p-2)^2 -4 e^2}{p -6 -2 e \cos \chi} \right]^{1/2} , \notag \\
\frac{d \varphi_p}{d\chi} &= \left[\frac{p}{p - 6 - 2 e \cos \chi}\right]^{1/2} .
\end{align}
There is an analytic solution for the azimuthal motion, 
\be
\label{eqn:vpChi}
\vp_p(\chi) = \left(\frac{4 p}{p - 6 - 2 e}\right)^{1/2} \, 
F\left(\frac{\chi}{2} \, \middle| \, -\frac{4 e}{p - 6 - 2 e}  \right) ,
\ee
where $F(\vp | m)$ is the incomplete elliptic integral of the first kind \cite{GradETC07}.
The time coordinate, meanwhile, is expanded in $1/p$ and $e$ before integrating.  The series begins
\begin{align}
t_p(\chi) &= \left(\chi -2 \sin (\chi ) e + \mathcal{O}\left(e^2\right)\right) p^{3/2}   \\ &
+\left(3 \chi - 3 \sin (\chi ) e + \mathcal{O}\left(e^2\right)\right) p^{1/2} + \mathcal{O}(p^{-1/2}).  \notag
\end{align}

This integration also provides the radial period and frequency:
\begin{align}
\label{eqn:TrOr}
T_r &= \int_{0}^{2 \pi} \l \frac{dt_p}{d\chi} \r d \chi =  t_p(2\pi) - t_p(0) =  \frac{2 \pi}{\O_r}.
\end{align}
The mean azimuthal frequency follows as
\begin{align}
\label{eqn:O_phi}
\O_\varphi = \frac{\vp(2 \pi)}{T_r} =  \frac{4}{T_r} \left(\frac{p}{p - 6 - 2 e}\right)^{1/2} \, 
K\left(-\frac{4 e}{p - 6 - 2 e}  \right) ,
\end{align}
where $K(m)$ is the complete elliptic integral of the first kind \cite{GradETC07}.  Finally, the compactness parameter 
$y$, which is a common (gauge-invariant) post-Newtonian expansion variable, is given by $y = (M \O_\vp)^{2/3}$.  
It is easy to transform any given PN expansion from $1/p$ to $y$ and vice versa.  Therefore, we will work with 
expansions in $1/p$ until the very end.  Expansions in $y$ for the source motion and normalization constants can be 
found in \cite{HoppKavaOtte16}.  Note that the PN series for the coordinates can be 
trivially applied to expand the source terms \eqref{eqn:OPFGsimp} and \eqref{eqn:EPFGsimp}.

\subsection{The $C_{lmn}^{\pm}$ integrals}
\label{sec:ClmnExps}

The inhomogeneous solutions are found by integrating the source motion for the constants $C_{lmn}^{\pm}$.  
This is most conveniently achieved in terms of $\chi$, using
\begin{align}
\label{eqn:ClmnIntChi}
C_{lmn}^{\pm} &= \frac{1}{W_{lmn} T_r} \int_0^{2 \pi} \left(\frac{dt}{d\chi} \right) 
 \bigg[  \frac{1}{f_p}G_{lm}(\chi) X^{\mp}_{lmn}   \\&
+\left( \frac{2 M}{r_p^2 f_p^2}  X^{\mp}_{lmn} - \frac{1}{f_p} \frac{d X^{\mp}_{lmn}}{dr} \right)  F_{lm}(\chi) \bigg] 
e^{i \o t(\chi)} d\chi \notag 
\end{align}
The homogeneous solutions are expressed as functions of $\chi$ by setting 
\begin{align}
z_p &= r_p \o = \frac{p M \o}{1 + e \cos(\chi)} = \frac{M \overline{\o}}{p^{1/2}(1 + e \cos(\chi))}, \notag \\
\e &= 2 M \o = \frac{2 M \overline{\o}}{p^{3/2}},
\end{align}
where we have introduced a PN-adjusted frequency $\overline{\o} = \o p^{3/2} = \mathcal{O}(1)$.  As with $\bar{z}$ 
and $\bar{\e}$ in Sec.~\ref{sec:MSTexp}, the use of the Newtonian-order $\overline{\o}$ implies that every quantity 
within the expansions for $X^{\pm}_{lmn}$ is Newtonian order except for the expansion variable, which in this case is 
$1/p$.  Thus, the PN order will now be tracked with $1/p$ alone, and the previous expansion parameter 
$\eta=1/c$ can be set to 1.  All series are now crafted to use the variables $1/p$ and $e$.  This also allows us to 
avoid evaluating $\overline{\o}$ in terms of $\O_r$ and $\O_\vp$ until the end, which saves computational time.

The last needed quantity is the Wronskian $W_{lmn}$, given by
\be
W_{lmn} = f \frac{d X^{+}_{lmn}}{d r} X^{-}_{lmn} - f \frac{d X^{-}_{lmn}}{d r} X^{+}_{lmn} .
\ee
Interestingly, this quantity is parity-independent.  This can be shown by direct evaluation using the 
Detweiler-Chandrasekar transformation, along with the RW equation and $z$-independence of the result.  

Overall, these integrals constitute the computational bottleneck in this analytic expansion procedure.  When reduced 
entirely to series in $1/p$ and $e$, the result is a large sum of complex exponentials, which are trivial to integrate but 
extremely time-consuming to handle.  However, the simplifications detailed above serve to reduce the size of the 
expanded integrand by multiple orders of magnitude.  This allows the procedure above to reach incredibly high PN 
orders in manageable time.  A representative sample of benchmarks is given in Table \ref{tab:Clmntimes}.

As an example, the expansion for the even-parity $2m1$ mode begins
\begin{align}
\label{eqn:C2m1Exp}
C^+_{2m1} &= \left[\left(\frac{16 \overline{\o}^2}{15}-\frac{8 \overline{\o}^3}{15}\right) e+\mathcal{O}\left(e^2\right)
\right] \frac{1}{p}  +  \bigg[\bigg(-\frac{20 \overline{\o}^2}{9}  \notag \\&
-\frac{8 m \overline{\o}^2}{45} -\frac{4 m^2 \overline{\o}^2}{45}+\frac{4 m^2 \overline{\o}^3}{45}
-\frac{16 \overline{\o}^4}{105} +\frac{16 \overline{\o}^5}{315}\bigg) e   \notag \\&      
+\mathcal{O}\left(e^2\right)\bigg] \frac{1}{p^2} +\bigg[\left(-\frac{136 i \overline{\o}^3}{45}
+\frac{68 i \overline{\o}^4}{45}\right) e   \notag \\&         +\mathcal{O}\left(e^2\right) \bigg] \frac{1}{p^{5/2}}
+\mathcal{O}\left(\frac{1}{p^{7/2}}\right)
\end{align}

\begin{table*}[t]
\begin{center}
\caption{Overview of the computational time needed for expansion of various even-parity normalization constants to 
high PN order.  Expansions were found for specific $l$ but general $m$ and $n$ on the UNC Longleaf cluster.  The 
third and fourth columns indicate the time and memory, respectively, needed for the calculation.  The fifth column 
gives the approximate size of a text file holding the output.  In each case the comparable odd-parity computation
is simpler and faster.  Note that only the infinity-side coefficients are needed for the fluxes at infinity.  Radiation to the 
larger black hole's horizon will be explored in a future paper 
\cite{MunnEvan20b}. \\}
\label{tab:Clmntimes}
\begin{tabular}{|| c | c | c | c | c ||}
\hline\hline
Coefficient & Relative Order & CPU time (hours) & Memory & Text File Size \\
\hline
$C^{+}_{2mn}$ &  19PN/$e^{10}$  &  173.5  &  5GB  &  60MB \\
\hline
$C^{+}_{4mn}$ &  18PN/$e^{10}$  &  41.1   &   4GB   &  15MB  \\
\hline
$C^{+}_{6mn}$ &  16PN/$e^{10}$   &  18.1   &   4GB  &  10MB  \\
\hline
$C^{+}_{2mn}$ &  10PN/$e^{20}$  &   8.2   &    3GB   &  40MB  \\
\hline \hline
\end{tabular}
\end{center}
\end{table*}

\subsection{Construction of the fluxes from the factored normalization constants}

With the (factored) constants $C_{lmn}^{\pm}$ analytically expanded, we can pursue the fluxes with the 
formulas given in Sec.~\ref{sec:fluxFormulas}:
\begin{align}
\bigg\langle\frac{dE}{dt}\bigg\rangle^{\infty} & \Longrightarrow \frac{1}{64 \pi} \sum_{lmn} (l+2)(l+1)(l)(l-1) \o^2 
| C^+_{lmn} |^2, \notag \\
\bigg\langle\frac{dL}{dt}\bigg\rangle^{\infty} & \Longrightarrow \frac{1}{64 \pi} \sum_{lmn} (l+2)(l+1)(l)(l-1) m \o 
| C^+_{lmn} |^2.
\end{align}
However, the flux expressions are still missing the $z$-independent factors that were removed in 
Sec.~\ref{sec:optExps}.  These must be multiplied back in to retrieve the fluxes.  

At infinity, the necessary term comes from the $z$-independent factors removed from $X^+_{lmn}$, 
as this function only appears in the Wronskian.  On the other hand, the $z$-independent factors for $X_{lmn}^-$ 
in $1/W_{lmn}$ will cancel with similar factors in the normalization integral, so those can be ignored.  We get
\begin{align}
\label{eqn:CfacToCflux}
C_{\rm flux}^{+} &= (-2 i \e)^{i \e} (-i \e p)^{\D \nu} \left(\frac{\G(1 + \D \nu - i \e)}{\G(1 + 2 \D\nu)}\right) 
C_{\rm fac}^{+},
\end{align}
where $C_{\rm fac}^{+} $ is the factorized normalization constant, while $C_{\rm flux}^{+}$ is the full constant
utilized in the flux formulas.  Then, the fluxes are found from 
\begin{align}
\label{eqn:CfacToCfluxSq}
| C_{\rm flux}^{+} |^2 &= e^{\pi \e} (\e p)^{2 \D \nu}
\frac{ |\G(1 + \D \nu - i \e) |^2}{\G(1 + 2 \D\nu)^2} | C_{\rm fac}^{+} |^2.
\end{align}
Note that this is identical to Johnson-McDaniel's $S_{lmn}$ factorization \cite{JohnMcDa14, MunnEvan19a}.  Similar 
factors appear in the fluxes at the larger black hole's horizon.  These will be described in a future paper 
\cite{MunnEvan20b}

The flux modes have different starting orders in $1/p$ and $e$.  
Specifically, mode $lmn$ will begin at relative PN order $l-1$ in the odd-parity sector and $l-2$ in the even parity 
sector.  The eccentricity series will begin at $e^{2|n|}$ in either case.  Therefore, once target orders are established, 
the exact (finite) number of required modes can be determined.  Computations can be 
separately made and stored for specific modes, which is a fast process on supercomputing clusters.  In 
practice, this generally works by making 2 full computations for each value of $l$ (one for each parity) while 
leaving $m$ and $n$ general until the end.  Then, the resulting contributions can be summed over $l$, $m$, and 
$n$ in straightforward fashion.

\section{The energy and angular momentum flux expansions}
\label{sec:fluxExps}

\subsection{Form of the expansions and past work}

When the expansions are completed, we find that the energy flux at infinity for eccentric-orbit 
Schwarzschild EMRIs can be written in the following form \cite{Blan14, Fuji12a, Fuji12b},
\begin{widetext}
\begin{align}
\label{eqn:energyfluxInf}
\left\langle \frac{dE}{dt} \right\rangle^\infty =&  
\frac{32}{5} \left(\frac{\mu}{M} \right)^2 y^5
\biggl[\mathcal{L}_0 + y\mathcal{L}_1
+y^{3/2}\mathcal{L}_{3/2}+y^2\mathcal{L}_2+y^{5/2}\mathcal{L}_{5/2}
+ y^3\left(\mathcal{L}_3 + \log(y)\mathcal{L}_{3L} \right)
+\mathcal{L}_{7/2}y^{7/2} +y^4\Bigl(\mathcal{L}_4  \notag\\&
+\log(y)\mathcal{L}_{4L}\Bigr)
+y^{9/2}\Bigl(\mathcal{L}_{9/2}+\log(y)\mathcal{L}_{9/2L}\Bigr)
+y^5\Bigl(\mathcal{L}_5
+\log(y)\mathcal{L}_{5L}\Bigr)
+y^{11/2}\Bigl(\mathcal{L}_{11/2}  \notag\\& 
+\log(y)\mathcal{L}_{11/2L}\Bigr)
+ y^6\Bigl(\mathcal{L}_6 + \log(y)\mathcal{L}_{6L}
+ \log^2(y)\mathcal{L}_{6L2} \Bigr)
+y^{13/2}\Bigl(\mathcal{L}_{13/2}+\log(y)\mathcal{L}_{13/2L}\Bigr)  +\cdots  \biggr],
\end{align}
where each PN term $\mathcal{L}_i = \mathcal{L}_i(e)$ is a general function of $e$.  The angular momentum flux 
has a nearly identical form \cite{MunnETC20}:
\begin{align}
\label{eqn:angmomfluxInf}
\left\langle \frac{dL}{dt} \right\rangle^\infty =&  
\frac{32}{5} \left(\frac{\mu^2}{M} \right) y^{7/2}
\biggl[\mathcal{J}_0 + y\mathcal{J}_1
+y^{3/2}\mathcal{J}_{3/2}+y^2\mathcal{J}_2+y^{5/2}\mathcal{J}_{5/2}
+ y^3\left(\mathcal{J}_3 + \log(y)\mathcal{J}_{3L} \right)
+\mathcal{J}_{7/2}y^{7/2} +y^4\Bigl(\mathcal{J}_4  \notag\\&
+\log(y)\mathcal{J}_{4L}\Bigr)
+y^{9/2}\Bigl(\mathcal{J}_{9/2}+\log(y)\mathcal{J}_{9/2L}\Bigr)
+y^5\Bigl(\mathcal{J}_5
+\log(y)\mathcal{J}_{5L}\Bigr)
+y^{11/2}\Bigl(\mathcal{J}_{11/2}  \notag\\& 
+\log(y)\mathcal{J}_{11/2L}\Bigr)
+ y^6\Bigl(\mathcal{J}_6 + \log(y)\mathcal{J}_{6L}
+ \log^2(y)\mathcal{J}_{6L2} \Bigr)
+y^{13/2}\Bigl(\mathcal{J}_{13/2}+\log(y)\mathcal{J}_{13/2L}\Bigr)  +\cdots  \biggr].
\end{align}
The $\mathcal{J}$ functions are similar in structure to their $\mathcal{L}$ counterparts, and all computations in this 
paper were made equally for both; therefore, from this point we primarily discuss the energy case but 
emphasize that the angular momentum is exactly analogous.

It is important to note that because the techniques described in Sections~\ref{sec:MSTexp} and \ref{sec:InhomExps} 
require expansion in $e$ at each PN order, each $\mathcal{L}_i(e)$ will be computed as a Taylor series about 
$e=0$.  However, in principle these flux terms can often be written as more compact functions of $e$.  Indeed, the 
full PN theory using the multipolar post-Minkowskian (MPM) PN formalism yields PN terms as simpler expressions 
involving source multipole moments \cite{Blan14}.  These can usually be evaluated to obtain any given PN term as
either a (closed-form) rational function, or as a compact Fourier summation that can be expanded to high order in 
$e$ \cite{ArunETC08a, MunnEvan19a, MunnEvan19b}.  For example, $\mathcal{L}_0$ and $\mathcal{L}_1$ can 
be found via PN theory to be \cite{PeteMath63, WagoWill76, ForsEvanHopp16}.
\begin{align}
\mathcal{L}_0 &= \frac{1}{(1-e^2)^{7/2}}
{\left(1+\frac{73}{24} e^2 + \frac{37}{96} e^4\right)}, \\
\mathcal{L}_1 &= \frac{1}{(1-e^2)^{9/2}}
\left(-\frac{1247}{336}-\frac{15901}{672} e^2-\frac{9253}{384} e^4 - \frac{4037}{1792} e^6 \right).
\end{align}
\end{widetext}
Interestingly, these closed forms for $\mathcal{L}_0$ and $\mathcal{L}_1$ can be extracted from their corresponding 
Taylor series simply by pulling out the initial eccentricity singular factors.  Eccentric singularities like these occur in 
all PN terms, though most do not reveal rational functions like $\mathcal{L}_0$ and $\mathcal{L}_1$ (see 
\cite{ForsEvanHopp16,MunnEvan19a,MunnETC20} for more details).  As a result, once the series in $e$ are 
found for each $\mathcal{L}_i$ using BHPT, we use knowledge from PN theory to resum the expansions 
in $e$ to improve convergence and, when possible, extract closed forms that would otherwise be much more
difficult to derive through PN theory alone \cite{MunnEvan19a, MunnEvan19b}.
\\

The expansions computed in this paper extend a recent sequence of advances on the eccentric-orbit fluxes.  In 
2009 Arun et al. completed derivation of the energy and angular momentum fluxes to 3PN for arbitrary-mass-ratio
binaries \cite{ArunETC08a, ArunETC08b, ArunETC09a}, continuing the work of \cite{PeteMath63, Pete64, 
WagoWill76, BlanScha93, BlanDamoIyer95, Blan96}.  Those efforts revealed that 
$\mathcal{L}_0, \mathcal{L}_1, \mathcal{L}_2, \mathcal{L}_{3L}$ all have closed forms.  
The remaining terms $\mathcal{L}_{3/2}, \mathcal{L}_{5/2}, \mathcal{L}_{3}$ do not, but the use of computational 
techniques laid out in \cite{ArunETC08a, ForsEvanHopp16, MunnETC20, MunnEvan19a, MunnEvan19b} 
permits their expansion to arbitrary order in $e$.  The angular momentum case is identical in form.

Beyond 3PN order, explicit eccentricity expansions have primarily been calculated using BHPT.  This was first 
pursued in 2016 in \cite{ForsEvanHopp16, Fors16}, which extracted coefficients in the flux expansions using a 
numeric-analytic fitting procedure.  Broadly speaking this worked as follows: First, full numeric BHPT fluxes were 
computed for a two-dimensional grid of orbits covering roughly 50 choices of $p$ and 35 choices of $e$ 
($\sim$1750 total orbits).  Then, these numeric results were fit to the double series.  By computing this fit to high 
precision (100s of significant digits), the authors were able in certain cases to determine analytic forms for the 
coefficients by applying an integer relation algorithm like PSLQ \cite{FergBailArno99}.  The result was the extraction
of varying numbers of new eccentricity coefficients in the two fluxes through 7PN.

More recently, the authors of \cite{MunnETC20} repeated and improved this endeavor by instead fitting the individual 
$lmn$ modes of the fluxes.  These modes are characterized by certain structures that simplify the fitting process and
greatly increase the output.  This permitted the extraction of many more eccentricity coefficients from 3.5PN to 9PN 
in both the energy and angular momentum regimes.  See \cite{MunnETC20} for additional details.

Finally, work in \cite{MunnEvan19a, MunnEvan19b} used complementary discoveries from BHPT and PN theory to 
find convenient forms for certain infinite sets of logarithmic terms in the fluxes.  In particular, closed-form eccentricity 
series were discovered for all flux terms of the form $\mathcal{L}_{(3k)L(k)}$ and $\mathcal{L}_{(3k+1)L(k)}$ for 
integers $k \ge 0$.  Simultaneously, methods were derived to determine to arbitrary order in $e$ all flux terms of the 
form $\mathcal{L}_{(3k+3/2)L(k)}$, $\mathcal{L}_{(3k+5/2)L(k)}$, $\mathcal{L}_{(3k+3)L(k)}$ and 
$\mathcal{L}_{(3k+4)L(k)}$ for $k \ge 0$.  From those, members of the first two sets can be computed to 
arbitrary order in $e$ immediately, while members of the second two sets require lengthy pre-computations using 
BHPT.  Additional simplifications were made in the sets $\mathcal{L}_{(3k+9/2)L(k)}$ and 
$\mathcal{L}_{(3k+11/2)L(k)}$.  The sets $\mathcal{L}_{(3k)L(k)}$ and $\mathcal{L}_{(3k+3/2)L(k)}$ are collectively 
referred to as the leading logarithm series \cite{GoldRoss10,MunnEvan19a}, and $\mathcal{L}_{(3k+1)L(k)}$ and 
$\mathcal{L}_{(3k+5/2)L(k)}$ as the 1PN logarithm series \cite{MunnEvan19b}.  $\mathcal{L}_{(3k+3)L(k)}$ and 
$\mathcal{L}_{(3k+9/2)L(k)}$ form the subleading or 3PN logarithm series, while $\mathcal{L}_{(3k+4)L(k)}$ and 
$\mathcal{L}_{(3k+11/2)L(k)}$) form the 4PN logarithm series \cite{MunnEvan19a, MunnEvan19b}.

The various past results for eccentric-orbit EMRI flux expansions are summarized and compared to the present
work in Table~\ref{tab:resTab}.  Of course, essentially all the energy flux terms in both this and past work 
were derived with an angular momentum counterpart, usually to the exact same order in $e$.  

\begin{table*}[htb]
\caption{Overview of past and present work on EMRI flux expansions through 19PN.  Terms from 0PN to 3PN were 
derived using the full PN theory.  The rest were found by \cite{ForsEvanHopp16} (``FEH16"), \cite{MunnETC20} 
(``MEHF20"), \cite{MunnEvan19a} (``ME19"), \cite{MunnEvan19b} (``ME20"), and the present work.  Boxes 
in the body of the table indicate the order in eccentricity extracted in the listed paper.  Boxes labeled ``CF" were
found in closed form, while those labeled ``AO" can be rapidly computed to arbitrary order.  Those labeled ``AO*" can
be found to arbitrary order only after (yet to be completed) lengthy pre-computations are made using BHPT.  
The columns labeled ``Max" take the highest power of $e$ found among all given sources.   A
comparable chart can be constructed for the angular momentum flux. }
\label{tab:resTab}
\begin{center}
\begin{tabular}{ || c | c | c | c | c | c | c || c | c | c | c | c | c | c ||}
\hline\hline
Term             & FEH16 & MEHF20 & ME19 & ME20 & This & Max & Term & FEH16 & MEHF20 & ME19 & ME20 & This & Max \\
\hline
$\mathcal{L}_{7/2}$ & $e^{24}$ & $e^{30}$ & --- & --- & $e^{20}$ &  $e^{30}$ &  $\mathcal{L}_{7L2}$ &  $e^2$  &  CF  & --- & CF & CF & CF \\
\hline
$\mathcal{L}_{4}$ &  $e^6$  &  $e^{30}$ & --- & AO & $e^{20}$ & AO & $\mathcal{L}_{15/2}$    & ---   &   $e^{12}$   & --- & --- & $e^{20}$ & $e^{20}$  \\
\hline
$\mathcal{L}_{4L}$ & CF &  CF  & --- & CF & CF  & CF &  $\mathcal{L}_{15/2L}$   & ---  &  $e^{26}$  & --- & --- &  $e^{20}$  &  $e^{26}$ \\
\hline
$\mathcal{L}_{9/2}$ & $e^2$ & $e^{30}$ & --- & --- & $e^{20}$ & $e^{30}$  &  $\mathcal{L}_{15/2L2}$   & --- &  $e^{28}$  & AO & --- & $e^{20}$ &  AO \\
\hline
$\mathcal{L}_{9/2L}$ & $e^{18}$  &  $e^{30}$ & AO & --- & $e^{20}$  &  AO &  $\mathcal{L}_{8}$   & --- &  $e^0$ & --- & --- & $e^{20}$ & $e^{20}$ \\
\hline
$\mathcal{L}_{5}$ &  $e^0$  & $e^{30}$  & --- & --- &  $e^{20}$ &  $e^{30}$ &  $\mathcal{L}_{8L}$   & --- &  $e^{18}$ & --- & --- & $e^{20}$  &  $e^{20}$ \\
\hline
$\mathcal{L}_{5L}$ & $e^{24}$  & CF & --- & --- & $e^{20}$ & CF &  $\mathcal{L}_{8L2}$   &  --- &  CF & --- & --- & $e^{20}$  &  CF  \\
\hline
$\mathcal{L}_{11/2}$ &  $e^2$ & $e^{30}$ & --- & --- &  $e^{20}$  & $e^{30}$ &   $\mathcal{L}_{17/2}$   & --- &  $e^2$ & --- & --- & $e^{20}$  &  $e^{20}$  \\
\hline
$\mathcal{L}_{11/2L}$ &  $e^{10}$  & $e^{30}$  & --- & AO & $e^{20}$ & AO &  $\mathcal{L}_{17/2L}$   & --- & $e^{16} $ & --- & --- & $e^{20}$  &  $e^{20}$ \\
\hline
$\mathcal{L}_{6}$ & $e^0$ & $e^{20}$ & --- & --- & $e^{20}$  &  $e^{20}$ &  $\mathcal{L}_{17/2L2}$   & --- &  $e^{20}$  & --- & AO & $e^{20}$  &  AO  \\
\hline
$\mathcal{L}_{6L}$ &  $e^{2}$  & $e^{30}$ & AO & --- & $e^{20}$ &  AO  &  $\mathcal{L}_{9}$   & --- & --- & --- & --- & $e^{20}$  &  $e^{20}$  \\
\hline
$\mathcal{L}_{6L2}$ & $e^{12}$  &  CF  & CF & --- &  CF  & CF &  $\mathcal{L}_{9L}$ & --- & --- & --- & --- & $e^{20}$  &  $e^{20}$  \\
\hline
$\mathcal{L}_{13/2}$ & $e^0$ & $e^{30}$ & --- & --- & $e^{20}$  &  $e^{30}$ &  $\mathcal{L}_{9L2}$  & --- & --- & AO* & --- & $e^{20}$  &  $e^{20}$ \\
\hline
$\mathcal{L}_{13/2L}$ & $e^2$  & $e^{30}$ & --- & --- & $e^{20}$  &  $e^{30}$ &  $\mathcal{L}_{9L3}$  & --- & CF  & CF & --- & CF &  CF  \\
\hline
$\mathcal{L}_{7}$ &  $e^0$  &   $e^{12}$  & --- & --- & $e^{20}$  &  $e^{20}$ & 9.5-10PN & --- &  --- & --- & --- & $e^{20}$  &  $e^{20}$ \\
\hline
$\mathcal{L}_{7L}$ &  $e^2$   &  $e^{26}$  & --- & AO* & $e^{20}$ & $e^{26}$  & 10.5-19PN & --- & --- & --- & --- &  $e^{10}$  &  $e^{10}$  \\
\hline \hline
\end{tabular}
\end{center}
\end{table*}

\subsection{Analytic expansion results for the fluxes}
\label{sec:FluxExpRes}

With previous efforts as a guide, the analytic expansion methods above were used to compute the two fluxes to
high PN order, extending the low-PN high-$e$ results of \cite{MunnETC20, MunnEvan19a, MunnEvan19b} 
to 19PN and $e^{10}$.  Note that because the orders in $y$ and $e$ must be fixed at the beginning of the 
procedure, it is not possible to obtain any individual terms to higher PN order as was possible with 
fitting \cite{MunnETC20}.  However, what can be done is the execution of the entire procedure multiple times in 
order to retrieve low-PN terms to higher order in $e$.  Therefore, in addition to obtaining the fluxes to 19PN and 
$e^{10}$, we also calculated them to 10PN and $e^{20}$. 

In total, all PN terms 10PN and below are now known to at least $e^{20}$, and all PN terms from 10.5PN to 20PN 
are known to at least $e^{10}$.  However, for many flux terms, particularly at low PN, $e$ power series computed in 
previous works remain the state of the art.  An optimal expansion can be formed by selecting the highest power of 
$e$ found at each order.  This is summarized in Table~\ref{tab:resTab}.  

It is interesting to evaluate the relative strengths of fitting and direct analytic expansions, two very different 
approaches to computing BHPT-PN series.  In particular, the fitting approach is particularly adept at reaching high 
orders in eccentricity but is computationally expensive and limited to fairly low PN order.  In contrast, the 
direct analytic method has some trouble calculating arbitrary orders in $e,$ but it is versatile and excellent at moving 
to high PN.  Thus, in some sense the two methods are complementary.  However, due to the known need for 
high-PN expressions, and the ability to still reach useful order in $e$, the analytic expansion techniques will 
likely be the preferred avenue in reproducing these results for other BHPT quantities (especially in the Kerr case), 
outside of a few niche scenarios. 

Explicit terms in the fluxes at infinity, and illustrations of the structure contained therein, are discussed at length in 
\cite{MunnETC20}.  Coefficients grow combinatorially in size with PN order, involving increasingly large combinations 
of transcendental numbers; therefore, we forego enumeration of higher-order analytic coefficients here.  The full 
series are all provided at \cite{BHPTK18} for convenient retrieval.  Instead, comparisons to numerical data are given 
below, allowing for assessment of the utility of these expansions.  

\subsection{Comparison to numerical calculations and convergence of the eccentric expansion}
\label{sec:numComp}

\subsubsection{Mode flux comparisons}

\begin{figure*}
\hspace{-1.5em}\includegraphics[scale=.71]{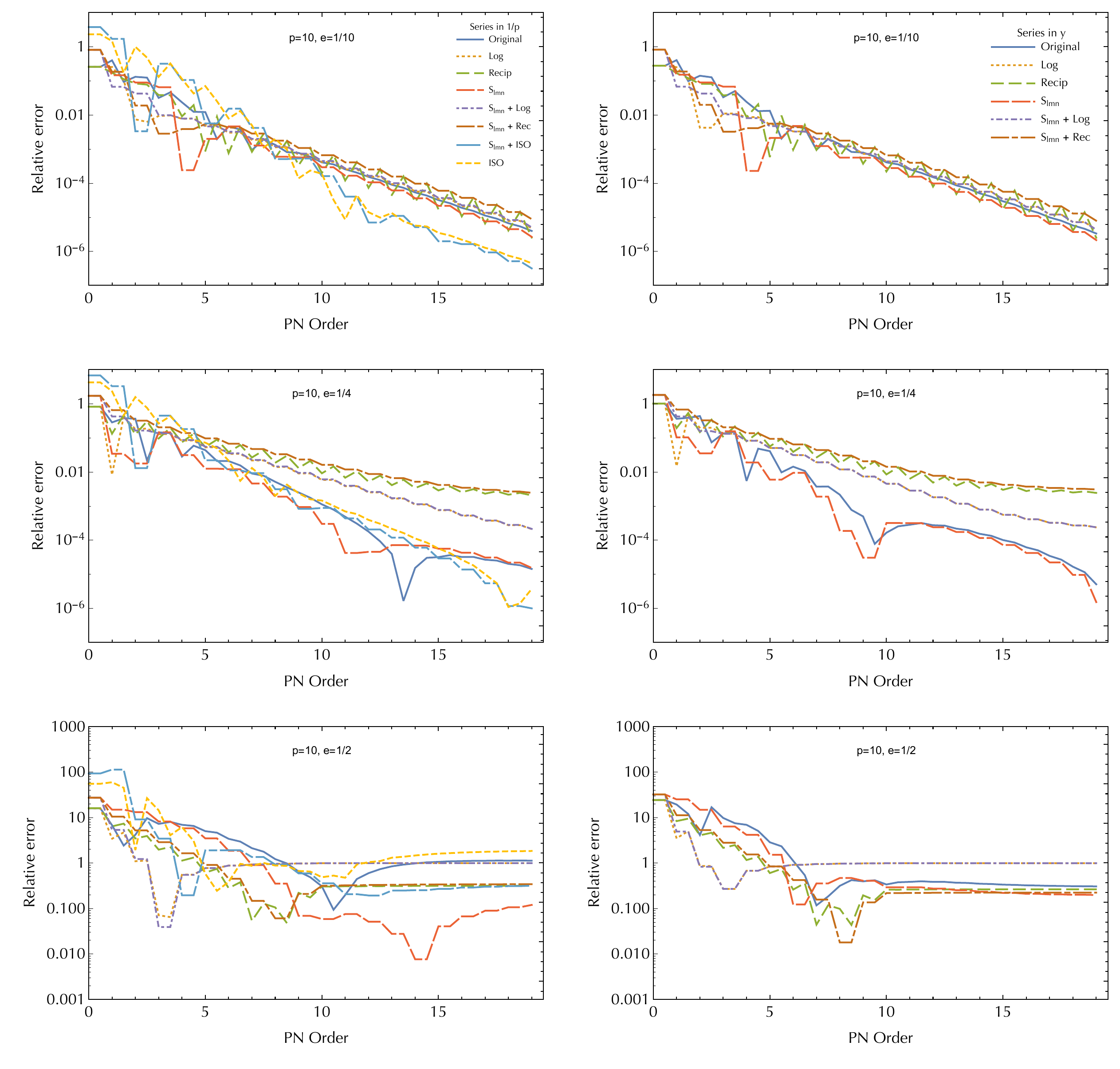}
\caption{Accuracy of the composite energy flux PN expansion and its resummations for the 220 mode for $p=10$.  
The left column plots expansions in $1/p$ and $e$, while the right column plots their analogous expansions in $y$ 
and $e$.  The $x$-axis denotes truncation of the series at the given PN order.  Factorization 
schemes include logarithmic and reciprocal re-expansions, with and without removal of the $S_{220}$ factor.  The 
$1/p$ expansion also includes re-expansion via the removal of the separatrix factor $1/(p-6-2e)$, labeled as ``ISO" 
or ``innermost stable orbit."  Note the change in vertical scaling for $e=1/2$.
\label{fig:220p10}}
\end{figure*}

\begin{figure*}
\hspace{-2em}\includegraphics[scale=.71]{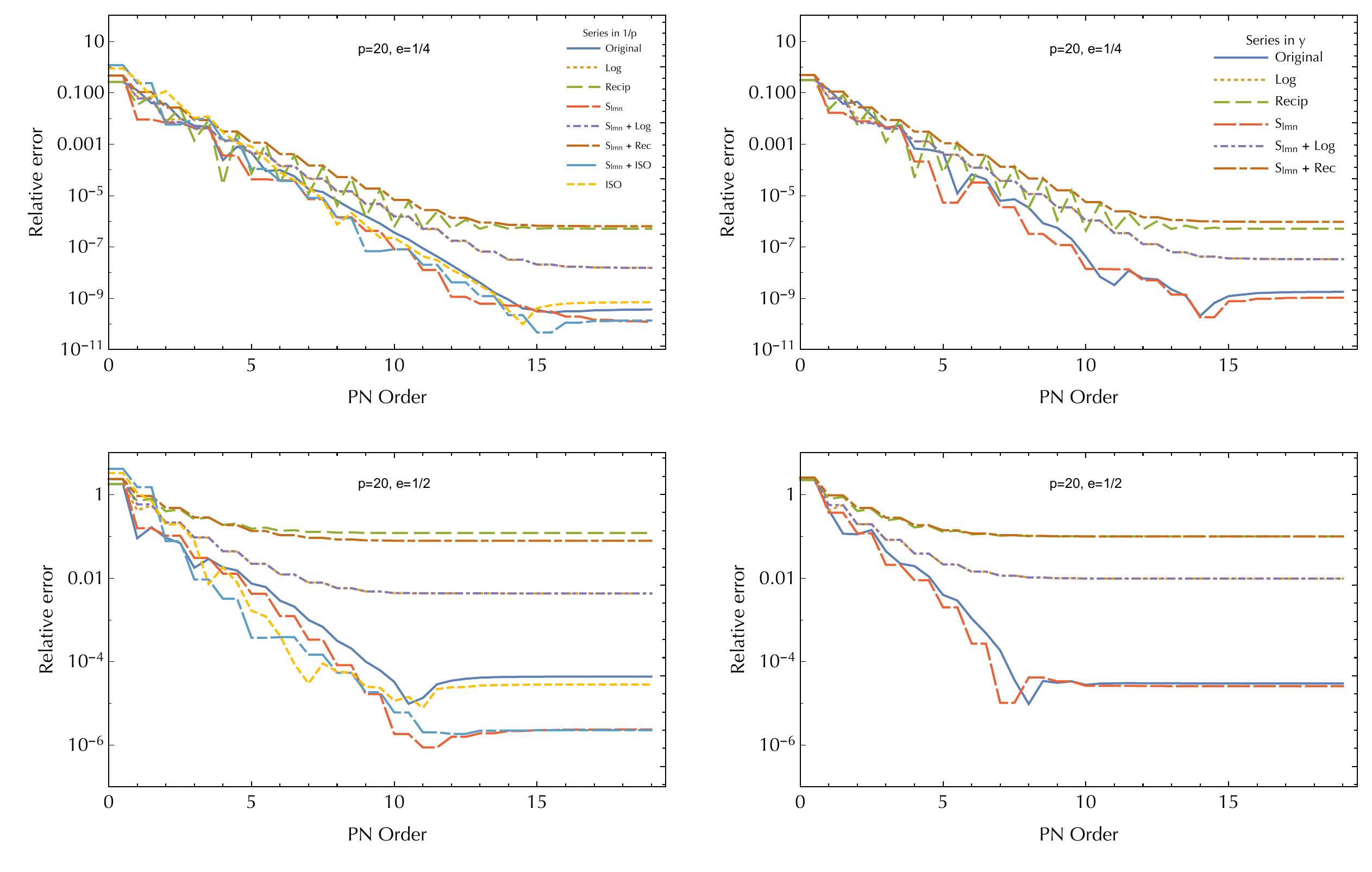}
\caption{Accuracy of the PN expansion and its resummations for the 220 mode for $p=20$.  The various labels and
factorization schemes are identical to those in Fig.~\ref{fig:220p10}.  Note the change in vertical scaling for $e=1/2$.
\label{fig:220p20}}
\end{figure*}

\begin{figure*}
\hspace{-1.5em}\includegraphics[scale=.71]{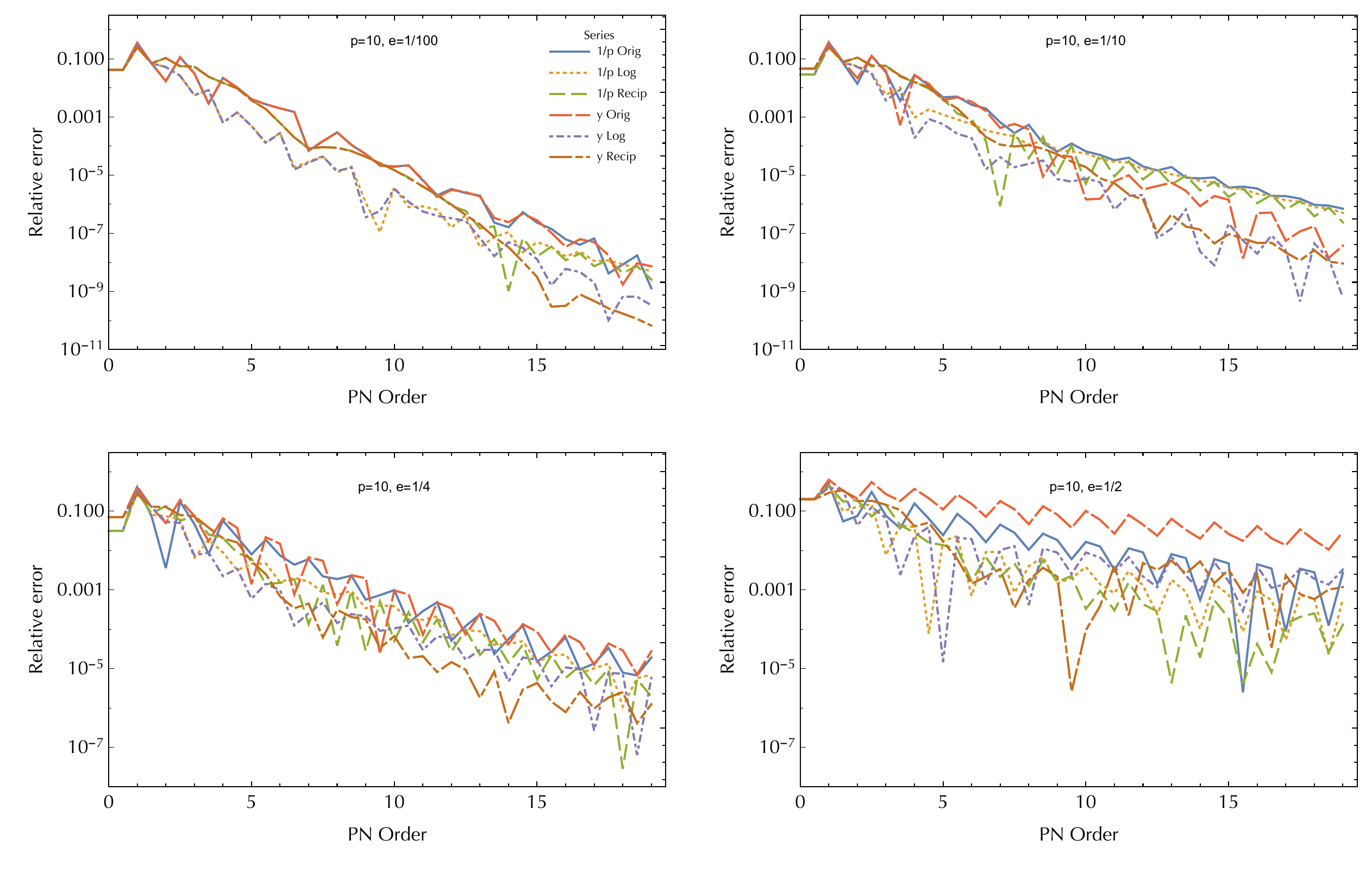}
\caption{Accuracy of the PN expansion and its resummations for the full flux for $p=10$.  Each plot corresponds to
a different value of $e$.  This time, in contrast to Figures~\ref{fig:220p10} and \ref{fig:220p20}, the $1/p$ and $y$ 
expansions are superimposed on the same plots.  Note the change in vertical scaling for the bottom two plots.
\label{fig:fullp10}}
\end{figure*}

With the high-order expansions computed, it is beneficial to assess their utility by comparing to numerical 
calculations for several specific orbits.  This is done in a few separate ways to evaluate the possibility of enhancing
convergence using factorization techniques.  Previous work on factorizations has primarily applied them on a 
mode-by-mode basis \cite{DamoIyerNaga09, JohnMcDa14, NagaShah16, MessNaga17, NagaETC19}.  
Therefore, we start by making comparisons for the individual 220 mode, proportional to $|C_{220}^+|^2$.  
Unfortunately, when working in this manner, low-order results and information from PN theory cannot be 
readily included.  Therefore, we utilize composite expansions constructed by joining only the 10PN/$e^{20}$ and 
19PN/$e^{10}$ results of this paper.  We do this for the $1/p$ expansion natural to BHPT, as well as the more 
standard expansion in $y$.

We then apply to these composite series several factorization schemes to check for improved convergence.  
Specifically, we try a logarithmic resummation (also referred to as the exponential resummation), in which a new
series is constructed from the logarithm of the flux, and then the numeric evaluation of the log series is 
exponentiated to obtain the result \cite{IsoyETC13, JohnMcDa14}.  Similar procedures are executed with a reciprocal 
resummation (inspired by \cite{NagaShah16}) and a singular factor resummation, the latter resulting from the 
removal of the separatrix $1/(p-6-2e)$ in the $1/p$ fit.  We also test the benefit of the $S_{lmn}$ factorization (see 
\eqref{eqn:CfacToCfluxSq} and \cite{JohnMcDa14}), both with and without the other resummations.  Note that the 
application of the factorizations here result in the generation of new double expansions (in PN and $e$).  In the 
case of the full flux analyzed below, resummations will only be applied at the PN level, with the eccentricity 
functions first evaluated numerically.  

Comparisons are made for $p=\{10, 20\}$, $e=\{1/10,1/4,1/2\}$, with the results summarized in Fig.~\ref{fig:220p10} 
and Fig.~\ref{fig:220p20}.  We find that the logarithmic and reciprocal factorization schemes begin to fail at relatively 
low $e$, implying that these approaches are likely not useful for eccentric binaries on an $lmn$ basis.  Additionally, 
the $S_{lmn}$ factorization seems to have little effect in the majority of cases, with close overlap between the 
$S_{lmn}$ and standard varieties of the resummation schemes.  However, it does provide noticeable benefit for the 
orbit $p=20, e=1/2$.  

It is noteworthy that the fit in $1/p$ seems consistently better than the fit in $y$.  This is particularly true in the 
low-$p$, low-$e$ regime, where the removal of the separatrix produces the best match.  Interestingly, though this 
separatrix (``ISO") factorization barely changes the series, it provides clear benefit for $p=10$ and $e=\{1/10, 1/4\}$, 
allowing for relative errors near $10^{-6}$.  A few other methods not depicted were tried as well (e.g., the 
$\tilde{\tilde{S}}_{lmn}$ factorization \cite{JohnMcDa14}), but none provided additional improvement.

Unfortunately, it is clear for $p=10$ that the PN approximation for the 220 mode rapidly loses validity beyond 
$e=1/4$, as the best matching series at $e=1/2$ produced by the $S_{220}$ factorization still yields $1\%$ error 
(with the rest much worse than that).  Better resummations and higher order series in $e$ will be 
required to produce faithful representations of the $lmn$ fluxes for $p \lesssim 10$ around this level.  However, the 
fidelity is markedly improved further into the PN regime, as the smallest relative error achieved for $p=20$ 
and $e=1/2$ is still near $10^{-6}$, as seen in Fig.~\ref{fig:220p20}.

We can roughly assess how the radius of convergence of this double series changes with $e$ by evaluating each
non-logarithmic PN coefficient numerically.  This leaves a single expansion in $y$ (or $1/p$) with coefficients 
$\mathcal{L}_n^{220}(e)$ (or something similar for the $1/p$ expansion).  The radius of convergence is given by 
$\lim_{n \rightarrow \infty} (\mathcal{L}_n^{220}(e))^{-1/n}$ \cite{JohnMcDaShahWhit15}.  For $e=0$, the high-order 
coefficients stabilize at a level that implies a minimal valid semi-latus rectum around $3 \lesssim p \lesssim 4$.  We 
find this rises to $p \sim 5$ for $1/10 \lesssim e \lesssim 1/4$, to $p \sim 6$ for $e = 1/2$, and to $p \sim 10$ at 
$e = 1$.  Of course, these numbers are very approximate, as the high-order PN terms are only expanded to 
$e^{10}$.  Nevertheless, a significant decrease in convergence with $e$ is apparent.

\subsubsection{Full flux comparisons}

For the full flux, we make comparisons using a composite PN series formed from four sources: results from PN 
theory through 3PN (involving closed forms or high-order $e$ expansions), an expansion to $e^{30}$ at 3.5PN (from 
fitting), expansions to $e^{20}$ from 4PN-10PN, and expansions to $e^{10}$ from 10.5PN-20PN.  We again 
construct two separate series in this fashion, one using $1/p$ as the PN variable and the other using $y$.  At each 
PN order eccentricity factors of $(1-e^2)^k$ for some appropriate $k$ are isolated to improve convergence.

We then apply to these composite series similar factorization methods to check for improved convergence.  
This time, the factorizations are only applied at the PN expansion level, meaning that the eccentricity functions are 
evaluated numerically before the re-expansion is executed.  This more easily preserves the closed forms and 
high-order expansions at low PN. 

Comparisons are made for $p=10$, $e=\{1/100, 1/10, 1/4, 1/2\}$, as depicted in Fig.~\ref{fig:fullp10}.  We find that 
the convergence is consistently better in the full-flux expansion than it was in the 220 mode, with the lowest error
reaching $10^{-7}$ for $e=1/4$ and $10^{-5}$ for $e=1/2$.  This is almost surely due to the use of closed forms 
and arbitrary-order expansions through 3PN, as well as the resummation of the eccentricity series at higher orders.
It is noteworthy that the 4PN flux is already known to arbitrary order \cite{MunnEvan19b} while the 3.5PN flux is not, 
implying that a higher-order expansion for the latter would be desirable in moving further into the high-$e$, low-$p$
regime.  

There was not much consistency on the best expansion form across the four orbits.  The $y$ expansions generally 
appear better than their $1/p$ counterparts at lower $e$, while the reverse seems to occur at higher $e$.  The two 
factorizations do not affect the convergence of the $1/p$ expansions at low $e$, but both provide clear benefit 
at $e=1/4$ and $e=1/2$.  In contrast, the $y$ expansion resummations prove better than the original in all 4 cases,
though the difference is fairly modest.  From this small sample of orbits, we can potentially speculate that the 
reciprocal and logarithmic factorizations of the $y$ series provide the best match for small $e$, while the 
reciprocal resummation of the $1/p$ series may begin to outpace those as $e$ increases. 

Despite the overall improved match over the 220 mode, the radius of convergence estimated through high-order 
coefficient magnitude appears worse in the full flux.  The same procedure used in the mode flux reveals a minimally
convergent $p \sim 4$ for $e=0$.  The eccentric cases yield $(e = 1/10, p \sim 5), (e = 1/4, p \sim 6), 
(e = 1/2, p \sim 8)$.  Again, the low order of the eccentric expansions implies that these results are highly imprecise.  
However, this is sufficient to infer that the PN expansion loses strong-field validity in the high-eccentricity regime.  
Thus, it appears unlikely that BHPT-PN expansions can replace numeric calculations at the separatrix for highly 
eccentric fluxes.  However, additional improvements are still possible through higher-order expansions.  Note that 
even at $e=1/2$, there is steady average improvement with increasing PN order in the full-flux expansion in 
Fig.~\ref{fig:fullp10}.  Thus, it will likely prove worthwhile to extend these series further and to continue to refine 
methods of factorization (perhaps by using Pad{\'e} or Chebyshev approximants).  Such explorations will 
be left to future work.

\section{Representation of EMRI expansions in harmonic gauge}
\label{sec:QKSchw}

\subsection{Gauge dependence of the flux expansions and the quasi-Keplerian formalism}
\label{sec:gauge}

The previous sections detailed high-order PN series for the energy and angular momentum radiated to infinity
by eccentric-orbit EMRIs.  These expansions were derived from first-order BHPT using the RWZ formalism, 
which involves the use of Schwarzschild-RW coordinates.  In particular, even though the fluxes themselves are 
gauge-invariant, the quantities $p$ and $e$ are defined within the Darwin parameterization of Schwarzschild 
coordinates.  Thus, the standard representation of BHPT flux expansions (and of all similar expansions) 
is dependent on those coordinates.

On the other hand, expansions found using the full PN theory are frequently derived in modified harmonic gauge, 
using quasi-Keplerian (QK) parameters like the so called time eccentricity $e_t$, whose definition will be given 
below.  It is possible to transform the fluxes from harmonic to Schwarzschild parameters by finding a PN expansion 
for $e_t$ in terms of $e$ (and vice versa).  One way to relate $e_t$ to $e$ is to compute the expansion of each in 
terms of gauge-invariant quantities like $\ve$ and $j$ (related to the energy and angular momentum, see below) and 
then compare.  In general, this can be done for $e_t$ only to the same PN order as the equations of motion, which 
have recently been completed to 4PN order \cite{MarcETC18}, though the expansion for $e_t$ has only been 
published explicitly to 3PN \cite{MemmGopaScha04, ArunETC08a, ArunETC08b, ForsEvanHopp16}.  

However, BHPT presently offers the fluxes only at lowest order in the mass ratio.  Thus, an expression 
for $e_t$ is similarly required only to lowest (zeroth) order in this mass ratio to enable transformation
to and from harmonic gauge. This is possible through analysis of Schwarzschild geodesic 
motion.  We show the procedure below and in the process derive the complete QK formalism
for Schwarzschild geodesic motion to higher PN order.  

We start by reviewing the current state of knowledge on the QK representation of non-spinning binary motion in 
general relativity.  This description is modeled off the standard Keplerian equations of motion for elliptical orbits, 
given by
\begin{align}
r &= a_r (1 - e_K \cos{u}), \notag \\
\O_r t &= u - e_K \sin{u}, \notag \\
\vp &= V, \notag \\
V &= 2 \arctan \left( \sqrt{ \frac{1+e_K}{1-e_K} } \tan{\frac{u}{2}} \right).
\end{align}
Here, $a_r$ is the semi-major axis, $e_K$ is the Keplerian eccentricity, $\O_r$ is the radial frequency, $u = u(t)$ is 
the eccentric anomaly, and $V$ is the true anomaly.  At Newtonian order the motion is periodic, meaning 
$\O_r=\O_\vp$ is the only frequency and the azimuthal coordinate $\vp$ matches the true anomaly.  $a_r$ and 
$e_K$ can be expressed in terms of other quantities as
\begin{align}
a_r &= \frac{r_{+} + r_{-}}{2} = \frac{M+\mu}{\ve}, \notag \\
e_K &= \frac{r_{+} - r_{-}}{r_{+} + r_{-}} = 1 - j .
\end{align}
Here, we have defined $r_{+} = r_{\rm max}, \, r_{-} = r_{\rm min}$ as the radii at apastron and periastron, 
respectively.  Additionally, $\ve = -2E, j = -2 E L^2/(M+\mu)^2$ are common parameters in PN work related 
to the energy and angular momentum of the system \cite{Blan14}.

In 1985 Damour and Deruelle derived the 1PN relativistic corrections to these equations \cite{DamoDeru85}, leading 
to the following:
\begin{align}
r_{\rm H} &= a_r (1 - e_r \cos{u}), \notag \\
\O_r t &= u - e_t \sin{u}, \notag \\
\vp &= \left(\frac{\O_\vp}{\O_r}\right) V = K V, \notag \\
V &= 2 \arctan \left( \sqrt{ \frac{1+e_\vp}{1-e_\vp} } \tan{\frac{u}{2}} \right). 
\label{eqn:1PNQK}
\end{align}
Though similar in form, these relations present a few complications over the Keplerian motion.  First, at 1PN order 
the motion no longer closes; thus, $\O_r \ne \O_\vp$ and $\vp \ne V$.  Next, the single Keplerian eccentricity $e_K$
is supplanted by the threefold set of the radial eccentricity $e_r$, the time eccentricity $e_t$, and the azimuthal 
eccentricity $e_\vp$, each of which has a different relationship to the energy and angular momentum of the system.  
Finally, the coordinates and parameters are all now defined in modified harmonic gauge \cite{Blan14}.  The subscript 
on $r_{\rm H}$ has been added to emphasize that fact, distinguishing it from the Schwarzschild radius (however, the 
other coordinates do not require explicit labels for our purposes; see the next subsection).

Later work at 2PN \cite{DamoScha88, SchaWex93} and then 3PN \cite{MemmGopaScha04} implied a 
model for an effectively generic QK representation.  This takes the form
\begin{align}
\label{eqn:GenQKeqns}
r_{\rm H} &= a_r (1 - e_r \cos{u}), \notag \\
\O_r t &= u - e_t \sin u + f_t \sin V + g_t (V - u) \notag \\&  \hspace{7.5em}
+ h_t \sin 2V + i_t \sin 3V + \cdots, \notag \\
\frac{\vp}{K} &= V + f_\vp \sin{2V} + g_\vp \sin{3V} + i_\vp \sin{4V} + \cdots, \notag \\
V &= 2 \arctan \left( \sqrt{ \frac{1+e_\vp}{1-e_\vp} } \tan{\frac{u}{2}} \right). 
\end{align}
We have explicitly listed only those terms that appear in the 3PN QK equations but indicate that the series of 
trigonometric functions are expected to continue with higher PN orders.

Thus, the form of the radial motion is valid to all orders, with $e_r$ and $a_r$ defined by 
\be
\label{eqn:deferar}
e_r = \frac{r_{\rm H +} - r_{\rm H -}}{r_{\rm H+} + r_{\rm H -}}, \q a_r = \frac{r_{\rm H +} + r_{\rm H -}}{2},
\ee
The $t$ and $\vp$ equations, meanwhile, pick up trigonometric functions of $V$.    In this generic representation, $e_\vp$ is defined order-by-order to eliminate $\sin{V}$ from 
the equation for $\vp$ \cite{MemmGopaScha04}.  The remaining parameters like $e_t$ or
$i_\vp$ are defined simply as the coefficients in front of their respective trigonometric functions.  Each is generally
obtained as an expansion in $\ve$ and $j$.  As such, these parameters can, in 
principle, only be extracted to the same order as the full equations of motion, both of which come
from iterating some formulation of the full PN formalism \cite{Blan14}.

However, in the small-mass-ratio limit, the situation reduces to geodesic motion of the smaller body on a 
Schwarzschild background.  Then, all the dynamics of the system are encoded in the geodesic equations of motion. 
We can thus apply the above definitions in this limit to generate the QK representation to all PN orders at lowest 
order in the mass ratio.

\subsection{Harmonic coordinates, Schwarzschild coordinates, and the Darwin parameterization}
\label{sec:harmToSchw}

We now extract the QK description by looking at geodesic motion on a Schwarzschild background.
First, the Schwarzschild metric can be expressed in harmonic gauge as \cite{FromPoisWill14}
\begin{align}
ds^2 &= - \frac{1 - M/r_{\rm H}}{1 + M/r_{\rm H}} dt_H^2 + \frac{1 + M/r_{\rm H}}{1 - M/r_{\rm H}} dr_{\rm H}^2 
\notag \\& \hspace{11em} + (r_{\rm H} + M)^2 d\O^2.
\end{align}
In fact, these coordinates are almost identical to the standard Schwarzschild coordinates 
$(t_S, r_S, \theta_S, \vp_S)$ with line element \eqref{eqn:SchwLE}.  The two are connected by
\begin{align}
t_{\rm H} &= t_{\rm S} = t, \notag \\ 
r_{\rm H} &= r_{\rm S} - M = r - M,  \notag \\
\th_{\rm H} &= \th_{\rm S} = \th, \notag \\  
\vp_{\rm H} &= \vp_{\rm S} = \vp.
\end{align}
Therefore, we can work directly with the motion in Schwarzschild coordinates and merely 
correct the radius when necessary.

As described in Sec.~\ref{sec:geoExps}, geodesic motion in Schwarzschild coordinates is conveniently described 
using the Darwin parameterization, which for bound orbits recasts the specific energy 
$\mathcal{E} = (1 - \ve/2)$ and angular momentum $\mathcal{L} = \sqrt{j M^2/\ve}$ in terms of 
semi-latus rectum $p$ and Darwin eccentricity $e$ \cite{Darw59, CutlKennPois94, BaraSago10}.  $p$ and $e$ are 
defined by
\begin{align}
\label{eqn:defpe}
p &= \frac{2 r_{+} r_{-}}{M (r_{+} + r_{-})}, \qq 
e =  \frac{r_{+} - r_{-}}{r_{+} + r_{-}}, 
\end{align}
with 
\begin{align}
r_{+} &= \frac{pM}{1-e},  \notag \\   
r_{-} &= \frac{pM}{1 + e} ,  \notag \\
a & = \frac{r_+ + r_-}{2} = \frac{pM}{1-e^2}
\end{align}
Note that with the expressions \eqref{eqn:defeandp} relating $\mathcal{E}$ and $\mathcal{L}$ to $p$ and $e$, 
$\ve$ and $j$ can be immediately expanded to arbitrary order in $1/p$ and $e$, and these can be inverted 
to give $p$ and $e$ in terms of $\ve$ and $j$.  The result to 6PN is given in App.~\ref{sec:highQKexps}.  

Next, recall the QK definitions of $e_r$ and $a_r$ \eqref{eqn:deferar}.  Expressing these in terms of the 
Schwarzschild radius gives
\begin{align}
e_r &= \frac{r_+ - r_-}{r_+ + r_- - 2M},   \notag \\  
a_r &= \frac{r_+ + r_-}{2} - M 
\end{align}
Then, these can be related to $e$ and $p$ simply by
\begin{align}
\label{eqn:erTope}
a_r &= a - M = \frac{pM}{1 - e^2} - M, \notag \\
e_r &= \frac{a}{a - M} \, e = \frac{p}{p - 1 + e^2} \, e.
\end{align}
This allows for the rapid expansion of $e_r$ and $a_r$ to arbitrary order in $p$ and $e$ and thus $\ve$ and $j$.  
The two series are given to 6PN in App.~\ref{sec:highQKexps}.  Note that \eqref{eqn:erTope} immediately allows 
for the transformation of our BHPT-PN flux expansions to harmonic gauge to arbitrary PN order, except using $e_r$ 
instead of the more common $e_t$.

\begin{widetext}
\subsection{Orbit integration and Kepler's equation}
\label{sec:KepEqn}

Further progress requires integration of the orbit.  As mentioned in Sec.~\ref{sec:geoExps} this is described in terms 
of the relativistic anomaly $\chi$, reducing the coordinates to
\begin{align}
r \l \chi \r &= \frac{pM}{1+ e \cos \chi}, \notag \\
\frac{dt}{d\chi} &= 
\frac{p^2 M}{(p-2 - 2e \cos(\chi)) (1+e\cos(\chi))^2} \left(\frac{(p-2)^2 - 4e^2}{p-6-2e \cos(\chi)} \right)^{1/2}, \notag \\
\vp(\chi) &= \left( \frac{4 p}{p- 6 - 2e} \right)^{1/2} F\left(\frac{\chi}{2} \bigg| - \frac{4e}{p-6-2e} \right)
\end{align}

Given the form of these equations, a reasonable general definition for an eccentric anomaly, call it $\tilde{u}$, 
could be constructed analogously to its Newtonian counterpart, with
\be
\label{eqn:chiTou}
\chi = 2 \arctan \left( \sqrt{ \frac{1+e}{1-e} } \tan{\frac{\tilde{u}}{2}} \right).
\ee
From this definition, it can be found that
\be
r = \left(\frac{p M}{1-e^2}\right) \left(1 - e \cos{\tilde{u}} \right) = a \left(1 - e \cos{\tilde{u}} \right).
\ee
But the corresponding QK equation is given by
\be
r-M = r_{\rm H} = a_r (1 - e_r \cos u) = (a - M) \left(1 - \frac{a}{a-M} e \cos{u} \right) = a (1 - e \cos{u}) - M.
\ee
Therefore, we observe that $u = \tilde{u}$ at lowest order in the mass ratio.  

The relation between $\chi$ and $u$ can then be used to find
\begin{align}
\label{eqn:dtdu}
\frac{d\chi}{d u} &= \frac{\sqrt{1-e^2}}{1 - e \cos{u}},   \notag \\
\frac{dt}{du} &= \frac{p^2 (1-e \cos u)^{5/2} \sqrt{(p-2)^2-4 e^2}}{\left(1-e^2\right)^{3/2} 
\left(p (1-e \cos u) - 2+2 e^2 \right) \sqrt{p-6+2 e^2-e (p-4) \cos u}} .
\end{align}
The righthand side of this equation can be expanded in $1/p$ (but left exact in $e$) and integrated to give $t(u)$ as a 
PN series to arbitrary order.  When done in this way, the series starts
\begin{align}
\label{eqn:tOfu}
t(u) &= \frac{u-e \sin u}{\left(1-e^2\right)^{3/2}} p^{3/2} + \frac{3 u}{\sqrt{1-e^2}} \sqrt{p} + 
\left[6 u - 2 e \sin u + \frac{15}{2} \sqrt{1 - e^2} \, \chi \right]  \left(\frac{1}{\sqrt{1-e^2} \sqrt{p}}\right) 
\notag \\ & + \left[12 \left(5-e^2\right) u -16 e \sin u + 75 \sqrt{1-e^2} \, \chi + 35 e \sqrt{1 - e^2} \sin \chi
\right] \left(\frac{1}{2 \sqrt{1-e^2} p^{3/2}}\right) + \mathcal{O}\left(\frac{1}{p^{5/2}} \right) 
\end{align}
where we used that $\sin \chi = (\sqrt{1-e^2} \sin u)/(1-e \cos u) $.
Then, Kepler's equation can be trivially recovered from \eqref{eqn:tOfu} through multiplication by $\O_r$.  
After rearranging terms, this gives
\begin{align}
\label{eqn:KepChip}
\O_r t &= u + 15 \left(1-e^2\right)^{3/2} \left[\frac{1}{2 p^2} + \frac{6+9 e^2}{2 p^3} \right](\chi - u)+\frac{35 
\left(1-e^2\right)^{3/2}}{2 p^3} e \sin \chi - e \sin{u} \bigg[1- \frac{3 \left(1-e^2\right)}{p}  \notag \\& 
+ \frac{\left(1-e^2\right) \left(10-18 e^2-15 \sqrt{1-e^2}\right)}{2 p^2}-\frac{\left(1-e^2\right) \left(38-60 e^2+54 e^4
- (15-90e^2) \sqrt{1-e^2} \right)}{2 p^3} \bigg]  + \mathcal{O}\left(\frac{1}{p^4} \right). 
\end{align}

The expression behind $\sin u$ in \eqref{eqn:KepChip} can be identified as an expansion for $e_t/e$ in terms of 
$p$ and $e$.  Transforming to $\ve$ and $j$ reveals that this matches the 3PN expression for $e_t$ in 
modified harmonic coordinates given in \cite{ArunETC08b}.  As with $e_r$ and $a_r$, $e_t$ can be found in this 
way to arbitrary PN order.  However, the procedure here --- with both the execution of the integral for $t(u)$ and the 
identification of the $\sin{n \chi}$ terms --- is far more cumbersome.  Here is the result to 5PN:
\begin{align}
\label{eqn:ete}
\frac{e_t}{e} &= 1-\frac{3 (1-e^2)}{p}+ \left(10-18 e^2-15 \sqrt{1-e^2}\right) \left( \frac{1-e^2}{2 p^2} \right)
-\left(38-60 e^2+54 e^4 - (15-90e^2) \sqrt{1-e^2} \right)\left(\frac{1-e^2}{2 p^3} \right) \notag \\ &  +
\left(4 (309-1006 e^2+765 e^4-324 e^6) - 3 \sqrt{1-e^2} (698-535 e^2+1080 e^4)\right)\left(\frac{1-e^2}{16 p^4}\right)
-\Big( 4 (299-2839 e^2+6777 e^4  \notag \\ &
-4185 e^6+972 e^8) + 3 \sqrt{1-e^2} (954  +6731 e^2-4050 e^4+4320 e^6)\Big)
\left(\frac{1-e^2}{16 p^5}\right)  +  \mathcal{O}\left(\frac{1}{p^6}\right)
\end{align}
Unfortunately, the completion of this procedure to 19PN would likely be difficult, implying that $e_r$ might be the
preferable choice of eccentricity when transforming high-order BHPT-PN series to harmonic gauge.  
We present the expansion for $e_t$ in $\ve$ and $j$ in App.~\ref{sec:highQKexps}.

The above results indicate that the coefficient of $(\chi - u)$ does not equal $g_t$, and the 
coefficient of $\sin \chi$ does not equal $f_t$.  This stems from the fact that $\chi \ne V$, as evidenced by comparing
\eqref{eqn:GenQKeqns} and \eqref{eqn:chiTou}.

\subsection{The azimuthal equation}
\label{sec:AzimuthEqn}

We can now pursue the rest of the QK parameterization, starting with the relationship between $\chi$ and $V$.  This
can be obtained using another equation of motion,
\be
\label{eqn:phiToChiV}
\frac{\vp}{K} = \chi + \tilde{a}_\vp \sin{\chi} + \tilde{f}_\vp \sin{2\chi} + \tilde{g}_\vp \sin{3\chi} + \tilde{i}_\vp \sin{4\chi} + \cdots = V + f_\vp \sin{2V} + g_\vp \sin{3V} + i_\vp \sin{4V} + \cdots,
\ee
where all given quantities are PN expanded to any desired order.  As mentioned above, we see that $V$ is 
defined order-by-order to eliminate the appearance of $\sin{V}$ in the representation for $\vp$.  
The expansion for $\vp/K$ in terms of $\chi$ is easily computed using the Darwin parameterization as
\be
\label{eqn:phiExpChi}
\frac{\vp}{K} = \chi +\frac{e \sin{\chi}}{p}+\frac{3 e (16 \sin{\chi}+e \sin{2 \chi})}{8 p^2}+\frac{\left(27e \left(32+e^2\right) 
\sin{\chi}+108 e^2 \sin{2 \chi}+5 e^3 \sin{3 \chi}\right)}{24 p^3}  +  \mathcal{O}\left(\frac{1}{p^4}\right)
\ee

The exact relationship between $\chi$ and $V$ is given by
\be
\chi = 2 \arctan \left( \sqrt{ \frac{1+e}{1-e} } \tan{\frac{u}{2}} \right) 
= 2 \arctan \left( \sqrt{ \frac{(1+e)(1-e_\vp)}{(1-e)(1+e_\vp)} } \tan{\frac{V}{2}} \right).
\ee
In order to eliminate $\sin{V}$ from \eqref{eqn:phiToChiV}, $\chi(V)$ is inserted.  Then, $\vp(\chi(V))/K$ is expanded 
using an ansatz for the PN series of $e_\vp$ in $1/p$.  The coefficients in this series are then exactly 
determined by the condition that $\sin{V}$ disappear from the representation for $\vp/K$.  

In this way, we obtain
\begin{align}
\label{eqn:ePhie}
\frac{e_\vp}{e} &= 1 + \frac{1 - e^2}{p} + \frac{(1 - e^2) (6 - e^2)}{p^2} + \frac{(1 - e^2) (36 - 11 e^2 + e^4)}{p^3}
+ \frac{(1 - e^2) (216 - 90 e^2 + 16 e^4 - e^6)}{p^4}  + (1296-648 e^2   \notag \\ & 
+170 e^4 -21 e^6+e^8) \left(\frac{1-e^2}{p^5} \right) + \left(7776-4320 e^2+1500 e^4-275 e^6+26 e^8-e^{10}\right)  
\left(\frac{1-e^2}{p^6} \right) + \mathcal{O}\left(\frac{1}{p^7}\right).
\end{align}
This method can be (fairly rapidly) extended to arbitrary order, and we cover the expansion in 
$\ve$ and $j$ in Appendix~\ref{sec:highQKexps}.

From here, the expansion for $\chi(V)$ can be substituted into \eqref{eqn:phiToChiV} to retrieve 
$f_\vp, g_\vp, \cdots$, and it can also be put into Kepler's equation to compute $f_t, g_t, \cdots$.  These are less 
useful than the eccentricities for the purposes of expansion transformations, but the full forms of these 
equations are given in Appendix~\ref{sec:otherParams}.
\end{widetext}

\section{Conclusions and outlook}
\label{sec:conclusions}

This paper has described the high-order analytic expansion of the total energy and angular momentum radiated to 
infinity by eccentric-orbit EMRIs.  By extending the methods of \cite{KavaOtteWard15, Fuji12b} to the eccentric
regime, we have computed both fluxes to 10PN and $e^{20}$, as well as to 19PN and $e^{10}$, a significant 
advance over previous work with numeric-analytic fitting \cite{ForsEvanHopp16, MunnETC20}.  
We thus conclude that the direct analytic expansion scheme is highly successful at reaching high PN order and 
moderate order in eccentricity for the energy and angular momentum fluxes at infinity.  

The high-order expansions in this work allow for a representation of the fluxes that is valid for small $p$ and 
moderate $e$ or large $p$ and fairly large $e$.  Unfortunately, it does appear to experience some trouble in the 
small-$p$ large-$e$ regime.  This is likely due at least in part to insufficient $n$ mode 
representation in the PN expansions.  Indeed, while the PN expansions only include $|n|$ up to half the maximum
eccentricity order, the numerical $(p=10, e=1/2)$ flux, for instance, accurate to 12 digits required $n$ higher than 20
for certain $lm$ modes.  Therefore, higher-order expansions in $e$ are likely necessary to ensure convergence at
higher $e$.  Insufficient representation of $l$ modes has also been noted as a limiting factor for small $p$ 
\cite{Fuji12b, Fuji15}. 

The bottleneck step in the procedure was the calculation of the even-parity normalization constant for $l=2$.  This 
calculation took about 7 days on a single core of the UNC cluster Longleaf, indicating that another PN term or 
another couple orders in $e^2$ could be obtained with a long runtime or faster core.  Nevertheless, significantly 
higher orders are probably out of reach with the current implementation of the code.  It is possible that additional 
simplifications are yet undiscovered in the construction of the homogeneous or inhomogeneous
solutions, which would allow for another large increase in attainable order.  A reformulation in another language
like \textsc{Python} or \textsc{C++} could also feasibly be advantageous.

However, more promising is the prospect of finding superior resummation schemes that will greatly increase the 
convergence to numerical calculations.  Unfortunately, it appears that some of the straightforward mode-based
factorizations applied successfully in the circular-orbit case will not be quite as fruitful in the high-eccentricity regime.  
Future work experimenting with more complex and unconventional factorization schemes (e.g., Pade or Chebyshev 
approximants) will be warranted. 

However, it is encouraging that the accuracy of the full-flux expansion was fairly strong even for the orbit
$(p=10, e=1/2)$, owing to the use of arbitrary-order eccentricity expansions at low PN and the use of eccentricity 
resummations throughout.  Increased validity at higher eccentricity can likely be obtained by extending these 
expansions to higher order in $e$, which is particularly important at lower PN order.  To that end, the techniques 
developed in \cite{MunnEvan19a,MunnEvan19b} can (in principle) be extended to derive expansions for the 3.5PN, 
4.5PN, and 5PN terms to arbitrary order in $e$, though with considerable difficulty (especially at 4.5PN).  This is 
achieved through intricate but manageable manipulations involving Fourier decomposition of source multipole 
moments (see \cite{MunnEvan19a,MunnEvan19b} for more details).  Beyond 5PN, further progress is likely more 
accessible to the MST analytic expansion approach of this paper.  For instance, it may be possible to obtain the 6PN 
and 7PN terms beyond $e^{30}$ using the methods of Sec.~\ref{sec:MSTexp} and Sec.~\ref{sec:InhomExps}, but 
this is not certain.  In addition, the $e^{20}$ calculation can potentially be extended to 11PN or 12PN.  These 
ideas will be explored in future work. 

In the meantime the methods developed in this paper can also be utilized to generate expansions for other BHPT 
quantities of interest.  The first and most obvious is the radiation at the larger black hole's horizon, found using the 
coefficients $C^-_{lmn}$.  We have already calculated these to 10PN and $e^{20}$ and 18PN and $e^{10}$ (relative 
order) using the techniques laid out above, and the results will be detailed in a follow-up paper \cite{MunnEvan20b}.  

Beyond that, direct analytic expansion techniques also have been successfully applied in the conservative sector of 
BHPT.  Conservative quantities supply crucial terms in EOB potentials (see, e.g., \cite{BaraDamoSago10, 
LetiBlanWhit12, BiniDamo14c, BiniDamoGera15, HoppKavaOtte16, KavaETC17, BiniDamoGera18, 
BiniDamoGera19, BiniDamoGera20a, BiniDamoGera20b}) and also contribute directly to the EMRI 
cumulative phase at post-1 adiabatic order \cite{HindFlan08}.  For instance, \cite{KavaOtteWard15} found the 
redshift invariant, spin-precession invariant, and tidal invariants to 21.5PN order for circular-orbit EMRIs on a 
Schwarzschild background.  Published results in the eccentric case are much more modest: For instance, the state 
of the art for the redshift invariant is 4PN and $e^{20}$ and 9.5PN and $e^8$ \cite{HoppKavaOtte16, 
BiniDamoGera16c, BiniDamoGera20b}, while the others are even less developed \cite{KavaETC17,BiniGera18a}.  
In general, expansions in the conservative sector are more complicated, as the leading PN order of individual modes 
does increase with $l$, meaning that expansions are required that remain general in $l$.  Nevertheless, techniques 
have been developed to handle this complication \cite{BiniDamo14a, BiniDamo14b, KavaOtteWard15, 
HoppKavaOtte16}, and we report that we have extended the present work to the conservative sector and 
found the redshift invariant to 8PN and $e^{20}$.  This will be discussed in a follow-up paper \cite{MunnEvan20c}.

With generic bound orbits on a Schwarzschild background analytically understood, it will be necessary to extend
these methods to the more intricate (but more astrophysically relevant) Kerr background.  There, analytic
expansions are possible using the Teukolsky formalism, which is similar to the RWZ formalism of this paper, though
more expensive by multiple orders of magnitude.  Past work has primarily focused on expanding the simpler case of 
circular equatorial orbits \cite{KavaOtteWard16, Fuji15, FujiSagoNaka18}, though flux series for generic (eccentric, 
inclined) orbits have been found to 4PN and $e^6$ \cite{SagoFuji15}.  The simplifications developed in this paper,
when properly adapted to the Kerr case, should allow for a significant improvement over the state of expansions for
generic orbits.

Finally, this paper has also presented a means to derive a quasi-Keplerian representation of Schwarzschild geodesic 
motion to high PN order.  This allows for the rapid transformation between certain high order PN series generated by 
BHPT and those derived through the full PN formalism in (modified) harmonic coordinates.  The QK results obtained 
in this manner provide a nice check on future developments in PN theory, as the small-mass-ratio limit of any new 
results should match the prescription laid out here.

It is of note that we sought the particular QK representation in harmonic coordinates, but this is not the 
only available choice.  By extracting the geodesic limit of some other gauge, we could repeat the above
procedure and ascertain the QK parameters in that gauge.  As an example, \cite{FromPoisWill14, Dese14} indicate
that the Schwarzschild limit of ADM gauge is given by isotropic coordinates:
\be
ds^2 = -\left( \frac{2r_{\rm I} - M}{2r_{\rm I} + M}\right)^2 dt^2 + \left(1 + \frac{M}{2r_{\rm I}}\right)^4 (dr_{\rm I}^2 + 
r_{\rm I}^2 d\O^2).
\ee
with $r_{\rm S} = r_{\rm I} (1 + M/(2r_{\rm I}))^2$.  This choice is amenable to the same techniques, though the more
complicated relationship between the two radii will make the process somewhat more cumbersome. 

In addition, Schwarzschild geodesic motion corresponds to the zeroth-order BHPT calculation; however, the 
first-order problem has also been (effectively) solved.  Thus, it is theoretically feasible to extend this procedure to first 
order in the mass ratio, obtaining all contributions at $\mathcal{O}(\nu)$ in the QK representation.  Deriving these 
corrections would be orders of magnitude more difficult, as geodesic motion on the first-order (regularized) metric
is complicated \cite{BaraSago11}.  Furthermore, the process of gauge transformation from first-order RW (or 
radiation) to harmonic coordinates is far more intricate than that from the simple Schwarzschild coordinates of 
geodesic motion \cite{HoppEvan13, PounMerlBara13, ThomWardWhit19}.  We will leave further exploration of this 
problem for future work.

\acknowledgments

The author thanks Charles R. Evans, Adrian Ottewill, Barry Wardell, Nathan Johnson-McDaniel, Niels Warburton, 
Seth Hopper, and Zachary Nasipak for many helpful discussions.  The author also thanks Jezreel Castillo for 
providing the numeric value of the flux for $(p=10,e=1/2)$ and Thomas Osburn for supplying additional flux data.  
This work makes use of the Black Hole Perturbation Toolkit.  This work was supported in part by NSF grants 
PHY-1506182 and PHY-1806447, the Bahnson Fund at the University of North Carolina-Chapel Hill, and the North 
Carolina Space Grant.

\appendix

\begin{widetext}
\section{The Kepler and azimuthal equations to 5PN}
\label{sec:otherParams}

The methods above can be used to generate higher order corrections to the full Kepler's equation.  In terms of
$p$ and $e$, we get
\begin{align}
\O_r t &= u - e_t \sin{u} + 3 \left(1-e^2\right)^{3/2} \bigg[\frac{5}{2 p^2}+\frac{5 \left(2+3 e^2\right)}{2 p^3}
+\frac{738+145 e^2+360 e^4-300 \left(1-e^2\right)^{3/2}}{16 p^4} +  \notag \\ &
\frac{3528+3512  e^2-165 e^4+1080 e^6-600 (1-e^2)^{3/2} \left(2+3 e^2\right)}{16 p^5} \bigg](V-u) 
+ e (1 - e^2)^{3/2}   
 \Bigg[\frac{10}{p^3} + \frac{5 \left(29+24 e^2\right)}{4 p^4}  \notag \\&
 +\frac{3 \left(722+267 e^2+240 e^4-200 \left(1-e^2\right)^{3/2}\right)}{8 p^5}  \Bigg] \sin V 
+ e^2 (1-e^2)^{3/2} \Bigg[  \frac{95}{32 p^4} + \frac{434+285 e^2}{32 p^5}    \Bigg] \sin2V       \notag \\ & 
+ e^3(1-e^2)^{3/2}\left(\frac{9}{8 p^5} \right)  \sin 3V
+ \mathcal{O}\left(\frac{1}{p^6} \right),
\end{align}
where $e_t$ is given in \eqref{eqn:ete}.

Likewise, $\chi(V)$ is plugged into the azimuthal equation to obtain 
\begin{align}
\frac{\vp}{K} &= V+\frac{e^2 \sin{2V}}{8 p^2}+\frac{3 e^2 \sin{2V}}{2 p^3}+\frac{\frac{1}{16} e^2 \left(216+5 e^2\right) 
\sin{2V}+\frac{3}{256} e^4 \sin{4V}}{p^4}  \notag \\ &  \hspace{22em}
+\frac{\frac{3}{2} e^2 \left(72+5 e^2\right) \sin{2V}+\frac{9}{32} e^4 \sin{4V}}{p^5}+ \mathcal{O}\left(\frac{1}{p^6}\right).
\end{align}

\section{Orbital parameters expanded in $\ve$ and $j$}
\label{sec:highQKexps}

We now present expansions for various QK quantities in terms of the gauge invariant quantities 
$\ve$ and $j$.  These are found by using the expansions for $p$ and $e$, given to 6PN by
\begin{align}
p &= \frac{j}{\ve}+(-4+j)+\left(4-\frac{16}{j}+\frac{3 j}{4}\right) \ve+\left(3-\frac{128}{j^2}+\frac{48}{j}
+\frac{j}{2}\right) \ve^2+\left(2-\frac{1280}{j^3}+\frac{640}{j^2}-\frac{12}{j}+\frac{5 j}{16}\right) \ve^3
\notag \\ & 
+\left(\frac{5}{4}-\frac{14336}{j^4}+\frac{8960}{j^3}-\frac{800}{j^2}+\frac{3 j}{16}\right) \ve^4
+\left(\frac{3}{4}-\frac{172032}{j^5}+\frac{129024}{j^4}-\frac{20160}{j^3}+\frac{320}{j^2}+\frac{7 j}{64}\right)
  \ve^5 + \mathcal{O} \left(\ve^6\right) \\
e^2 &= (1-j)+\left(4-\frac{7 j}{4}\right) \ve - \left(5 - \frac{16}{j} + 2 j\right) \ve^2
- \left(10 - \frac{128}{j^2} + \frac{52}{j} + \frac{15 j}{8}\right) \ve^3  
- \left(\frac{45}{4} - \frac{1280}{j^3} + \frac{672}{j^2} + \frac{40}{j} + \frac{25 j}{16}\right) \ve^4 \notag \\&
-\left(\frac{41}{4} - \frac{14336}{j^4} + \frac{9280}{j^3} - \frac{320}{j^2} + \frac{35}{j} +\frac{77 j}{64}\right) \ve^5
-\left(\frac{133}{16} - \frac{172032}{j^5} + \frac{132608}{j^4} - \frac{15232}{j^3} + \frac{40}{j^2} + \frac{28}{j}
+\frac{7 j}{8}\right) \ve^6 + \mathcal{O} \left(\ve^7 \right)  \notag
\end{align}
First, the harmonic semi-major axis, $a_r = pM/(1-e^2) - M$, takes the form
\begin{align}
\frac{a_r}{M} &= \frac{1}{\ve}-\frac{7}{4}+\left(\frac{1}{16}-\frac{4}{j}\right) \ve+\left(\frac{1}{64}
-\frac{32}{j^2}+\frac{4}{j}\right) \ve^2 +\left(\frac{1}{256}-\frac{320}{j^3}+\frac{80}{j^2}-\frac{1}{j}\right) 
\ve^3 \notag \\&  +\left(\frac{1}{1024}-\frac{3584}{j^4}+\frac{1344}{j^3}-\frac{68}{j^2}\right) \ve^4
+\left(\frac{1}{4096}-\frac{43008}{j^5}+\frac{21504}{j^4}-\frac{2128}{j^3}+\frac{24}{j^2}\right) 
\ve^5 + \mathcal{O} \left(\ve^6 \right) 
\end{align}
Next, the three eccentricities.  $e_r^2 = (a/a_r)^2 e^2$ is given by
\begin{align}
e_r^2 &= 1-j+\left(6-\frac{15 j}{4}\right) \ve+\left(\frac{15}{2}+\frac{16}{j}-10 j\right) \ve^2
-\left(\frac{1}{2} - \frac{128}{j^2} + \frac{12}{j} + \frac{93 j}{4}\right) \ve^3  \notag \\&
-\left(\frac{615}{16} - \frac{1280}{j^3} + \frac{352}{j^2} + \frac{76}{j} + \frac{201 j}{4}\right)  \ve^4
+\left(-\frac{621}{4}+\frac{14336}{j^4}-\frac{6080}{j^3}-\frac{544}{j^2}-\frac{633}{2 j}-\frac{1661 j}{16}\right)
\ve^5  \notag \\&  
+\left(-\frac{7385}{16}+\frac{172032}{j^5}-\frac{96768}{j^4}+\frac{576}{j^3}-\frac{2196}{j^2}
-\frac{1047}{j}-208  j\right) \ve^6 + \mathcal{O} \left(\ve^7 \right) 
\end{align}
The azimuthal eccentricity is similarly simple, giving
\begin{align}
e_\vp^2 &= 1-j+\left(6-\frac{15 j}{4}\right) \ve+\left(-\frac{5}{2}+\frac{26}{j}-10 j\right)\ve^2
+\left(-\frac{87}{2}+\frac{220}{j^2}-\frac{77}{2 j}-\frac{93 j}{4}\right)\ve^3 \notag \\&
+\Big(-\frac{2737}{16}+\frac{2298}{j^3} -\frac{646}{j^2}  
-\frac{313}{j} 
-\frac{201 j}{4}\Big)\ve^4+\left(-\frac{2033}{4}+\frac{26676}{j^4}-\frac{20981}{2 j^3}-\frac{5021}{2 j^2}
-\frac{5373}{4 j}-\frac{1661 j}{16}\right) \ve^5  \notag \\&
+\left(-\frac{21181}{16}+\frac{330020}{j^5}-\frac{167759}{j^4}-\frac{16342}{j^3}-\frac{177879}{16 j^2}
-\frac{72027}{16 j}-208 j\right) \ve^6
+\mathcal{O} \left(\ve^7\right).
\end{align}
The time eccentricity is more complicated, containing half powers of $j$, and also more tedious to construct.  
We give it to 5PN:
\begin{align}
\label{eqn:etepsj}
e_t^2 &= 1-j - \left(2 - \frac{17 j}{4}\right) \ve
+\left(\frac{3}{2}+\frac{8}{j}-\frac{15}{\sqrt{j}}+15 \sqrt{j}-14 j\right) \ve^2
+\left(\frac{7}{2}+\frac{64}{j^2}-\frac{105}{j^{3/2}}-\frac{66}{j}+\frac{1365}{8 \sqrt{j}}-\frac{795\sqrt{j}}{8}
+\frac{165 j}{4}\right) \ve^3  \notag \\&
+\left(-\frac{3067}{16}+\frac{640}{j^3}-\frac{9009}{8 j^{5/2}}-\frac{672}{j^2}+\frac{12879}{8 j^{3/2}}+\frac{1795}{4 j}
-\frac{129645}{128 \sqrt{j}}+\frac{56385 \sqrt{j}}{128}-\frac{457 j}{4}\right) \ve^4 
+\Big(\frac{3207}{2}+\frac{7168}{j^4}-\frac{109395}{8 j^{7/2}}   \notag \\&  -\frac{8096}{j^3}+\frac{1291491}{64 j^{5/2}}
+\frac{11461}{2 j^2}-\frac{1464531}{128 j^{3/2}}-\frac{70841}{16 j}+\frac{4517145}{1024 \sqrt{j}}
-\frac{1668795 \sqrt{j}}{1024}+\frac{4867 j}{16}\Big) \ve^5 + \mathcal{O}\left(\ve^6\right).
\end{align}
Note that these expansions match the expressions in \cite{ArunETC08b} to 3PN at lowest order in $\nu$. 

\end{widetext}

\clearpage

\bibliography{inffluxes}

\begin{thebibliography}{146}%
\makeatletter
\providecommand \@ifxundefined [1]{%
 \@ifx{#1\undefined}
}%
\providecommand \@ifnum [1]{%
 \ifnum #1\expandafter \@firstoftwo
 \else \expandafter \@secondoftwo
 \fi
}%
\providecommand \@ifx [1]{%
 \ifx #1\expandafter \@firstoftwo
 \else \expandafter \@secondoftwo
 \fi
}%
\providecommand \natexlab [1]{#1}%
\providecommand \enquote  [1]{``#1''}%
\providecommand \bibnamefont  [1]{#1}%
\providecommand \bibfnamefont [1]{#1}%
\providecommand \citenamefont [1]{#1}%
\providecommand \href@noop [0]{\@secondoftwo}%
\providecommand \href [0]{\begingroup \@sanitize@url \@href}%
\providecommand \@href[1]{\@@startlink{#1}\@@href}%
\providecommand \@@href[1]{\endgroup#1\@@endlink}%
\providecommand \@sanitize@url [0]{\catcode `\\12\catcode `\$12\catcode
  `\&12\catcode `\#12\catcode `\^12\catcode `\_12\catcode `\%12\relax}%
\providecommand \@@startlink[1]{}%
\providecommand \@@endlink[0]{}%
\providecommand \url  [0]{\begingroup\@sanitize@url \@url }%
\providecommand \@url [1]{\endgroup\@href {#1}{\urlprefix }}%
\providecommand \urlprefix  [0]{URL }%
\providecommand \Eprint [0]{\href }%
\providecommand \doibase [0]{http://dx.doi.org/}%
\providecommand \selectlanguage [0]{\@gobble}%
\providecommand \bibinfo  [0]{\@secondoftwo}%
\providecommand \bibfield  [0]{\@secondoftwo}%
\providecommand \translation [1]{[#1]}%
\providecommand \BibitemOpen [0]{}%
\providecommand \bibitemStop [0]{}%
\providecommand \bibitemNoStop [0]{.\EOS\space}%
\providecommand \EOS [0]{\spacefactor3000\relax}%
\providecommand \BibitemShut  [1]{\csname bibitem#1\endcsname}%
\let\auto@bib@innerbib\@empty
\bibitem [{\citenamefont {{Barack}}\ \emph {et~al.}(2019)\citenamefont
  {{Barack}}, \citenamefont {{Cardoso}}, \citenamefont {{Nissanke}},
  \citenamefont {{Sotiriou}},\ and\ \citenamefont {{~\textit{et
  al.}}}}]{BaraETC18}%
  \BibitemOpen
  \bibfield  {author} {\bibinfo {author} {\bibfnamefont {L.}~\bibnamefont
  {{Barack}}}, \bibinfo {author} {\bibfnamefont {V.}~\bibnamefont {{Cardoso}}},
  \bibinfo {author} {\bibfnamefont {S.}~\bibnamefont {{Nissanke}}}, \bibinfo
  {author} {\bibfnamefont {T.~P.}\ \bibnamefont {{Sotiriou}}}, \ and\ \bibinfo
  {author} {\bibnamefont {{~\textit{et al.}}}},\ }\href {\doibase
  10.1088/1361-6382/ab0587} {\bibfield  {journal} {\bibinfo  {journal}
  {Classical and Quantum Gravity}\ }\textbf {\bibinfo {volume} {36}},\ \bibinfo
  {eid} {143001} (\bibinfo {year} {2019})},\ \Eprint
  {http://arxiv.org/abs/1806.05195} {arXiv:1806.05195 [gr-qc]} \BibitemShut
  {NoStop}%
\bibitem [{\citenamefont {{Amaro-Seoane}}\ \emph {et~al.}(2007)\citenamefont
  {{Amaro-Seoane}}, \citenamefont {{Gair}}, \citenamefont {{Freitag}},
  \citenamefont {{Miller}}, \citenamefont {{Mandel}}, \citenamefont
  {{Cutler}},\ and\ \citenamefont {{Babak}}}]{AmarETC07}%
  \BibitemOpen
  \bibfield  {author} {\bibinfo {author} {\bibfnamefont {P.}~\bibnamefont
  {{Amaro-Seoane}}}, \bibinfo {author} {\bibfnamefont {J.~R.}\ \bibnamefont
  {{Gair}}}, \bibinfo {author} {\bibfnamefont {M.}~\bibnamefont {{Freitag}}},
  \bibinfo {author} {\bibfnamefont {M.~C.}\ \bibnamefont {{Miller}}}, \bibinfo
  {author} {\bibfnamefont {I.}~\bibnamefont {{Mandel}}}, \bibinfo {author}
  {\bibfnamefont {C.~J.}\ \bibnamefont {{Cutler}}}, \ and\ \bibinfo {author}
  {\bibfnamefont {S.}~\bibnamefont {{Babak}}},\ }\href {\doibase
  10.1088/0264-9381/24/17/R01} {\bibfield  {journal} {\bibinfo  {journal}
  {Classical and Quantum Gravity}\ }\textbf {\bibinfo {volume} {24}},\ \bibinfo
  {pages} {R113} (\bibinfo {year} {2007})},\ \Eprint
  {http://arxiv.org/abs/astro-ph/0703495} {astro-ph/0703495} \BibitemShut
  {NoStop}%
\bibitem [{\citenamefont {{Berry}}\ \emph {et~al.}(2019)\citenamefont
  {{Berry}}, \citenamefont {{Hughes}}, \citenamefont {{Sopuerta}},
  \citenamefont {{Chua}}, \citenamefont {{Heffernan}}, \citenamefont
  {{Holley-Bockelmann}}, \citenamefont {{Mihaylov}}, \citenamefont {{Miller}},\
  and\ \citenamefont {{Sesana}}}]{BerrETC19}%
  \BibitemOpen
  \bibfield  {author} {\bibinfo {author} {\bibfnamefont {C.}~\bibnamefont
  {{Berry}}}, \bibinfo {author} {\bibfnamefont {S.}~\bibnamefont {{Hughes}}},
  \bibinfo {author} {\bibfnamefont {C.}~\bibnamefont {{Sopuerta}}}, \bibinfo
  {author} {\bibfnamefont {A.}~\bibnamefont {{Chua}}}, \bibinfo {author}
  {\bibfnamefont {A.}~\bibnamefont {{Heffernan}}}, \bibinfo {author}
  {\bibfnamefont {K.}~\bibnamefont {{Holley-Bockelmann}}}, \bibinfo {author}
  {\bibfnamefont {D.}~\bibnamefont {{Mihaylov}}}, \bibinfo {author}
  {\bibfnamefont {C.}~\bibnamefont {{Miller}}}, \ and\ \bibinfo {author}
  {\bibfnamefont {A.}~\bibnamefont {{Sesana}}},\ }\href@noop {} {\bibfield
  {journal} {\bibinfo  {journal} {Bull. Am. Astron. Soc.}\ }\textbf {\bibinfo
  {volume} {51}},\ \bibinfo {eid} {42} (\bibinfo {year} {2019})},\ \Eprint
  {http://arxiv.org/abs/1903.03686} {arXiv:1903.03686 [astro-ph.HE]}
  \BibitemShut {NoStop}%
\bibitem [{LIS()}]{LISA}%
  \BibitemOpen
  \href@noop {} {\enquote {\bibinfo {title} {Lisa home page},}\ }\bibinfo
  {note} {\texttt{http://sci.esa.int/lisa/}}\BibitemShut {NoStop}%
\bibitem [{\citenamefont {{Barack}}\ and\ \citenamefont
  {{Pound}}(2018)}]{BaraPoun18}%
  \BibitemOpen
  \bibfield  {author} {\bibinfo {author} {\bibfnamefont {L.}~\bibnamefont
  {{Barack}}}\ and\ \bibinfo {author} {\bibfnamefont {A.}~\bibnamefont
  {{Pound}}},\ }\href {\doibase 10.1088/1361-6633/aae552} {\bibfield  {journal}
  {\bibinfo  {journal} {Reports on Progress in Physics}\ }\textbf {\bibinfo
  {volume} {82}} (\bibinfo {year} {2018}),\ 10.1088/1361-6633/aae552},\ \Eprint
  {http://arxiv.org/abs/1805.10385} {arXiv:1805.10385 [gr-qc]} \BibitemShut
  {NoStop}%
\bibitem [{\citenamefont {Regge}\ and\ \citenamefont
  {Wheeler}(1957)}]{ReggWhee57}%
  \BibitemOpen
  \bibfield  {author} {\bibinfo {author} {\bibfnamefont {T.}~\bibnamefont
  {Regge}}\ and\ \bibinfo {author} {\bibfnamefont {J.}~\bibnamefont
  {Wheeler}},\ }\href@noop {} {\bibfield  {journal} {\bibinfo  {journal} {Phys.
  Rev.}\ }\textbf {\bibinfo {volume} {108}},\ \bibinfo {pages} {1063} (\bibinfo
  {year} {1957})}\BibitemShut {NoStop}%
\bibitem [{\citenamefont {Zerilli}(1970)}]{Zeri70}%
  \BibitemOpen
  \bibfield  {author} {\bibinfo {author} {\bibfnamefont {F.}~\bibnamefont
  {Zerilli}},\ }\href@noop {} {\bibfield  {journal} {\bibinfo  {journal} {Phys.
  Rev. D}\ }\textbf {\bibinfo {volume} {2}},\ \bibinfo {pages} {2141} (\bibinfo
  {year} {1970})}\BibitemShut {NoStop}%
\bibitem [{\citenamefont {{Martel}}\ and\ \citenamefont
  {{Poisson}}(2005)}]{MartPois05}%
  \BibitemOpen
  \bibfield  {author} {\bibinfo {author} {\bibfnamefont {K.}~\bibnamefont
  {{Martel}}}\ and\ \bibinfo {author} {\bibfnamefont {E.}~\bibnamefont
  {{Poisson}}},\ }\href {\doibase 10.1103/PhysRevD.71.104003} {\bibfield
  {journal} {\bibinfo  {journal} {Phys. Rev. D}\ }\textbf {\bibinfo {volume}
  {71}},\ \bibinfo {pages} {104003} (\bibinfo {year} {2005})},\ \Eprint
  {http://arxiv.org/abs/arXiv:gr-qc/0502028} {arXiv:gr-qc/0502028} \BibitemShut
  {NoStop}%
\bibitem [{\citenamefont {Hopper}\ and\ \citenamefont
  {Evans}(2010)}]{HoppEvan10}%
  \BibitemOpen
  \bibfield  {author} {\bibinfo {author} {\bibfnamefont {S.}~\bibnamefont
  {Hopper}}\ and\ \bibinfo {author} {\bibfnamefont {C.~R.}\ \bibnamefont
  {Evans}},\ }\href {\doibase 10.1103/PhysRevD.82.084010} {\bibfield  {journal}
  {\bibinfo  {journal} {Phys. Rev. D}\ }\textbf {\bibinfo {volume} {82}},\
  \bibinfo {pages} {084010} (\bibinfo {year} {2010})}\BibitemShut {NoStop}%
\bibitem [{\citenamefont {{Hopper}}\ and\ \citenamefont
  {{Evans}}(2013)}]{HoppEvan13}%
  \BibitemOpen
  \bibfield  {author} {\bibinfo {author} {\bibfnamefont {S.}~\bibnamefont
  {{Hopper}}}\ and\ \bibinfo {author} {\bibfnamefont {C.~R.}\ \bibnamefont
  {{Evans}}},\ }\href {\doibase 10.1103/PhysRevD.87.064008} {\bibfield
  {journal} {\bibinfo  {journal} {Phys. Rev. D}\ }\textbf {\bibinfo {volume}
  {87}},\ \bibinfo {eid} {064008} (\bibinfo {year} {2013})},\ \Eprint
  {http://arxiv.org/abs/1210.7969} {arXiv:1210.7969 [gr-qc]} \BibitemShut
  {NoStop}%
\bibitem [{\citenamefont {Teukolsky}(1973)}]{Teuk73}%
  \BibitemOpen
  \bibfield  {author} {\bibinfo {author} {\bibfnamefont {S.}~\bibnamefont
  {Teukolsky}},\ }\href@noop {} {\bibfield  {journal} {\bibinfo  {journal}
  {Astrophys. J.}\ }\textbf {\bibinfo {volume} {185}},\ \bibinfo {pages} {635}
  (\bibinfo {year} {1973})}\BibitemShut {NoStop}%
\bibitem [{\citenamefont {Chrzanowski}(1975)}]{Chrz75}%
  \BibitemOpen
  \bibfield  {author} {\bibinfo {author} {\bibfnamefont {P.~L.}\ \bibnamefont
  {Chrzanowski}},\ }\href {\doibase 10.1103/PhysRevD.11.2042} {\bibfield
  {journal} {\bibinfo  {journal} {Phys. Rev. D}\ }\textbf {\bibinfo {volume}
  {11}},\ \bibinfo {pages} {2042} (\bibinfo {year} {1975})}\BibitemShut
  {NoStop}%
\bibitem [{\citenamefont {Kegeles}\ and\ \citenamefont
  {Cohen}(1979)}]{KegeCohe79}%
  \BibitemOpen
  \bibfield  {author} {\bibinfo {author} {\bibfnamefont {L.~S.}\ \bibnamefont
  {Kegeles}}\ and\ \bibinfo {author} {\bibfnamefont {J.~M.}\ \bibnamefont
  {Cohen}},\ }\href {\doibase 10.1103/PhysRevD.19.1641} {\bibfield  {journal}
  {\bibinfo  {journal} {Phys. Rev. D}\ }\textbf {\bibinfo {volume} {19}},\
  \bibinfo {pages} {1641} (\bibinfo {year} {1979})}\BibitemShut {NoStop}%
\bibitem [{\citenamefont {{van de Meent}}(2018)}]{Vand18}%
  \BibitemOpen
  \bibfield  {author} {\bibinfo {author} {\bibfnamefont {M.}~\bibnamefont {{van
  de Meent}}},\ }\href {\doibase 10.1103/PhysRevD.97.104033} {\bibfield
  {journal} {\bibinfo  {journal} {Phys. Rev. D}\ }\textbf {\bibinfo {volume}
  {97}},\ \bibinfo {eid} {104033} (\bibinfo {year} {2018}),\
  10.1103/PhysRevD.97.104033},\ \Eprint {http://arxiv.org/abs/1711.09607}
  {arXiv:1711.09607 [gr-qc]} \BibitemShut {NoStop}%
\bibitem [{\citenamefont {Pound}(2012)}]{Poun12a}%
  \BibitemOpen
  \bibfield  {author} {\bibinfo {author} {\bibfnamefont {A.}~\bibnamefont
  {Pound}},\ }\href {\doibase 10.1103/PhysRevLett.109.051101} {\bibfield
  {journal} {\bibinfo  {journal} {Physical Review Letters}\ }\textbf {\bibinfo
  {volume} {109}},\ \bibinfo {pages} {051101} (\bibinfo {year} {2012})},\
  \Eprint {http://arxiv.org/abs/1201.5089} {arXiv:1201.5089 [gr-qc]}
  \BibitemShut {NoStop}%
\bibitem [{\citenamefont {{Pound}}\ and\ \citenamefont
  {{Miller}}(2014)}]{PounMill14}%
  \BibitemOpen
  \bibfield  {author} {\bibinfo {author} {\bibfnamefont {A.}~\bibnamefont
  {{Pound}}}\ and\ \bibinfo {author} {\bibfnamefont {J.}~\bibnamefont
  {{Miller}}},\ }\href {\doibase 10.1103/PhysRevD.89.104020} {\bibfield
  {journal} {\bibinfo  {journal} {Phys. Rev. D}\ }\textbf {\bibinfo {volume}
  {89}},\ \bibinfo {eid} {104020} (\bibinfo {year} {2014})},\ \Eprint
  {http://arxiv.org/abs/1403.1843} {arXiv:1403.1843 [gr-qc]} \BibitemShut
  {NoStop}%
\bibitem [{\citenamefont {{Miller}}\ \emph {et~al.}(2016)\citenamefont
  {{Miller}}, \citenamefont {{Wardell}},\ and\ \citenamefont
  {{Pound}}}]{MillWardPoun16}%
  \BibitemOpen
  \bibfield  {author} {\bibinfo {author} {\bibfnamefont {J.}~\bibnamefont
  {{Miller}}}, \bibinfo {author} {\bibfnamefont {B.}~\bibnamefont {{Wardell}}},
  \ and\ \bibinfo {author} {\bibfnamefont {A.}~\bibnamefont {{Pound}}},\ }\href
  {\doibase 10.1103/PhysRevD.94.104018} {\bibfield  {journal} {\bibinfo
  {journal} {Phys. Rev. D}\ }\textbf {\bibinfo {volume} {94}},\ \bibinfo {eid}
  {104018} (\bibinfo {year} {2016})},\ \Eprint
  {http://arxiv.org/abs/1608.06783} {arXiv:1608.06783 [gr-qc]} \BibitemShut
  {NoStop}%
\bibitem [{\citenamefont {{Pound}}(2017)}]{Poun17}%
  \BibitemOpen
  \bibfield  {author} {\bibinfo {author} {\bibfnamefont {A.}~\bibnamefont
  {{Pound}}},\ }\href {\doibase 10.1103/PhysRevD.95.104056} {\bibfield
  {journal} {\bibinfo  {journal} {Phys. Rev. D}\ }\textbf {\bibinfo {volume}
  {95}},\ \bibinfo {eid} {104056} (\bibinfo {year} {2017})},\ \Eprint
  {http://arxiv.org/abs/1703.02836} {arXiv:1703.02836 [gr-qc]} \BibitemShut
  {NoStop}%
\bibitem [{\citenamefont {{Pound}}\ \emph {et~al.}(2020)\citenamefont
  {{Pound}}, \citenamefont {{Wardell}}, \citenamefont {{Warburton}},\ and\
  \citenamefont {{Miller}}}]{PounETC20}%
  \BibitemOpen
  \bibfield  {author} {\bibinfo {author} {\bibfnamefont {A.}~\bibnamefont
  {{Pound}}}, \bibinfo {author} {\bibfnamefont {B.}~\bibnamefont {{Wardell}}},
  \bibinfo {author} {\bibfnamefont {N.}~\bibnamefont {{Warburton}}}, \ and\
  \bibinfo {author} {\bibfnamefont {J.}~\bibnamefont {{Miller}}},\ }\href
  {\doibase 10.1103/PhysRevLett.124.021101} {\bibfield  {journal} {\bibinfo
  {journal} {\prl}\ }\textbf {\bibinfo {volume} {124}},\ \bibinfo {eid}
  {021101} (\bibinfo {year} {2020})},\ \Eprint
  {http://arxiv.org/abs/1908.07419} {arXiv:1908.07419 [gr-qc]} \BibitemShut
  {NoStop}%
\bibitem [{\citenamefont {{Blanchet}}(2014)}]{Blan14}%
  \BibitemOpen
  \bibfield  {author} {\bibinfo {author} {\bibfnamefont {L.}~\bibnamefont
  {{Blanchet}}},\ }\href {\doibase 10.12942/lrr-2014-2} {\bibfield  {journal}
  {\bibinfo  {journal} {Living Reviews in Relativity}\ }\textbf {\bibinfo
  {volume} {17}},\ \bibinfo {pages} {2} (\bibinfo {year} {2014})},\ \Eprint
  {http://arxiv.org/abs/1310.1528} {arXiv:1310.1528 [gr-qc]} \BibitemShut
  {NoStop}%
\bibitem [{\citenamefont {{Buonanno}}\ and\ \citenamefont
  {{Damour}}(1999)}]{BuonDamo99}%
  \BibitemOpen
  \bibfield  {author} {\bibinfo {author} {\bibfnamefont {A.}~\bibnamefont
  {{Buonanno}}}\ and\ \bibinfo {author} {\bibfnamefont {T.}~\bibnamefont
  {{Damour}}},\ }\href {\doibase 10.1103/PhysRevD.59.084006} {\bibfield
  {journal} {\bibinfo  {journal} {Phys. Rev. D}\ }\textbf {\bibinfo {volume}
  {59}},\ \bibinfo {eid} {084006} (\bibinfo {year} {1999})},\ \Eprint
  {http://arxiv.org/abs/gr-qc/9811091} {gr-qc/9811091} \BibitemShut {NoStop}%
\bibitem [{\citenamefont {{Damour}}(2001)}]{Damo01}%
  \BibitemOpen
  \bibfield  {author} {\bibinfo {author} {\bibfnamefont {T.}~\bibnamefont
  {{Damour}}},\ }\href {\doibase 10.1103/PhysRevD.64.124013} {\bibfield
  {journal} {\bibinfo  {journal} {Phys. Rev. D}\ }\textbf {\bibinfo {volume}
  {64}},\ \bibinfo {eid} {124013} (\bibinfo {year} {2001})},\ \Eprint
  {http://arxiv.org/abs/0103018} {arXiv:0103018 [gr-qc]} \BibitemShut {NoStop}%
\bibitem [{\citenamefont {{Damour}}(2010)}]{Damo10}%
  \BibitemOpen
  \bibfield  {author} {\bibinfo {author} {\bibfnamefont {T.}~\bibnamefont
  {{Damour}}},\ }\href {\doibase 10.1103/PhysRevD.81.024017} {\bibfield
  {journal} {\bibinfo  {journal} {Phys. Rev. D}\ }\textbf {\bibinfo {volume}
  {81}},\ \bibinfo {eid} {024017} (\bibinfo {year} {2010})},\ \Eprint
  {http://arxiv.org/abs/0910.5533} {arXiv:0910.5533 [gr-qc]} \BibitemShut
  {NoStop}%
\bibitem [{\citenamefont {{{Abbott}, B.~P.~et al.~(The LIGO Scientific
  Collaboration and VIRGO Collaboration)}}(2016{\natexlab{a}})}]{AbboETC16a}%
  \BibitemOpen
  \bibfield  {author} {\bibinfo {author} {\bibnamefont {{{Abbott}, B.~P.~et
  al.~(The LIGO Scientific Collaboration and VIRGO Collaboration)}}},\ }\href
  {\doibase 10.1103/PhysRevLett.116.061102} {\bibfield  {journal} {\bibinfo
  {journal} {Physical Review Letters}\ }\textbf {\bibinfo {volume} {116}},\
  \bibinfo {eid} {061102} (\bibinfo {year} {2016}{\natexlab{a}})},\ \Eprint
  {http://arxiv.org/abs/1602.03837} {arXiv:1602.03837 [gr-qc]} \BibitemShut
  {NoStop}%
\bibitem [{\citenamefont {{{Abbott}, B.~P.~et al.~(The LIGO Scientific
  Collaboration and VIRGO Collaboration)}}(2016{\natexlab{b}})}]{AbboETC16b}%
  \BibitemOpen
  \bibfield  {author} {\bibinfo {author} {\bibnamefont {{{Abbott}, B.~P.~et
  al.~(The LIGO Scientific Collaboration and VIRGO Collaboration)}}},\ }\href
  {\doibase 10.1103/PhysRevLett.116.241103} {\bibfield  {journal} {\bibinfo
  {journal} {Physical Review Letters}\ }\textbf {\bibinfo {volume} {116}},\
  \bibinfo {eid} {241103} (\bibinfo {year} {2016}{\natexlab{b}})},\ \Eprint
  {http://arxiv.org/abs/1606.04855} {arXiv:1606.04855 [gr-qc]} \BibitemShut
  {NoStop}%
\bibitem [{\citenamefont {{{Abbott}, B.~P.~et al.~(The LIGO Scientific
  Collaboration and VIRGO Collaboration)}}(2016{\natexlab{c}})}]{AbboETC16c}%
  \BibitemOpen
  \bibfield  {author} {\bibinfo {author} {\bibnamefont {{{Abbott}, B.~P.~et
  al.~(The LIGO Scientific Collaboration and VIRGO Collaboration)}}},\ }\href
  {\doibase 10.1103/PhysRevLett.116.221101} {\bibfield  {journal} {\bibinfo
  {journal} {Physical Review Letters}\ }\textbf {\bibinfo {volume} {116}},\
  \bibinfo {pages} {221101} (\bibinfo {year} {2016}{\natexlab{c}})}\BibitemShut
  {NoStop}%
\bibitem [{\citenamefont {{{Abbott}, B.~P.~et al.~(The LIGO Scientific
  Collaboration and VIRGO
  Collaboration)}}(2016{\natexlab{d}})}]{LVC1606.01210}%
  \BibitemOpen
  \bibfield  {author} {\bibinfo {author} {\bibnamefont {{{Abbott}, B.~P.~et
  al.~(The LIGO Scientific Collaboration and VIRGO Collaboration)}}} (\bibinfo
  {collaboration} {LIGO Scientific Collaboration and Virgo Collaboration}),\
  }\href {\doibase 10.1103/PhysRevX.6.041014} {\bibfield  {journal} {\bibinfo
  {journal} {Phys. Rev. X}\ }\textbf {\bibinfo {volume} {6}},\ \bibinfo {pages}
  {041014} (\bibinfo {year} {2016}{\natexlab{d}})}\BibitemShut {NoStop}%
\bibitem [{\citenamefont {{{Abbott}, B.~P.~et al.~(The LIGO Scientific
  Collaboration and VIRGO
  Collaboration)}}(2017{\natexlab{a}})}]{LVC1706.01812}%
  \BibitemOpen
  \bibfield  {author} {\bibinfo {author} {\bibnamefont {{{Abbott}, B.~P.~et
  al.~(The LIGO Scientific Collaboration and VIRGO Collaboration)}}} (\bibinfo
  {collaboration} {LIGO Scientific and Virgo Collaboration}),\ }\href {\doibase
  10.1103/PhysRevLett.118.221101} {\bibfield  {journal} {\bibinfo  {journal}
  {Physical Review Letters}\ }\textbf {\bibinfo {volume} {118}},\ \bibinfo
  {pages} {221101} (\bibinfo {year} {2017}{\natexlab{a}})}\BibitemShut
  {NoStop}%
\bibitem [{\citenamefont {{{Abbott}, B.~P.~et al.~(The LIGO Scientific
  Collaboration and VIRGO Collaboration)}}(2019)}]{LVC1811.12907}%
  \BibitemOpen
  \bibfield  {author} {\bibinfo {author} {\bibnamefont {{{Abbott}, B.~P.~et
  al.~(The LIGO Scientific Collaboration and VIRGO Collaboration)}}} (\bibinfo
  {collaboration} {LIGO Scientific Collaboration and Virgo Collaboration}),\
  }\href {\doibase 10.1103/PhysRevX.9.031040} {\bibfield  {journal} {\bibinfo
  {journal} {Phys. Rev. X}\ }\textbf {\bibinfo {volume} {9}},\ \bibinfo {pages}
  {031040} (\bibinfo {year} {2019})},\ \Eprint
  {http://arxiv.org/abs/1811.12907} {arXiv:1811.12907 [astro-ph.HE]}
  \BibitemShut {NoStop}%
\bibitem [{\citenamefont {{{Abbott}, B.~P.~et al.~(The LIGO Scientific
  Collaboration and VIRGO Collaboration)}}(2017{\natexlab{b}})}]{LVC171016}%
  \BibitemOpen
  \bibfield  {author} {\bibinfo {author} {\bibnamefont {{{Abbott}, B.~P.~et
  al.~(The LIGO Scientific Collaboration and VIRGO Collaboration)}}} (\bibinfo
  {collaboration} {LIGO Scientific Collaboration and Virgo Collaboration}),\
  }\href {\doibase 10.1103/PhysRevLett.119.161101} {\bibfield  {journal}
  {\bibinfo  {journal} {Physical Review Letters}\ }\textbf {\bibinfo {volume}
  {119}},\ \bibinfo {pages} {161101} (\bibinfo {year}
  {2017}{\natexlab{b}})}\BibitemShut {NoStop}%
\bibitem [{\citenamefont {{{Abbott}, B.~P.~et al.~(The LIGO Scientific
  Collaboration and VIRGO
  Collaboration)}}(2017{\natexlab{c}})}]{LVC1710.05833}%
  \BibitemOpen
  \bibfield  {author} {\bibinfo {author} {\bibnamefont {{{Abbott}, B.~P.~et
  al.~(The LIGO Scientific Collaboration and VIRGO Collaboration)}}} (\bibinfo
  {collaboration} {LIGO Scientific and Virgo Collaboration}),\ }\href {\doibase
  10.3847/2041-8213/aa91c9} {\bibfield  {journal} {\bibinfo  {journal} {The
  Astrophysical Journal Letters}\ }\textbf {\bibinfo {volume} {848}},\ \bibinfo
  {eid} {L12} (\bibinfo {year} {2017}{\natexlab{c}})},\ \Eprint
  {http://arxiv.org/abs/1710.05833} {arXiv:1710.05833 [astro-ph.HE]}
  \BibitemShut {NoStop}%
\bibitem [{\citenamefont {{{Abbott}, B.~P.~et al.~(The LIGO Scientific
  Collaboration and VIRGO
  Collaboration)}}(2017{\natexlab{d}})}]{LVC1710.05834}%
  \BibitemOpen
  \bibfield  {author} {\bibinfo {author} {\bibnamefont {{{Abbott}, B.~P.~et
  al.~(The LIGO Scientific Collaboration and VIRGO Collaboration)}}} (\bibinfo
  {collaboration} {LIGO Scientific and Virgo Collaboration}),\ }\href {\doibase
  10.3847/2041-8213/aa920c} {\bibfield  {journal} {\bibinfo  {journal} {The
  Astrophysical Journal Letters}\ }\textbf {\bibinfo {volume} {848}},\ \bibinfo
  {eid} {L13} (\bibinfo {year} {2017}{\natexlab{d}})},\ \Eprint
  {http://arxiv.org/abs/1710.05834} {arXiv:1710.05834 [astro-ph.HE]}
  \BibitemShut {NoStop}%
\bibitem [{\citenamefont {{The LIGO Scientific Collaboration}}\ and\
  \citenamefont {{the Virgo Collaboration}}(2020)}]{LVC2004.08342}%
  \BibitemOpen
  \bibfield  {author} {\bibinfo {author} {\bibnamefont {{The LIGO Scientific
  Collaboration}}}\ and\ \bibinfo {author} {\bibnamefont {{the Virgo
  Collaboration}}},\ }\href@noop {} {\bibfield  {journal} {\bibinfo  {journal}
  {arXiv e-prints}\ ,\ \bibinfo {eid} {arXiv:2004.08342}} (\bibinfo {year}
  {2020})},\ \Eprint {http://arxiv.org/abs/2004.08342} {arXiv:2004.08342
  [astro-ph.HE]} \BibitemShut {NoStop}%
\bibitem [{\citenamefont {{{Abbott}, B.~P.~et al.~(The LIGO Scientific
  Collaboration and VIRGO Collaboration)}}(2020)}]{LVC2006.12611}%
  \BibitemOpen
  \bibfield  {author} {\bibinfo {author} {\bibnamefont {{{Abbott}, B.~P.~et
  al.~(The LIGO Scientific Collaboration and VIRGO Collaboration)}}} (\bibinfo
  {collaboration} {LIGO Scientific and Virgo Collaboration}),\ }\href {\doibase
  10.3847/2041-8213/ab960f} {\bibfield  {journal} {\bibinfo  {journal} {The
  Astrophysical Journal Letters}\ }\textbf {\bibinfo {volume} {896}},\ \bibinfo
  {eid} {L44} (\bibinfo {year} {2020})},\ \Eprint
  {http://arxiv.org/abs/2006.12611} {arXiv:2006.12611 [astro-ph.HE]}
  \BibitemShut {NoStop}%
\bibitem [{\citenamefont {{Boh{\'e}}}\ \emph {et~al.}(2017)\citenamefont
  {{Boh{\'e}}}, \citenamefont {{Shao}}, \citenamefont {{Taracchini}},
  \citenamefont {{Buonanno}}, \citenamefont {{Babak}}, \citenamefont {{Harry}},
  \citenamefont {{Hinder}}, \citenamefont {{Ossokine}}, \citenamefont
  {{P{\"u}rrer}}, \citenamefont {{Raymond}}, \citenamefont {{Chu}},
  \citenamefont {{Fong}}, \citenamefont {{Kumar}}, \citenamefont {{Pfeiffer}},
  \citenamefont {{Boyle}}, \citenamefont {{Hemberger}}, \citenamefont
  {{Kidder}}, \citenamefont {{Lovelace}}, \citenamefont {{Scheel}},\ and\
  \citenamefont {{Szil{\'a}gyi}}}]{BoheETC17}%
  \BibitemOpen
  \bibfield  {author} {\bibinfo {author} {\bibfnamefont {A.}~\bibnamefont
  {{Boh{\'e}}}}, \bibinfo {author} {\bibfnamefont {L.}~\bibnamefont {{Shao}}},
  \bibinfo {author} {\bibfnamefont {A.}~\bibnamefont {{Taracchini}}}, \bibinfo
  {author} {\bibfnamefont {A.}~\bibnamefont {{Buonanno}}}, \bibinfo {author}
  {\bibfnamefont {S.}~\bibnamefont {{Babak}}}, \bibinfo {author} {\bibfnamefont
  {I.~W.}\ \bibnamefont {{Harry}}}, \bibinfo {author} {\bibfnamefont
  {I.}~\bibnamefont {{Hinder}}}, \bibinfo {author} {\bibfnamefont
  {S.}~\bibnamefont {{Ossokine}}}, \bibinfo {author} {\bibfnamefont
  {M.}~\bibnamefont {{P{\"u}rrer}}}, \bibinfo {author} {\bibfnamefont
  {V.}~\bibnamefont {{Raymond}}}, \bibinfo {author} {\bibfnamefont
  {T.}~\bibnamefont {{Chu}}}, \bibinfo {author} {\bibfnamefont
  {H.}~\bibnamefont {{Fong}}}, \bibinfo {author} {\bibfnamefont
  {P.}~\bibnamefont {{Kumar}}}, \bibinfo {author} {\bibfnamefont {H.~P.}\
  \bibnamefont {{Pfeiffer}}}, \bibinfo {author} {\bibfnamefont
  {M.}~\bibnamefont {{Boyle}}}, \bibinfo {author} {\bibfnamefont {D.~A.}\
  \bibnamefont {{Hemberger}}}, \bibinfo {author} {\bibfnamefont {L.~E.}\
  \bibnamefont {{Kidder}}}, \bibinfo {author} {\bibfnamefont {G.}~\bibnamefont
  {{Lovelace}}}, \bibinfo {author} {\bibfnamefont {M.~A.}\ \bibnamefont
  {{Scheel}}}, \ and\ \bibinfo {author} {\bibfnamefont {B.}~\bibnamefont
  {{Szil{\'a}gyi}}},\ }\href {\doibase 10.1103/PhysRevD.95.044028} {\bibfield
  {journal} {\bibinfo  {journal} {Phys. Rev. D}\ }\textbf {\bibinfo {volume}
  {95}},\ \bibinfo {eid} {044028} (\bibinfo {year} {2017})},\ \Eprint
  {http://arxiv.org/abs/1611.03703} {arXiv:1611.03703 [gr-qc]} \BibitemShut
  {NoStop}%
\bibitem [{\citenamefont {{Cotesta}}\ \emph {et~al.}(2018)\citenamefont
  {{Cotesta}}, \citenamefont {{Buonanno}}, \citenamefont {{Boh{\'e}}},
  \citenamefont {{Taracchini}},\ and\ \citenamefont {{Ossokine}}}]{CoteETC18}%
  \BibitemOpen
  \bibfield  {author} {\bibinfo {author} {\bibfnamefont {R.}~\bibnamefont
  {{Cotesta}}}, \bibinfo {author} {\bibfnamefont {A.}~\bibnamefont
  {{Buonanno}}}, \bibinfo {author} {\bibfnamefont {A.}~\bibnamefont
  {{Boh{\'e}}}}, \bibinfo {author} {\bibfnamefont {I.}~\bibnamefont
  {{Taracchini}}, \bibfnamefont {A.~{Hinder}}}, \ and\ \bibinfo {author}
  {\bibfnamefont {S.}~\bibnamefont {{Ossokine}}},\ }\href {\doibase
  10.1103/PhysRevD.98.084028} {\bibfield  {journal} {\bibinfo  {journal} {Phys.
  Rev. D}\ }\textbf {\bibinfo {volume} {98}},\ \bibinfo {eid} {084028}
  (\bibinfo {year} {2018})},\ \Eprint {http://arxiv.org/abs/1803.10701}
  {arXiv:1803.10701 [gr-qc]} \BibitemShut {NoStop}%
\bibitem [{\citenamefont {{Ossokine}}\ \emph {et~al.}(2020)\citenamefont
  {{Ossokine}}, \citenamefont {{Buonanno}}, \citenamefont {{Marsat}},
  \citenamefont {{Cotesta}}, \citenamefont {{Babak}}, \citenamefont
  {{Dietrich}}, \citenamefont {{Haas}}, \citenamefont {{Hinder}}, \citenamefont
  {{Pfeiffer}}, \citenamefont {{P{\"u}rrer}}, \citenamefont {{Woodford}},
  \citenamefont {{Boyle}}, \citenamefont {{Kidder}}, \citenamefont {{Scheel}},\
  and\ \citenamefont {{Szil{\'a}gyi}}}]{OssoETC20}%
  \BibitemOpen
  \bibfield  {author} {\bibinfo {author} {\bibfnamefont {S.}~\bibnamefont
  {{Ossokine}}}, \bibinfo {author} {\bibfnamefont {A.}~\bibnamefont
  {{Buonanno}}}, \bibinfo {author} {\bibfnamefont {S.}~\bibnamefont
  {{Marsat}}}, \bibinfo {author} {\bibfnamefont {R.}~\bibnamefont {{Cotesta}}},
  \bibinfo {author} {\bibfnamefont {S.}~\bibnamefont {{Babak}}}, \bibinfo
  {author} {\bibfnamefont {T.}~\bibnamefont {{Dietrich}}}, \bibinfo {author}
  {\bibfnamefont {R.}~\bibnamefont {{Haas}}}, \bibinfo {author} {\bibfnamefont
  {I.}~\bibnamefont {{Hinder}}}, \bibinfo {author} {\bibfnamefont {H.~P.}\
  \bibnamefont {{Pfeiffer}}}, \bibinfo {author} {\bibfnamefont
  {M.}~\bibnamefont {{P{\"u}rrer}}}, \bibinfo {author} {\bibfnamefont {C.~J.}\
  \bibnamefont {{Woodford}}}, \bibinfo {author} {\bibfnamefont
  {M.}~\bibnamefont {{Boyle}}}, \bibinfo {author} {\bibfnamefont {L.~E.}\
  \bibnamefont {{Kidder}}}, \bibinfo {author} {\bibfnamefont {M.~A.}\
  \bibnamefont {{Scheel}}}, \ and\ \bibinfo {author} {\bibfnamefont
  {B.}~\bibnamefont {{Szil{\'a}gyi}}},\ }\href@noop {} {\bibfield  {journal}
  {\bibinfo  {journal} {Phys. Rev. D}\ } (\bibinfo {year} {2020})},\ \Eprint
  {http://arxiv.org/abs/2004.09442} {arXiv:2004.09442 [gr-qc]} \BibitemShut
  {NoStop}%
\bibitem [{\citenamefont {Detweiler}(2008)}]{Detw08}%
  \BibitemOpen
  \bibfield  {author} {\bibinfo {author} {\bibfnamefont {S.}~\bibnamefont
  {Detweiler}},\ }\href {\doibase 10.1103/PhysRevD.77.124026} {\bibfield
  {journal} {\bibinfo  {journal} {Phys. Rev. D}\ }\textbf {\bibinfo {volume}
  {77}},\ \bibinfo {pages} {124026} (\bibinfo {year} {2008})},\ \Eprint
  {http://arxiv.org/abs/0804.3529} {arXiv:0804.3529 [gr-qc]} \BibitemShut
  {NoStop}%
\bibitem [{\citenamefont {Sago}\ \emph {et~al.}(2008)\citenamefont {Sago},
  \citenamefont {Barack},\ and\ \citenamefont {Detweiler}}]{SagoBaraDetw08}%
  \BibitemOpen
  \bibfield  {author} {\bibinfo {author} {\bibfnamefont {N.}~\bibnamefont
  {Sago}}, \bibinfo {author} {\bibfnamefont {L.}~\bibnamefont {Barack}}, \ and\
  \bibinfo {author} {\bibfnamefont {S.~L.}\ \bibnamefont {Detweiler}},\ }\href
  {\doibase 10.1103/PhysRevD.78.124024} {\bibfield  {journal} {\bibinfo
  {journal} {Phys. Rev. D}\ }\textbf {\bibinfo {volume} {78}},\ \bibinfo
  {pages} {124024} (\bibinfo {year} {2008})},\ \Eprint
  {http://arxiv.org/abs/0810.2530} {arXiv:0810.2530 [gr-qc]} \BibitemShut
  {NoStop}%
\bibitem [{\citenamefont {Barack}\ and\ \citenamefont
  {Sago}(2009)}]{BaraSago09}%
  \BibitemOpen
  \bibfield  {author} {\bibinfo {author} {\bibfnamefont {L.}~\bibnamefont
  {Barack}}\ and\ \bibinfo {author} {\bibfnamefont {N.}~\bibnamefont {Sago}},\
  }\href {\doibase 10.1103/PhysRevLett.102.191101} {\bibfield  {journal}
  {\bibinfo  {journal} {Physical Review Letters}\ }\textbf {\bibinfo {volume}
  {102}},\ \bibinfo {pages} {191101} (\bibinfo {year} {2009})},\ \Eprint
  {http://arxiv.org/abs/0902.0573} {arXiv:0902.0573 [gr-qc]} \BibitemShut
  {NoStop}%
\bibitem [{\citenamefont {{Blanchet}}\ \emph
  {et~al.}(2010{\natexlab{a}})\citenamefont {{Blanchet}}, \citenamefont
  {{Detweiler}}, \citenamefont {{Le Tiec}},\ and\ \citenamefont
  {{Whiting}}}]{BlanETC09}%
  \BibitemOpen
  \bibfield  {author} {\bibinfo {author} {\bibfnamefont {L.}~\bibnamefont
  {{Blanchet}}}, \bibinfo {author} {\bibfnamefont {S.}~\bibnamefont
  {{Detweiler}}}, \bibinfo {author} {\bibfnamefont {A.}~\bibnamefont {{Le
  Tiec}}}, \ and\ \bibinfo {author} {\bibfnamefont {B.~F.}\ \bibnamefont
  {{Whiting}}},\ }\href {\doibase 10.1103/PhysRevD.81.064004} {\bibfield
  {journal} {\bibinfo  {journal} {Phys. Rev. D}\ }\textbf {\bibinfo {volume}
  {81}},\ \bibinfo {eid} {064004} (\bibinfo {year} {2010}{\natexlab{a}})},\
  \Eprint {http://arxiv.org/abs/0910.0207} {arXiv:0910.0207 [gr-qc]}
  \BibitemShut {NoStop}%
\bibitem [{\citenamefont {{Blanchet}}\ \emph
  {et~al.}(2010{\natexlab{b}})\citenamefont {{Blanchet}}, \citenamefont
  {{Detweiler}}, \citenamefont {{Le Tiec}},\ and\ \citenamefont
  {{Whiting}}}]{BlanETC10}%
  \BibitemOpen
  \bibfield  {author} {\bibinfo {author} {\bibfnamefont {L.}~\bibnamefont
  {{Blanchet}}}, \bibinfo {author} {\bibfnamefont {S.}~\bibnamefont
  {{Detweiler}}}, \bibinfo {author} {\bibfnamefont {A.}~\bibnamefont {{Le
  Tiec}}}, \ and\ \bibinfo {author} {\bibfnamefont {B.~F.}\ \bibnamefont
  {{Whiting}}},\ }\href {\doibase 10.1103/PhysRevD.81.084033} {\bibfield
  {journal} {\bibinfo  {journal} {Phys. Rev. D}\ }\textbf {\bibinfo {volume}
  {81}},\ \bibinfo {eid} {084033} (\bibinfo {year} {2010}{\natexlab{b}})},\
  \Eprint {http://arxiv.org/abs/1002.0726} {arXiv:1002.0726 [gr-qc]}
  \BibitemShut {NoStop}%
\bibitem [{\citenamefont {{Fujita}}(2012{\natexlab{a}})}]{Fuji12a}%
  \BibitemOpen
  \bibfield  {author} {\bibinfo {author} {\bibfnamefont {R.}~\bibnamefont
  {{Fujita}}},\ }\href {\doibase 10.1143/PTP.127.583} {\bibfield  {journal}
  {\bibinfo  {journal} {Progress of Theoretical Physics}\ }\textbf {\bibinfo
  {volume} {127}},\ \bibinfo {pages} {583} (\bibinfo {year}
  {2012}{\natexlab{a}})},\ \Eprint {http://arxiv.org/abs/1104.5615}
  {arXiv:1104.5615 [gr-qc]} \BibitemShut {NoStop}%
\bibitem [{\citenamefont {{Fujita}}(2012{\natexlab{b}})}]{Fuji12b}%
  \BibitemOpen
  \bibfield  {author} {\bibinfo {author} {\bibfnamefont {R.}~\bibnamefont
  {{Fujita}}},\ }\href@noop {} {\bibfield  {journal} {\bibinfo  {journal}
  {Progress of Theoretical Physics}\ }\textbf {\bibinfo {volume} {128}},\
  \bibinfo {pages} {971} (\bibinfo {year} {2012}{\natexlab{b}})},\ \Eprint
  {http://arxiv.org/abs/1211.5535} {arXiv:1211.5535 [gr-qc]} \BibitemShut
  {NoStop}%
\bibitem [{\citenamefont {{Shah}}\ \emph {et~al.}(2014)\citenamefont {{Shah}},
  \citenamefont {{Friedman}},\ and\ \citenamefont
  {{Whiting}}}]{ShahFrieWhit14}%
  \BibitemOpen
  \bibfield  {author} {\bibinfo {author} {\bibfnamefont {A.~G.}\ \bibnamefont
  {{Shah}}}, \bibinfo {author} {\bibfnamefont {J.~L.}\ \bibnamefont
  {{Friedman}}}, \ and\ \bibinfo {author} {\bibfnamefont {B.~F.}\ \bibnamefont
  {{Whiting}}},\ }\href {\doibase 10.1103/PhysRevD.89.064042} {\bibfield
  {journal} {\bibinfo  {journal} {Phys. Rev. D}\ }\textbf {\bibinfo {volume}
  {89}},\ \bibinfo {eid} {064042} (\bibinfo {year} {2014})},\ \Eprint
  {http://arxiv.org/abs/1312.1952} {arXiv:1312.1952 [gr-qc]} \BibitemShut
  {NoStop}%
\bibitem [{\citenamefont {Shah}(2014)}]{Shah14}%
  \BibitemOpen
  \bibfield  {author} {\bibinfo {author} {\bibfnamefont {A.~G.}\ \bibnamefont
  {Shah}},\ }\href {\doibase 10.1103/PhysRevD.90.044025} {\bibfield  {journal}
  {\bibinfo  {journal} {Phys. Rev. D}\ }\textbf {\bibinfo {volume} {90}},\
  \bibinfo {pages} {044025} (\bibinfo {year} {2014})},\ \Eprint
  {http://arxiv.org/abs/1403.2697} {arXiv:1403.2697 [gr-qc]} \BibitemShut
  {NoStop}%
\bibitem [{\citenamefont {{Dolan}}\ \emph {et~al.}(2015)\citenamefont
  {{Dolan}}, \citenamefont {{Nolan}}, \citenamefont {{Ottewill}}, \citenamefont
  {{Warburton}},\ and\ \citenamefont {{Wardell}}}]{DolaETC14b}%
  \BibitemOpen
  \bibfield  {author} {\bibinfo {author} {\bibfnamefont {S.~R.}\ \bibnamefont
  {{Dolan}}}, \bibinfo {author} {\bibfnamefont {P.}~\bibnamefont {{Nolan}}},
  \bibinfo {author} {\bibfnamefont {A.~C.}\ \bibnamefont {{Ottewill}}},
  \bibinfo {author} {\bibfnamefont {N.}~\bibnamefont {{Warburton}}}, \ and\
  \bibinfo {author} {\bibfnamefont {B.}~\bibnamefont {{Wardell}}},\ }\href
  {\doibase 10.1103/PhysRevD.91.023009} {\bibfield  {journal} {\bibinfo
  {journal} {Phys. Rev. D}\ }\textbf {\bibinfo {volume} {91}},\ \bibinfo {eid}
  {023009} (\bibinfo {year} {2015})},\ \Eprint {http://arxiv.org/abs/1406.4890}
  {arXiv:1406.4890 [gr-qc]} \BibitemShut {NoStop}%
\bibitem [{\citenamefont {{Johnson-McDaniel}}\ \emph
  {et~al.}(2015)\citenamefont {{Johnson-McDaniel}}, \citenamefont {{Shah}},\
  and\ \citenamefont {{Whiting}}}]{JohnMcDaShahWhit15}%
  \BibitemOpen
  \bibfield  {author} {\bibinfo {author} {\bibfnamefont {N.~K.}\ \bibnamefont
  {{Johnson-McDaniel}}}, \bibinfo {author} {\bibfnamefont {A.~G.}\ \bibnamefont
  {{Shah}}}, \ and\ \bibinfo {author} {\bibfnamefont {B.~F.}\ \bibnamefont
  {{Whiting}}},\ }\href {\doibase 10.1103/PhysRevD.92.044007} {\bibfield
  {journal} {\bibinfo  {journal} {Phys. Rev. D}\ }\textbf {\bibinfo {volume}
  {92}},\ \bibinfo {eid} {044007} (\bibinfo {year} {2015})},\ \Eprint
  {http://arxiv.org/abs/1503.02638} {arXiv:1503.02638 [gr-qc]} \BibitemShut
  {NoStop}%
\bibitem [{\citenamefont {Bini}\ \emph {et~al.}(2016)\citenamefont {Bini},
  \citenamefont {Damour},\ and\ \citenamefont {Geralico}}]{BiniDamoGera15}%
  \BibitemOpen
  \bibfield  {author} {\bibinfo {author} {\bibfnamefont {D.}~\bibnamefont
  {Bini}}, \bibinfo {author} {\bibfnamefont {T.}~\bibnamefont {Damour}}, \ and\
  \bibinfo {author} {\bibfnamefont {A.}~\bibnamefont {Geralico}},\ }\href@noop
  {} {\bibfield  {journal} {\bibinfo  {journal} {Phys. Rev. D}\ }\textbf
  {\bibinfo {volume} {93}} (\bibinfo {year} {2016})},\ \Eprint
  {http://arxiv.org/abs/1511.04533} {arXiv:1511.04533 [gr-qc]} \BibitemShut
  {NoStop}%
\bibitem [{\citenamefont {Kavanagh}\ \emph {et~al.}(2015)\citenamefont
  {Kavanagh}, \citenamefont {Ottewill},\ and\ \citenamefont
  {Wardell}}]{KavaOtteWard15}%
  \BibitemOpen
  \bibfield  {author} {\bibinfo {author} {\bibfnamefont {C.}~\bibnamefont
  {Kavanagh}}, \bibinfo {author} {\bibfnamefont {A.~C.}\ \bibnamefont
  {Ottewill}}, \ and\ \bibinfo {author} {\bibfnamefont {B.}~\bibnamefont
  {Wardell}},\ }\href {\doibase 10.1103/PhysRevD.92.084025} {\bibfield
  {journal} {\bibinfo  {journal} {Phys. Rev. D}\ }\textbf {\bibinfo {volume}
  {92}},\ \bibinfo {pages} {084025} (\bibinfo {year} {2015})},\ \Eprint
  {http://arxiv.org/abs/1503.02334} {arXiv:1503.02334 [gr-qc]} \BibitemShut
  {NoStop}%
\bibitem [{\citenamefont {{Akcay}}\ \emph {et~al.}(2015)\citenamefont
  {{Akcay}}, \citenamefont {{Le Tiec}}, \citenamefont {{Barack}}, \citenamefont
  {{Sago}},\ and\ \citenamefont {{Warburton}}}]{AkcaETC15}%
  \BibitemOpen
  \bibfield  {author} {\bibinfo {author} {\bibfnamefont {S.}~\bibnamefont
  {{Akcay}}}, \bibinfo {author} {\bibfnamefont {A.}~\bibnamefont {{Le Tiec}}},
  \bibinfo {author} {\bibfnamefont {L.}~\bibnamefont {{Barack}}}, \bibinfo
  {author} {\bibfnamefont {N.}~\bibnamefont {{Sago}}}, \ and\ \bibinfo {author}
  {\bibfnamefont {N.}~\bibnamefont {{Warburton}}},\ }\href {\doibase
  10.1103/PhysRevD.91.124014} {\bibfield  {journal} {\bibinfo  {journal} {Phys.
  Rev. D}\ }\textbf {\bibinfo {volume} {91}},\ \bibinfo {eid} {124014}
  (\bibinfo {year} {2015})},\ \Eprint {http://arxiv.org/abs/1503.01374}
  {arXiv:1503.01374 [gr-qc]} \BibitemShut {NoStop}%
\bibitem [{\citenamefont {{Sago}}\ and\ \citenamefont
  {{Fujita}}(2015)}]{SagoFuji15}%
  \BibitemOpen
  \bibfield  {author} {\bibinfo {author} {\bibfnamefont {N.}~\bibnamefont
  {{Sago}}}\ and\ \bibinfo {author} {\bibfnamefont {R.}~\bibnamefont
  {{Fujita}}},\ }\href {\doibase 10.1093/ptep/ptv092} {\bibfield  {journal}
  {\bibinfo  {journal} {Progress of Theoretical and Experimental Physics}\
  }\textbf {\bibinfo {volume} {2015}},\ \bibinfo {eid} {073E03} (\bibinfo
  {year} {2015})},\ \Eprint {http://arxiv.org/abs/1505.01600} {arXiv:1505.01600
  [gr-qc]} \BibitemShut {NoStop}%
\bibitem [{\citenamefont {Forseth}\ \emph {et~al.}(2016)\citenamefont
  {Forseth}, \citenamefont {Evans},\ and\ \citenamefont
  {Hopper}}]{ForsEvanHopp16}%
  \BibitemOpen
  \bibfield  {author} {\bibinfo {author} {\bibfnamefont {E.}~\bibnamefont
  {Forseth}}, \bibinfo {author} {\bibfnamefont {C.~R.}\ \bibnamefont {Evans}},
  \ and\ \bibinfo {author} {\bibfnamefont {S.}~\bibnamefont {Hopper}},\ }\href
  {\doibase 10.1103/PhysRevD.93.064058} {\bibfield  {journal} {\bibinfo
  {journal} {Phys. Rev. D}\ }\textbf {\bibinfo {volume} {93}},\ \bibinfo
  {pages} {064058} (\bibinfo {year} {2016})}\BibitemShut {NoStop}%
\bibitem [{\citenamefont {{Hopper}}\ \emph {et~al.}(2016)\citenamefont
  {{Hopper}}, \citenamefont {{Kavanagh}},\ and\ \citenamefont
  {{Ottewill}}}]{HoppKavaOtte16}%
  \BibitemOpen
  \bibfield  {author} {\bibinfo {author} {\bibfnamefont {S.}~\bibnamefont
  {{Hopper}}}, \bibinfo {author} {\bibfnamefont {C.}~\bibnamefont
  {{Kavanagh}}}, \ and\ \bibinfo {author} {\bibfnamefont {A.~C.}\ \bibnamefont
  {{Ottewill}}},\ }\href {\doibase 10.1103/PhysRevD.93.044010} {\bibfield
  {journal} {\bibinfo  {journal} {Phys. Rev. D}\ }\textbf {\bibinfo {volume}
  {93}},\ \bibinfo {eid} {044010} (\bibinfo {year} {2016})},\ \Eprint
  {http://arxiv.org/abs/1512.01556} {arXiv:1512.01556 [gr-qc]} \BibitemShut
  {NoStop}%
\bibitem [{\citenamefont {{Kavanagh}}\ \emph {et~al.}(2016)\citenamefont
  {{Kavanagh}}, \citenamefont {{Ottewill}},\ and\ \citenamefont
  {{Wardell}}}]{KavaOtteWard16}%
  \BibitemOpen
  \bibfield  {author} {\bibinfo {author} {\bibfnamefont {C.}~\bibnamefont
  {{Kavanagh}}}, \bibinfo {author} {\bibfnamefont {A.~C.}\ \bibnamefont
  {{Ottewill}}}, \ and\ \bibinfo {author} {\bibfnamefont {B.}~\bibnamefont
  {{Wardell}}},\ }\href {\doibase 10.1103/PhysRevD.93.124038} {\bibfield
  {journal} {\bibinfo  {journal} {Phys. Rev. D}\ }\textbf {\bibinfo {volume}
  {93}},\ \bibinfo {eid} {124038} (\bibinfo {year} {2016})},\ \Eprint
  {http://arxiv.org/abs/1601.03394} {arXiv:1601.03394 [gr-qc]} \BibitemShut
  {NoStop}%
\bibitem [{\citenamefont {{Bini}}\ \emph
  {et~al.}(2016{\natexlab{a}})\citenamefont {{Bini}}, \citenamefont
  {{Damour}},\ and\ \citenamefont {{Geralico}}}]{BiniDamoGera16a}%
  \BibitemOpen
  \bibfield  {author} {\bibinfo {author} {\bibfnamefont {D.}~\bibnamefont
  {{Bini}}}, \bibinfo {author} {\bibfnamefont {T.}~\bibnamefont {{Damour}}}, \
  and\ \bibinfo {author} {\bibfnamefont {A.}~\bibnamefont {{Geralico}}},\
  }\href {\doibase 10.1103/PhysRevD.93.064023} {\bibfield  {journal} {\bibinfo
  {journal} {Physical Review D}\ }\textbf {\bibinfo {volume} {93}},\ \bibinfo
  {eid} {064023} (\bibinfo {year} {2016}{\natexlab{a}})},\ \Eprint
  {http://arxiv.org/abs/1511.04533} {arXiv:1511.04533 [gr-qc]} \BibitemShut
  {NoStop}%
\bibitem [{\citenamefont {{Bini}}\ \emph
  {et~al.}(2016{\natexlab{b}})\citenamefont {{Bini}}, \citenamefont
  {{Damour}},\ and\ \citenamefont {{Geralico}}}]{BiniDamoGera16b}%
  \BibitemOpen
  \bibfield  {author} {\bibinfo {author} {\bibfnamefont {D.}~\bibnamefont
  {{Bini}}}, \bibinfo {author} {\bibfnamefont {T.}~\bibnamefont {{Damour}}}, \
  and\ \bibinfo {author} {\bibfnamefont {A.}~\bibnamefont {{Geralico}}},\
  }\href {\doibase 10.1103/PhysRevD.93.104017} {\bibfield  {journal} {\bibinfo
  {journal} {Physical Review D}\ }\textbf {\bibinfo {volume} {93}},\ \bibinfo
  {eid} {104017} (\bibinfo {year} {2016}{\natexlab{b}})},\ \Eprint
  {http://arxiv.org/abs/1601.02988} {arXiv:1601.02988 [gr-qc]} \BibitemShut
  {NoStop}%
\bibitem [{\citenamefont {{Bini}}\ \emph
  {et~al.}(2016{\natexlab{c}})\citenamefont {{Bini}}, \citenamefont
  {{Damour}},\ and\ \citenamefont {{Geralico}}}]{BiniDamoGera16c}%
  \BibitemOpen
  \bibfield  {author} {\bibinfo {author} {\bibfnamefont {D.}~\bibnamefont
  {{Bini}}}, \bibinfo {author} {\bibfnamefont {T.}~\bibnamefont {{Damour}}}, \
  and\ \bibinfo {author} {\bibfnamefont {A.}~\bibnamefont {{Geralico}}},\
  }\href {\doibase 10.1103/PhysRevD.93.124058} {\bibfield  {journal} {\bibinfo
  {journal} {Physical Review D}\ }\textbf {\bibinfo {volume} {93}},\ \bibinfo
  {eid} {124058} (\bibinfo {year} {2016}{\natexlab{c}})},\ \Eprint
  {http://arxiv.org/abs/1602.08282} {arXiv:1602.08282 [gr-qc]} \BibitemShut
  {NoStop}%
\bibitem [{\citenamefont {{Nagar}}\ and\ \citenamefont
  {{Shah}}(2016)}]{NagaShah16}%
  \BibitemOpen
  \bibfield  {author} {\bibinfo {author} {\bibfnamefont {A.}~\bibnamefont
  {{Nagar}}}\ and\ \bibinfo {author} {\bibfnamefont {A.}~\bibnamefont
  {{Shah}}},\ }\href {\doibase 10.1103/PhysRevD.94.104017} {\bibfield
  {journal} {\bibinfo  {journal} {Phys. Rev. D}\ }\textbf {\bibinfo {volume}
  {94}},\ \bibinfo {eid} {104017} (\bibinfo {year} {2016})},\ \Eprint
  {http://arxiv.org/abs/1606.00207} {arXiv:1606.00207 [gr-qc]} \BibitemShut
  {NoStop}%
\bibitem [{\citenamefont {{Sago}}\ \emph {et~al.}(2016)\citenamefont {{Sago}},
  \citenamefont {{Fujita}},\ and\ \citenamefont {{Nakano}}}]{SagoFujiNaka16}%
  \BibitemOpen
  \bibfield  {author} {\bibinfo {author} {\bibfnamefont {N.}~\bibnamefont
  {{Sago}}}, \bibinfo {author} {\bibfnamefont {R.}~\bibnamefont {{Fujita}}}, \
  and\ \bibinfo {author} {\bibfnamefont {H.}~\bibnamefont {{Nakano}}},\ }\href
  {\doibase 10.1103/PhysRevD.93.104023} {\bibfield  {journal} {\bibinfo
  {journal} {Phys. Rev. D}\ }\textbf {\bibinfo {volume} {93}},\ \bibinfo {eid}
  {104023} (\bibinfo {year} {2016})},\ \Eprint
  {http://arxiv.org/abs/1601.02174} {arXiv:1601.02174 [gr-qc]} \BibitemShut
  {NoStop}%
\bibitem [{\citenamefont {Kavanagh}\ \emph {et~al.}(2017)\citenamefont
  {Kavanagh}, \citenamefont {Bini}, \citenamefont {Damour}, \citenamefont
  {Hopper}, \citenamefont {Ottewil},\ and\ \citenamefont
  {Wardell}}]{KavaETC17}%
  \BibitemOpen
  \bibfield  {author} {\bibinfo {author} {\bibfnamefont {C.}~\bibnamefont
  {Kavanagh}}, \bibinfo {author} {\bibfnamefont {D.}~\bibnamefont {Bini}},
  \bibinfo {author} {\bibfnamefont {T.}~\bibnamefont {Damour}}, \bibinfo
  {author} {\bibfnamefont {S.}~\bibnamefont {Hopper}}, \bibinfo {author}
  {\bibfnamefont {A.}~\bibnamefont {Ottewil}}, \ and\ \bibinfo {author}
  {\bibfnamefont {B.}~\bibnamefont {Wardell}},\ }\href {\doibase
  10.1103/PhysRevD.96.064012} {\bibfield  {journal} {\bibinfo  {journal} {Phys.
  Rev. D}\ }\textbf {\bibinfo {volume} {96}},\ \bibinfo {eid} {064012}
  (\bibinfo {year} {2017}),\ 10.1103/PhysRevD.96.064012},\ \Eprint
  {http://arxiv.org/abs/1706.00459} {arXiv:1706.00459 [gr-qc]} \BibitemShut
  {NoStop}%
\bibitem [{\citenamefont {{Bini}}\ \emph
  {et~al.}(2018{\natexlab{a}})\citenamefont {{Bini}}, \citenamefont {{Damour}},
  \citenamefont {{Geralico}}, \citenamefont {{Kavanagh}},\ and\ \citenamefont
  {{van de Meent}}}]{BiniETC18}%
  \BibitemOpen
  \bibfield  {author} {\bibinfo {author} {\bibfnamefont {D.}~\bibnamefont
  {{Bini}}}, \bibinfo {author} {\bibfnamefont {T.}~\bibnamefont {{Damour}}},
  \bibinfo {author} {\bibfnamefont {A.}~\bibnamefont {{Geralico}}}, \bibinfo
  {author} {\bibfnamefont {C.}~\bibnamefont {{Kavanagh}}}, \ and\ \bibinfo
  {author} {\bibfnamefont {M.}~\bibnamefont {{van de Meent}}},\ }\href
  {\doibase 10.1103/PhysRevD.98.104062} {\bibfield  {journal} {\bibinfo
  {journal} {Phys. Rev. D}\ }\textbf {\bibinfo {volume} {98}},\ \bibinfo {eid}
  {104062} (\bibinfo {year} {2018}{\natexlab{a}})},\ \Eprint
  {http://arxiv.org/abs/1809.02516} {arXiv:1809.02516 [gr-qc]} \BibitemShut
  {NoStop}%
\bibitem [{\citenamefont {{Bini}}\ \emph
  {et~al.}(2018{\natexlab{b}})\citenamefont {{Bini}}, \citenamefont {{Damour}},
  \citenamefont {{Geralico}}, \citenamefont {{Kavanagh}},\ and\ \citenamefont
  {{van de Meent}}}]{BiniETC18b}%
  \BibitemOpen
  \bibfield  {author} {\bibinfo {author} {\bibfnamefont {D.}~\bibnamefont
  {{Bini}}}, \bibinfo {author} {\bibfnamefont {T.}~\bibnamefont {{Damour}}},
  \bibinfo {author} {\bibfnamefont {A.}~\bibnamefont {{Geralico}}}, \bibinfo
  {author} {\bibfnamefont {C.}~\bibnamefont {{Kavanagh}}}, \ and\ \bibinfo
  {author} {\bibfnamefont {M.}~\bibnamefont {{van de Meent}}},\ }\href
  {\doibase 10.1103/PhysRevD.97.104022} {\bibfield  {journal} {\bibinfo
  {journal} {Phys. Rev. D}\ }\textbf {\bibinfo {volume} {97}},\ \bibinfo {eid}
  {104022} (\bibinfo {year} {2018}{\natexlab{b}})},\ \Eprint
  {http://arxiv.org/abs/1801.09616} {arXiv:1801.09616 [gr-qc]} \BibitemShut
  {NoStop}%
\bibitem [{\citenamefont {{Bini}}\ and\ \citenamefont
  {{Geralico}}(2018{\natexlab{a}})}]{BiniGera18a}%
  \BibitemOpen
  \bibfield  {author} {\bibinfo {author} {\bibfnamefont {D.}~\bibnamefont
  {{Bini}}}\ and\ \bibinfo {author} {\bibfnamefont {A.}~\bibnamefont
  {{Geralico}}},\ }\href {\doibase 10.1103/PhysRevD.98.064026} {\bibfield
  {journal} {\bibinfo  {journal} {Phys. Rev. D}\ }\textbf {\bibinfo {volume}
  {98}},\ \bibinfo {eid} {064026} (\bibinfo {year} {2018}{\natexlab{a}})},\
  \Eprint {http://arxiv.org/abs/1806.06635} {arXiv:1806.06635 [gr-qc]}
  \BibitemShut {NoStop}%
\bibitem [{\citenamefont {{Bini}}\ and\ \citenamefont
  {{Geralico}}(2018{\natexlab{b}})}]{BiniGera18b}%
  \BibitemOpen
  \bibfield  {author} {\bibinfo {author} {\bibfnamefont {D.}~\bibnamefont
  {{Bini}}}\ and\ \bibinfo {author} {\bibfnamefont {A.}~\bibnamefont
  {{Geralico}}},\ }\href {\doibase 10.1103/PhysRevD.98.064040} {\bibfield
  {journal} {\bibinfo  {journal} {Phys. Rev. D}\ }\textbf {\bibinfo {volume}
  {98}},\ \bibinfo {eid} {064040} (\bibinfo {year} {2018}{\natexlab{b}})},\
  \Eprint {http://arxiv.org/abs/1806.08765} {arXiv:1806.08765 [gr-qc]}
  \BibitemShut {NoStop}%
\bibitem [{\citenamefont {{Bini}}\ and\ \citenamefont
  {{Geralico}}(2018{\natexlab{c}})}]{BiniGera18c}%
  \BibitemOpen
  \bibfield  {author} {\bibinfo {author} {\bibfnamefont {D.}~\bibnamefont
  {{Bini}}}\ and\ \bibinfo {author} {\bibfnamefont {A.}~\bibnamefont
  {{Geralico}}},\ }\href {\doibase 10.1103/PhysRevD.98.084021} {\bibfield
  {journal} {\bibinfo  {journal} {Phys. Rev. D}\ }\textbf {\bibinfo {volume}
  {98}},\ \bibinfo {eid} {084021} (\bibinfo {year} {2018}{\natexlab{c}})},\
  \Eprint {http://arxiv.org/abs/1806.03495} {arXiv:1806.03495 [gr-qc]}
  \BibitemShut {NoStop}%
\bibitem [{\citenamefont {{Bini}}\ \emph
  {et~al.}(2018{\natexlab{c}})\citenamefont {{Bini}}, \citenamefont
  {{Damour}},\ and\ \citenamefont {{Geralico}}}]{BiniDamoGera18}%
  \BibitemOpen
  \bibfield  {author} {\bibinfo {author} {\bibfnamefont {D.}~\bibnamefont
  {{Bini}}}, \bibinfo {author} {\bibfnamefont {T.}~\bibnamefont {{Damour}}}, \
  and\ \bibinfo {author} {\bibfnamefont {A.}~\bibnamefont {{Geralico}}},\
  }\href {\doibase 10.1103/PhysRevD.97.104046} {\bibfield  {journal} {\bibinfo
  {journal} {Phys. Rev. D}\ }\textbf {\bibinfo {volume} {97}},\ \bibinfo {eid}
  {104046} (\bibinfo {year} {2018}{\natexlab{c}})},\ \Eprint
  {http://arxiv.org/abs/1801.03704} {arXiv:1801.03704 [gr-qc]} \BibitemShut
  {NoStop}%
\bibitem [{\citenamefont {{Bini}}\ and\ \citenamefont
  {{Geralico}}(2019{\natexlab{a}})}]{BiniGera19a}%
  \BibitemOpen
  \bibfield  {author} {\bibinfo {author} {\bibfnamefont {D.}~\bibnamefont
  {{Bini}}}\ and\ \bibinfo {author} {\bibfnamefont {A.}~\bibnamefont
  {{Geralico}}},\ }\href {\doibase 10.1103/PhysRevD.100.104002} {\bibfield
  {journal} {\bibinfo  {journal} {Phys. Rev. D}\ }\textbf {\bibinfo {volume}
  {100}},\ \bibinfo {eid} {104002} (\bibinfo {year} {2019}{\natexlab{a}})},\
  \Eprint {http://arxiv.org/abs/1907.11080} {arXiv:1907.11080 [gr-qc]}
  \BibitemShut {NoStop}%
\bibitem [{\citenamefont {{Bini}}\ and\ \citenamefont
  {{Geralico}}(2019{\natexlab{b}})}]{BiniGera19b}%
  \BibitemOpen
  \bibfield  {author} {\bibinfo {author} {\bibfnamefont {D.}~\bibnamefont
  {{Bini}}}\ and\ \bibinfo {author} {\bibfnamefont {A.}~\bibnamefont
  {{Geralico}}},\ }\href {\doibase 10.1103/PhysRevD.100.104003} {\bibfield
  {journal} {\bibinfo  {journal} {Phys. Rev. D}\ }\textbf {\bibinfo {volume}
  {100}},\ \bibinfo {eid} {104003} (\bibinfo {year} {2019}{\natexlab{b}})},\
  \Eprint {http://arxiv.org/abs/1907.11082} {arXiv:1907.11082 [gr-qc]}
  \BibitemShut {NoStop}%
\bibitem [{\citenamefont {{Bini}}\ and\ \citenamefont
  {{Geralico}}(2019{\natexlab{c}})}]{BiniGera19c}%
  \BibitemOpen
  \bibfield  {author} {\bibinfo {author} {\bibfnamefont {D.}~\bibnamefont
  {{Bini}}}\ and\ \bibinfo {author} {\bibfnamefont {A.}~\bibnamefont
  {{Geralico}}},\ }\href {\doibase 10.1103/PhysRevD.100.121502} {\bibfield
  {journal} {\bibinfo  {journal} {Phys. Rev. D}\ }\textbf {\bibinfo {volume}
  {100}},\ \bibinfo {eid} {121502} (\bibinfo {year} {2019}{\natexlab{c}})},\
  \Eprint {http://arxiv.org/abs/1907.11083} {arXiv:1907.11083 [gr-qc]}
  \BibitemShut {NoStop}%
\bibitem [{\citenamefont {{Nagar}}\ \emph {et~al.}(2019)\citenamefont
  {{Nagar}}, \citenamefont {{Messina}}, \citenamefont {{Kavanagh}},
  \citenamefont {{Lukes-Gerakopoulos}}, \citenamefont {{Warburton}},
  \citenamefont {{Bernuzzi}},\ and\ \citenamefont {{Harms}}}]{NagaETC19}%
  \BibitemOpen
  \bibfield  {author} {\bibinfo {author} {\bibfnamefont {A.}~\bibnamefont
  {{Nagar}}}, \bibinfo {author} {\bibfnamefont {F.}~\bibnamefont {{Messina}}},
  \bibinfo {author} {\bibfnamefont {C.}~\bibnamefont {{Kavanagh}}}, \bibinfo
  {author} {\bibfnamefont {G.}~\bibnamefont {{Lukes-Gerakopoulos}}}, \bibinfo
  {author} {\bibfnamefont {N.}~\bibnamefont {{Warburton}}}, \bibinfo {author}
  {\bibfnamefont {S.}~\bibnamefont {{Bernuzzi}}}, \ and\ \bibinfo {author}
  {\bibfnamefont {E.}~\bibnamefont {{Harms}}},\ }\href {\doibase
  10.1103/PhysRevD.100.104056} {\bibfield  {journal} {\bibinfo  {journal}
  {Phys. Rev. D}\ }\textbf {\bibinfo {volume} {100}},\ \bibinfo {eid} {104056}
  (\bibinfo {year} {2019})},\ \Eprint {http://arxiv.org/abs/1907.12233}
  {arXiv:1907.12233 [gr-qc]} \BibitemShut {NoStop}%
\bibitem [{\citenamefont {{Munna}}\ and\ \citenamefont
  {{Evans}}(2019)}]{MunnEvan19a}%
  \BibitemOpen
  \bibfield  {author} {\bibinfo {author} {\bibfnamefont {C.}~\bibnamefont
  {{Munna}}}\ and\ \bibinfo {author} {\bibfnamefont {C.~R.}\ \bibnamefont
  {{Evans}}},\ }\href {\doibase 10.1103/PhysRevD.100.104060} {\bibfield
  {journal} {\bibinfo  {journal} {Phys. Rev. D}\ }\textbf {\bibinfo {volume}
  {100}},\ \bibinfo {eid} {104060} (\bibinfo {year} {2019})},\ \Eprint
  {http://arxiv.org/abs/1909.05877} {arXiv:1909.05877 [gr-qc]} \BibitemShut
  {NoStop}%
\bibitem [{\citenamefont {Munna}\ and\ \citenamefont
  {Evans}({\natexlab{a}})}]{MunnEvan19b}%
  \BibitemOpen
  \bibfield  {author} {\bibinfo {author} {\bibfnamefont {C.}~\bibnamefont
  {Munna}}\ and\ \bibinfo {author} {\bibfnamefont {C.~R.}\ \bibnamefont
  {Evans}},\ }\href@noop {} {\bibfield  {journal} {\bibinfo  {journal} {to be
  submitted to Phys. Rev. D}\ } ({\natexlab{a}})}\BibitemShut {NoStop}%
\bibitem [{\citenamefont {{Antonelli}}\ \emph {et~al.}(2020)\citenamefont
  {{Antonelli}}, \citenamefont {{Kavanagh}}, \citenamefont {{Khalil}},
  \citenamefont {{Steinhoff}},\ and\ \citenamefont {{Vines}}}]{AntoETC20}%
  \BibitemOpen
  \bibfield  {author} {\bibinfo {author} {\bibfnamefont {A.}~\bibnamefont
  {{Antonelli}}}, \bibinfo {author} {\bibfnamefont {C.}~\bibnamefont
  {{Kavanagh}}}, \bibinfo {author} {\bibfnamefont {M.}~\bibnamefont
  {{Khalil}}}, \bibinfo {author} {\bibfnamefont {J.}~\bibnamefont
  {{Steinhoff}}}, \ and\ \bibinfo {author} {\bibfnamefont {J.}~\bibnamefont
  {{Vines}}},\ }\href@noop {} {\bibfield  {journal} {\bibinfo  {journal} {Phys.
  Rev. Lett.}\ }\textbf {\bibinfo {volume} {125}},\ \bibinfo {pages} {011103}
  (\bibinfo {year} {2020})},\ \Eprint {http://arxiv.org/abs/2003.11391}
  {arXiv:2003.11391 [gr-qc]} \BibitemShut {NoStop}%
\bibitem [{\citenamefont {{Bini}}\ \emph
  {et~al.}(2020{\natexlab{a}})\citenamefont {{Bini}}, \citenamefont
  {{Damour}},\ and\ \citenamefont {{Geralico}}}]{BiniDamoGera20a}%
  \BibitemOpen
  \bibfield  {author} {\bibinfo {author} {\bibfnamefont {D.}~\bibnamefont
  {{Bini}}}, \bibinfo {author} {\bibfnamefont {T.}~\bibnamefont {{Damour}}}, \
  and\ \bibinfo {author} {\bibfnamefont {A.}~\bibnamefont {{Geralico}}},\
  }\href@noop {} {\  (\bibinfo {year} {2020}{\natexlab{a}})},\ \Eprint
  {http://arxiv.org/abs/2003.11891} {arXiv:2003.11891 [gr-qc]} \BibitemShut
  {NoStop}%
\bibitem [{\citenamefont {{Bini}}\ \emph
  {et~al.}(2020{\natexlab{b}})\citenamefont {{Bini}}, \citenamefont
  {{Damour}},\ and\ \citenamefont {{Geralico}}}]{BiniDamoGera20b}%
  \BibitemOpen
  \bibfield  {author} {\bibinfo {author} {\bibfnamefont {D.}~\bibnamefont
  {{Bini}}}, \bibinfo {author} {\bibfnamefont {T.}~\bibnamefont {{Damour}}}, \
  and\ \bibinfo {author} {\bibfnamefont {A.}~\bibnamefont {{Geralico}}},\
  }\href@noop {} {\  (\bibinfo {year} {2020}{\natexlab{b}})},\ \Eprint
  {http://arxiv.org/abs/2004.05407} {arXiv:2004.05407 [gr-qc]} \BibitemShut
  {NoStop}%
\bibitem [{\citenamefont {{Mano}}\ \emph {et~al.}(1996)\citenamefont {{Mano}},
  \citenamefont {{Suzuki}},\ and\ \citenamefont
  {{Takasugi}}}]{ManoSuzuTaka96a}%
  \BibitemOpen
  \bibfield  {author} {\bibinfo {author} {\bibfnamefont {S.}~\bibnamefont
  {{Mano}}}, \bibinfo {author} {\bibfnamefont {H.}~\bibnamefont {{Suzuki}}}, \
  and\ \bibinfo {author} {\bibfnamefont {E.}~\bibnamefont {{Takasugi}}},\
  }\href {\doibase 10.1143/PTP.96.549} {\bibfield  {journal} {\bibinfo
  {journal} {Progress of Theoretical Physics}\ }\textbf {\bibinfo {volume}
  {96}},\ \bibinfo {pages} {549} (\bibinfo {year} {1996})},\ \Eprint
  {http://arxiv.org/abs/gr-qc/9605057} {gr-qc/9605057} \BibitemShut {NoStop}%
\bibitem [{\citenamefont {{Bini}}\ and\ \citenamefont
  {{Damour}}(2014)}]{BiniDamo14a}%
  \BibitemOpen
  \bibfield  {author} {\bibinfo {author} {\bibfnamefont {D.}~\bibnamefont
  {{Bini}}}\ and\ \bibinfo {author} {\bibfnamefont {T.}~\bibnamefont
  {{Damour}}},\ }\href {\doibase 10.1103/PhysRevD.89.064063} {\bibfield
  {journal} {\bibinfo  {journal} {Phys. Rev. D}\ }\textbf {\bibinfo {volume}
  {89}},\ \bibinfo {eid} {064063} (\bibinfo {year} {2014})},\ \Eprint
  {http://arxiv.org/abs/1312.2503} {arXiv:1312.2503 [gr-qc]} \BibitemShut
  {NoStop}%
\bibitem [{\citenamefont {Bini}\ and\ \citenamefont
  {Damour}(2014{\natexlab{a}})}]{BiniDamo14b}%
  \BibitemOpen
  \bibfield  {author} {\bibinfo {author} {\bibfnamefont {D.}~\bibnamefont
  {Bini}}\ and\ \bibinfo {author} {\bibfnamefont {T.}~\bibnamefont {Damour}},\
  }\href {\doibase 10.1103/PhysRevD.90.024039} {\bibfield  {journal} {\bibinfo
  {journal} {Phys. Rev.}\ }\textbf {\bibinfo {volume} {D90}},\ \bibinfo {pages}
  {024039} (\bibinfo {year} {2014}{\natexlab{a}})},\ \Eprint
  {http://arxiv.org/abs/1404.2747} {arXiv:1404.2747 [gr-qc]} \BibitemShut
  {NoStop}%
\bibitem [{\citenamefont {Bini}\ and\ \citenamefont
  {Damour}(2014{\natexlab{b}})}]{BiniDamo14c}%
  \BibitemOpen
  \bibfield  {author} {\bibinfo {author} {\bibfnamefont {D.}~\bibnamefont
  {Bini}}\ and\ \bibinfo {author} {\bibfnamefont {T.}~\bibnamefont {Damour}},\
  }\href {\doibase 10.1103/PhysRevD.90.124037} {\bibfield  {journal} {\bibinfo
  {journal} {Phys. Rev. D}\ }\textbf {\bibinfo {volume} {90}},\ \bibinfo
  {pages} {124037} (\bibinfo {year} {2014}{\natexlab{b}})},\ \Eprint
  {http://arxiv.org/abs/1409.6933} {arXiv:1409.6933 [gr-qc]} \BibitemShut
  {NoStop}%
\bibitem [{\citenamefont {{Fujita}}(2015)}]{Fuji15}%
  \BibitemOpen
  \bibfield  {author} {\bibinfo {author} {\bibfnamefont {R.}~\bibnamefont
  {{Fujita}}},\ }\href {\doibase 10.1093/ptep/ptv012} {\bibfield  {journal}
  {\bibinfo  {journal} {Prog. Theor. Exp. Phys.}\ }\textbf {\bibinfo {volume}
  {2015}} (\bibinfo {year} {2015}),\ 10.1093/ptep/ptv012},\ \Eprint
  {http://arxiv.org/abs/1412.5689} {arXiv:1412.5689 [gr-qc]} \BibitemShut
  {NoStop}%
\bibitem [{\citenamefont {Hinderer}\ and\ \citenamefont
  {Flanagan}(2008)}]{HindFlan08}%
  \BibitemOpen
  \bibfield  {author} {\bibinfo {author} {\bibfnamefont {T.}~\bibnamefont
  {Hinderer}}\ and\ \bibinfo {author} {\bibfnamefont {E.~E.}\ \bibnamefont
  {Flanagan}},\ }\href {\doibase 10.1103/PhysRevD.78.064028} {\bibfield
  {journal} {\bibinfo  {journal} {Phys. Rev. D}\ }\textbf {\bibinfo {volume}
  {78}},\ \bibinfo {pages} {064028} (\bibinfo {year} {2008})},\ \Eprint
  {http://arxiv.org/abs/0805.3337} {arXiv:0805.3337 [gr-qc]} \BibitemShut
  {NoStop}%
\bibitem [{\citenamefont {Osburn}\ \emph {et~al.}(2016)\citenamefont {Osburn},
  \citenamefont {Warburton},\ and\ \citenamefont {Evans}}]{OsbuWarbEvan16}%
  \BibitemOpen
  \bibfield  {author} {\bibinfo {author} {\bibfnamefont {T.}~\bibnamefont
  {Osburn}}, \bibinfo {author} {\bibfnamefont {N.}~\bibnamefont {Warburton}}, \
  and\ \bibinfo {author} {\bibfnamefont {C.~R.}\ \bibnamefont {Evans}},\ }\href
  {\doibase 10.1103/PhysRevD.93.064024} {\bibfield  {journal} {\bibinfo
  {journal} {Phys. Rev. D}\ }\textbf {\bibinfo {volume} {93}},\ \bibinfo
  {pages} {064024} (\bibinfo {year} {2016})}\BibitemShut {NoStop}%
\bibitem [{\citenamefont {Barack}\ and\ \citenamefont
  {Cutler}(2007)}]{BaraCutl07}%
  \BibitemOpen
  \bibfield  {author} {\bibinfo {author} {\bibfnamefont {L.}~\bibnamefont
  {Barack}}\ and\ \bibinfo {author} {\bibfnamefont {C.}~\bibnamefont
  {Cutler}},\ }\href {\doibase 10.1103/PhysRevD.75.042003} {\bibfield
  {journal} {\bibinfo  {journal} {Phys. Rev. D}\ }\textbf {\bibinfo {volume}
  {75}},\ \bibinfo {pages} {042003} (\bibinfo {year} {2007})},\ \Eprint
  {http://arxiv.org/abs/gr-qc/0612029} {arXiv:gr-qc/0612029} \BibitemShut
  {NoStop}%
\bibitem [{\citenamefont {{Hopman}}\ and\ \citenamefont
  {{Alexander}}(2005)}]{HopmAlex05}%
  \BibitemOpen
  \bibfield  {author} {\bibinfo {author} {\bibfnamefont {C.}~\bibnamefont
  {{Hopman}}}\ and\ \bibinfo {author} {\bibfnamefont {T.}~\bibnamefont
  {{Alexander}}},\ }\href {\doibase 10.1086/431475} {\bibfield  {journal}
  {\bibinfo  {journal} {Astrophysical Journal}\ }\textbf {\bibinfo {volume}
  {629}},\ \bibinfo {pages} {362} (\bibinfo {year} {2005})},\ \Eprint
  {http://arxiv.org/abs/arXiv:astro-ph/0503672} {arXiv:astro-ph/0503672}
  \BibitemShut {NoStop}%
\bibitem [{\citenamefont {{Munna}}\ \emph {et~al.}(2020)\citenamefont
  {{Munna}}, \citenamefont {{Evans}}, \citenamefont {{Hopper}},\ and\
  \citenamefont {{Forseth}}}]{MunnETC20}%
  \BibitemOpen
  \bibfield  {author} {\bibinfo {author} {\bibfnamefont {C.}~\bibnamefont
  {{Munna}}}, \bibinfo {author} {\bibfnamefont {C.~R.}\ \bibnamefont
  {{Evans}}}, \bibinfo {author} {\bibfnamefont {S.}~\bibnamefont {{Hopper}}}, \
  and\ \bibinfo {author} {\bibfnamefont {E.}~\bibnamefont {{Forseth}}},\
  }\href@noop {} {\bibfield  {journal} {\bibinfo  {journal} {arXiv e-prints}\
  ,\ \bibinfo {eid} {arXiv:2005.03044}} (\bibinfo {year} {2020})},\ \Eprint
  {http://arxiv.org/abs/2005.03044} {arXiv:2005.03044 [gr-qc]} \BibitemShut
  {NoStop}%
\bibitem [{\citenamefont {Ferguson}\ \emph {et~al.}(1999)\citenamefont
  {Ferguson}, \citenamefont {Bailey},\ and\ \citenamefont
  {Arno}}]{FergBailArno99}%
  \BibitemOpen
  \bibfield  {author} {\bibinfo {author} {\bibfnamefont {H.~R.~P.}\
  \bibnamefont {Ferguson}}, \bibinfo {author} {\bibfnamefont {D.~H.}\
  \bibnamefont {Bailey}}, \ and\ \bibinfo {author} {\bibfnamefont
  {S.}~\bibnamefont {Arno}},\ }\href@noop {} {\bibfield  {journal} {\bibinfo
  {journal} {Journal of Mathematics of Computation}\ }\textbf {\bibinfo
  {volume} {68}},\ \bibinfo {pages} {351} (\bibinfo {year} {1999})}\BibitemShut
  {NoStop}%
\bibitem [{\citenamefont {{Johnson-McDaniel}}(2014)}]{JohnMcDa14}%
  \BibitemOpen
  \bibfield  {author} {\bibinfo {author} {\bibfnamefont {N.~K.}\ \bibnamefont
  {{Johnson-McDaniel}}},\ }\href {\doibase 10.1103/PhysRevD.90.024043}
  {\bibfield  {journal} {\bibinfo  {journal} {Phys. Rev. D}\ }\textbf {\bibinfo
  {volume} {90}},\ \bibinfo {eid} {024043} (\bibinfo {year} {2014})},\ \Eprint
  {http://arxiv.org/abs/1405.1572} {arXiv:1405.1572 [gr-qc]} \BibitemShut
  {NoStop}%
\bibitem [{BHP()}]{BHPTK18}%
  \BibitemOpen
  \href@noop {} {\enquote {\bibinfo {title} {{Black Hole Perturbation
  Toolkit}},}\ }\bibinfo {note} {\url{bhptoolkit.org}}\BibitemShut {NoStop}%
\bibitem [{\citenamefont {Marchand}\ \emph {et~al.}(2018)\citenamefont
  {Marchand}, \citenamefont {Bernard}, \citenamefont {Blanchet},\ and\
  \citenamefont {Faye}}]{MarcETC18}%
  \BibitemOpen
  \bibfield  {author} {\bibinfo {author} {\bibfnamefont {T.}~\bibnamefont
  {Marchand}}, \bibinfo {author} {\bibfnamefont {L.}~\bibnamefont {Bernard}},
  \bibinfo {author} {\bibfnamefont {L.}~\bibnamefont {Blanchet}}, \ and\
  \bibinfo {author} {\bibfnamefont {G.}~\bibnamefont {Faye}},\ }\href {\doibase
  10.1103/PhysRevD.97.044023} {\bibfield  {journal} {\bibinfo  {journal} {Phys.
  Rev. D}\ }\textbf {\bibinfo {volume} {97}},\ \bibinfo {pages} {044023}
  (\bibinfo {year} {2018})},\ \Eprint {http://arxiv.org/abs/1707.09289}
  {arXiv:1707.09289 [gr-qc]} \BibitemShut {NoStop}%
\bibitem [{\citenamefont {{Arun}}\ \emph
  {et~al.}(2008{\natexlab{a}})\citenamefont {{Arun}}, \citenamefont
  {{Blanchet}}, \citenamefont {{Iyer}},\ and\ \citenamefont
  {{Qusailah}}}]{ArunETC08a}%
  \BibitemOpen
  \bibfield  {author} {\bibinfo {author} {\bibfnamefont {K.~G.}\ \bibnamefont
  {{Arun}}}, \bibinfo {author} {\bibfnamefont {L.}~\bibnamefont {{Blanchet}}},
  \bibinfo {author} {\bibfnamefont {B.~R.}\ \bibnamefont {{Iyer}}}, \ and\
  \bibinfo {author} {\bibfnamefont {M.~S.~S.}\ \bibnamefont {{Qusailah}}},\
  }\href {\doibase 10.1103/PhysRevD.77.064034} {\bibfield  {journal} {\bibinfo
  {journal} {Phys. Rev. D}\ }\textbf {\bibinfo {volume} {77}},\ \bibinfo {eid}
  {064034} (\bibinfo {year} {2008}{\natexlab{a}})},\ \Eprint
  {http://arxiv.org/abs/0711.0250} {arXiv:0711.0250 [gr-qc]} \BibitemShut
  {NoStop}%
\bibitem [{\citenamefont {{Arun}}\ \emph
  {et~al.}(2008{\natexlab{b}})\citenamefont {{Arun}}, \citenamefont
  {{Blanchet}}, \citenamefont {{Iyer}},\ and\ \citenamefont
  {{Qusailah}}}]{ArunETC08b}%
  \BibitemOpen
  \bibfield  {author} {\bibinfo {author} {\bibfnamefont {K.~G.}\ \bibnamefont
  {{Arun}}}, \bibinfo {author} {\bibfnamefont {L.}~\bibnamefont {{Blanchet}}},
  \bibinfo {author} {\bibfnamefont {B.~R.}\ \bibnamefont {{Iyer}}}, \ and\
  \bibinfo {author} {\bibfnamefont {M.~S.~S.}\ \bibnamefont {{Qusailah}}},\
  }\href {\doibase 10.1103/PhysRevD.77.064035} {\bibfield  {journal} {\bibinfo
  {journal} {Phys. Rev. D}\ }\textbf {\bibinfo {volume} {77}},\ \bibinfo {eid}
  {064035} (\bibinfo {year} {2008}{\natexlab{b}})},\ \Eprint
  {http://arxiv.org/abs/0711.0302} {arXiv:0711.0302 [gr-qc]} \BibitemShut
  {NoStop}%
\bibitem [{\citenamefont {Memmesheimer}\ \emph {et~al.}(2004)\citenamefont
  {Memmesheimer}, \citenamefont {Gopakumar},\ and\ \citenamefont
  {Sch{\"a}fer}}]{MemmGopaScha04}%
  \BibitemOpen
  \bibfield  {author} {\bibinfo {author} {\bibfnamefont {R.-M.}\ \bibnamefont
  {Memmesheimer}}, \bibinfo {author} {\bibfnamefont {A.}~\bibnamefont
  {Gopakumar}}, \ and\ \bibinfo {author} {\bibfnamefont {G.}~\bibnamefont
  {Sch{\"a}fer}},\ }\href {\doibase 10.1103/PhysRevD.70.104011} {\bibfield
  {journal} {\bibinfo  {journal} {Phys. Rev. D}\ }\textbf {\bibinfo {volume}
  {70}},\ \bibinfo {eid} {104011} (\bibinfo {year} {2004})},\ \Eprint
  {http://arxiv.org/abs/0407049} {arXiv:0407049 [gr-qc]} \BibitemShut {NoStop}%
\bibitem [{\citenamefont {{Le Tiec}}\ \emph {et~al.}(2012)\citenamefont {{Le
  Tiec}}, \citenamefont {{Blanchet}},\ and\ \citenamefont
  {{Whiting}}}]{LetiBlanWhit12}%
  \BibitemOpen
  \bibfield  {author} {\bibinfo {author} {\bibfnamefont {A.}~\bibnamefont {{Le
  Tiec}}}, \bibinfo {author} {\bibfnamefont {L.}~\bibnamefont {{Blanchet}}}, \
  and\ \bibinfo {author} {\bibfnamefont {B.}~\bibnamefont {{Whiting}}},\ }\href
  {\doibase 10.1103/PhysRevD.85.064039} {\bibfield  {journal} {\bibinfo
  {journal} {Phys. Rev. D}\ }\textbf {\bibinfo {volume} {85}},\ \bibinfo {eid}
  {064039} (\bibinfo {year} {2012})},\ \Eprint {http://arxiv.org/abs/1111.5378}
  {arXiv:1111.5378} \BibitemShut {NoStop}%
\bibitem [{\citenamefont {Damour}\ \emph {et~al.}(2013)\citenamefont {Damour},
  \citenamefont {Nagar},\ and\ \citenamefont {Bernuzzi}}]{DamoNagaBern13}%
  \BibitemOpen
  \bibfield  {author} {\bibinfo {author} {\bibfnamefont {T.}~\bibnamefont
  {Damour}}, \bibinfo {author} {\bibfnamefont {A.}~\bibnamefont {Nagar}}, \
  and\ \bibinfo {author} {\bibfnamefont {S.}~\bibnamefont {Bernuzzi}},\ }\href
  {\doibase 10.1103/PhysRevD.87.084035} {\bibfield  {journal} {\bibinfo
  {journal} {Phys. Rev. D}\ }\textbf {\bibinfo {volume} {87}},\ \bibinfo {eid}
  {084035} (\bibinfo {year} {2013})},\ \Eprint {http://arxiv.org/abs/1212.4357}
  {arXiv:1212.4357 [gr-qc]} \BibitemShut {NoStop}%
\bibitem [{\citenamefont {Misner}\ \emph {et~al.}(1973)\citenamefont {Misner},
  \citenamefont {Thorne},\ and\ \citenamefont {Wheeler}}]{MisnThorWhee73}%
  \BibitemOpen
  \bibfield  {author} {\bibinfo {author} {\bibfnamefont {C.}~\bibnamefont
  {Misner}}, \bibinfo {author} {\bibfnamefont {K.}~\bibnamefont {Thorne}}, \
  and\ \bibinfo {author} {\bibfnamefont {J.}~\bibnamefont {Wheeler}},\
  }\href@noop {} {\emph {\bibinfo {title} {{Gravitation}}}}\ (\bibinfo
  {publisher} {Freeman},\ \bibinfo {address} {San Francisco, CA, U.S.A.},\
  \bibinfo {year} {1973})\BibitemShut {NoStop}%
\bibitem [{\citenamefont {{Sasaki}}\ and\ \citenamefont
  {{Tagoshi}}(2003)}]{SasaTago03}%
  \BibitemOpen
  \bibfield  {author} {\bibinfo {author} {\bibfnamefont {M.}~\bibnamefont
  {{Sasaki}}}\ and\ \bibinfo {author} {\bibfnamefont {H.}~\bibnamefont
  {{Tagoshi}}},\ }\href {\doibase 10.12942/lrr-2003-6} {\bibfield  {journal}
  {\bibinfo  {journal} {Living Reviews in Relativity}\ }\textbf {\bibinfo
  {volume} {6}},\ \bibinfo {pages} {6} (\bibinfo {year} {2003})},\ \Eprint
  {http://arxiv.org/abs/gr-qc/0306120} {gr-qc/0306120} \BibitemShut {NoStop}%
\bibitem [{\citenamefont {{Chandrasekhar}}(1975)}]{Chan75}%
  \BibitemOpen
  \bibfield  {author} {\bibinfo {author} {\bibfnamefont {S.}~\bibnamefont
  {{Chandrasekhar}}},\ }\href {\doibase 10.1098/rspa.1975.0066} {\bibfield
  {journal} {\bibinfo  {journal} {Royal Society of London Proceedings Series
  A}\ }\textbf {\bibinfo {volume} {343}},\ \bibinfo {pages} {289} (\bibinfo
  {year} {1975})}\BibitemShut {NoStop}%
\bibitem [{\citenamefont {{Chandrasekhar}}\ and\ \citenamefont
  {{Detweiler}}(1975)}]{ChanDetw75}%
  \BibitemOpen
  \bibfield  {author} {\bibinfo {author} {\bibfnamefont {S.}~\bibnamefont
  {{Chandrasekhar}}}\ and\ \bibinfo {author} {\bibfnamefont {S.}~\bibnamefont
  {{Detweiler}}},\ }\href {\doibase 10.1098/rspa.1975.0130} {\bibfield
  {journal} {\bibinfo  {journal} {Royal Society of London Proceedings Series
  A}\ }\textbf {\bibinfo {volume} {345}},\ \bibinfo {pages} {145} (\bibinfo
  {year} {1975})}\BibitemShut {NoStop}%
\bibitem [{\citenamefont {Chandrasekhar}(1983)}]{Chan83}%
  \BibitemOpen
  \bibfield  {author} {\bibinfo {author} {\bibfnamefont {S.}~\bibnamefont
  {Chandrasekhar}},\ }\href@noop {} {\emph {\bibinfo {title} {{The Mathematical
  Theory of Black Holes}}}},\ \bibinfo {series} {The International Series of
  Monographs on Physics}, Vol.~\bibinfo {volume} {69}\ (\bibinfo  {publisher}
  {Clarendon},\ \bibinfo {address} {Oxford},\ \bibinfo {year}
  {1983})\BibitemShut {NoStop}%
\bibitem [{\citenamefont {Berndston}(2007)}]{Bern07}%
  \BibitemOpen
  \bibfield  {author} {\bibinfo {author} {\bibfnamefont {M.}~\bibnamefont
  {Berndston}},\ }\emph {\bibinfo {title} {Harmonic Gauge Perturbations of the
  Schwarzschild Metric}},\ \href@noop {} {Ph.D. thesis},\ \bibinfo  {school}
  {University of Colorado} (\bibinfo {year} {2007}),\ \Eprint
  {http://arxiv.org/abs/0904.0033v1} {arXiv:0904.0033v1} \BibitemShut {NoStop}%
\bibitem [{\citenamefont {{Barack}}\ \emph {et~al.}(2008)\citenamefont
  {{Barack}}, \citenamefont {{Ori}},\ and\ \citenamefont
  {{Sago}}}]{BaraOriSago08}%
  \BibitemOpen
  \bibfield  {author} {\bibinfo {author} {\bibfnamefont {L.}~\bibnamefont
  {{Barack}}}, \bibinfo {author} {\bibfnamefont {A.}~\bibnamefont {{Ori}}}, \
  and\ \bibinfo {author} {\bibfnamefont {N.}~\bibnamefont {{Sago}}},\ }\href
  {\doibase 10.1103/PhysRevD.78.084021} {\bibfield  {journal} {\bibinfo
  {journal} {Phys. Rev. D}\ }\textbf {\bibinfo {volume} {78}},\ \bibinfo
  {pages} {084021} (\bibinfo {year} {2008})},\ \Eprint
  {http://arxiv.org/abs/0808.2315} {arXiv:0808.2315} \BibitemShut {NoStop}%
\bibitem [{\citenamefont {Flanagan}\ and\ \citenamefont
  {Hinderer}(2012)}]{FlanHind12}%
  \BibitemOpen
  \bibfield  {author} {\bibinfo {author} {\bibfnamefont {E.~E.}\ \bibnamefont
  {Flanagan}}\ and\ \bibinfo {author} {\bibfnamefont {T.}~\bibnamefont
  {Hinderer}},\ }\href {\doibase 10.1103/PhysRevLett.109.071102} {\bibfield
  {journal} {\bibinfo  {journal} {Physical Review Letters}\ }\textbf {\bibinfo
  {volume} {109}},\ \bibinfo {pages} {071102} (\bibinfo {year} {2012})},\
  \Eprint {http://arxiv.org/abs/1009.4923} {arXiv:1009.4923 [gr-qc]}
  \BibitemShut {NoStop}%
\bibitem [{\citenamefont {{Drasco}}\ \emph {et~al.}(2005)\citenamefont
  {{Drasco}}, \citenamefont {{Flanagan}},\ and\ \citenamefont
  {{Hughes}}}]{DrasFlanHugh05}%
  \BibitemOpen
  \bibfield  {author} {\bibinfo {author} {\bibfnamefont {S.}~\bibnamefont
  {{Drasco}}}, \bibinfo {author} {\bibfnamefont {{\'E}.~{\'E}.}\ \bibnamefont
  {{Flanagan}}}, \ and\ \bibinfo {author} {\bibfnamefont {S.~A.}\ \bibnamefont
  {{Hughes}}},\ }\href {\doibase 10.1088/0264-9381/22/15/011} {\bibfield
  {journal} {\bibinfo  {journal} {Classical and Quantum Gravity}\ }\textbf
  {\bibinfo {volume} {22}},\ \bibinfo {pages} {S801} (\bibinfo {year}
  {2005})},\ \Eprint {http://arxiv.org/abs/gr-qc/0505075} {gr-qc/0505075}
  \BibitemShut {NoStop}%
\bibitem [{\citenamefont {Hughes}\ \emph {et~al.}(2005)\citenamefont {Hughes},
  \citenamefont {Drasco}, \citenamefont {Flanagan},\ and\ \citenamefont
  {Franklin}}]{HughETC05}%
  \BibitemOpen
  \bibfield  {author} {\bibinfo {author} {\bibfnamefont {S.~A.}\ \bibnamefont
  {Hughes}}, \bibinfo {author} {\bibfnamefont {S.}~\bibnamefont {Drasco}},
  \bibinfo {author} {\bibfnamefont {E.~E.}\ \bibnamefont {Flanagan}}, \ and\
  \bibinfo {author} {\bibfnamefont {J.}~\bibnamefont {Franklin}},\ }\href
  {\doibase 10.1103/PhysRevLett.94.221101} {\bibfield  {journal} {\bibinfo
  {journal} {Physical Review Letters}\ }\textbf {\bibinfo {volume} {94}},\
  \bibinfo {pages} {221101} (\bibinfo {year} {2005})},\ \Eprint
  {http://arxiv.org/abs/gr-qc/0504015} {arXiv:gr-qc/0504015} \BibitemShut
  {NoStop}%
\bibitem [{\citenamefont {{Drasco}}\ and\ \citenamefont
  {{Hughes}}(2006)}]{DrasHugh06}%
  \BibitemOpen
  \bibfield  {author} {\bibinfo {author} {\bibfnamefont {S.}~\bibnamefont
  {{Drasco}}}\ and\ \bibinfo {author} {\bibfnamefont {S.~A.}\ \bibnamefont
  {{Hughes}}},\ }\href {\doibase 10.1103/PhysRevD.73.024027} {\bibfield
  {journal} {\bibinfo  {journal} {Phys. Rev. D}\ }\textbf {\bibinfo {volume}
  {73}},\ \bibinfo {eid} {024027} (\bibinfo {year} {2006})},\ \Eprint
  {http://arxiv.org/abs/gr-qc/0509101} {gr-qc/0509101} \BibitemShut {NoStop}%
\bibitem [{\citenamefont {Fujita}\ \emph {et~al.}(2009)\citenamefont {Fujita},
  \citenamefont {Hikida},\ and\ \citenamefont {Tagoshi}}]{FujiHikiTago09}%
  \BibitemOpen
  \bibfield  {author} {\bibinfo {author} {\bibfnamefont {R.}~\bibnamefont
  {Fujita}}, \bibinfo {author} {\bibfnamefont {W.}~\bibnamefont {Hikida}}, \
  and\ \bibinfo {author} {\bibfnamefont {H.}~\bibnamefont {Tagoshi}},\ }\href
  {\doibase 10.1143/PTP.121.843} {\bibfield  {journal} {\bibinfo  {journal}
  {Prog. Theor. Phys.}\ }\textbf {\bibinfo {volume} {121}},\ \bibinfo {pages}
  {843} (\bibinfo {year} {2009})},\ \Eprint {http://arxiv.org/abs/0904.3810}
  {arXiv:0904.3810 [gr-qc]} \BibitemShut {NoStop}%
\bibitem [{\citenamefont {{Taracchini}}\ \emph {et~al.}(2013)\citenamefont
  {{Taracchini}}, \citenamefont {{Buonanno}}, \citenamefont {{Hughes}},\ and\
  \citenamefont {{Khanna}}}]{TaraETC13}%
  \BibitemOpen
  \bibfield  {author} {\bibinfo {author} {\bibfnamefont {A.}~\bibnamefont
  {{Taracchini}}}, \bibinfo {author} {\bibfnamefont {A.}~\bibnamefont
  {{Buonanno}}}, \bibinfo {author} {\bibfnamefont {S.}~\bibnamefont
  {{Hughes}}}, \ and\ \bibinfo {author} {\bibfnamefont {G.}~\bibnamefont
  {{Khanna}}},\ }\href {\doibase 10.1103/PhysRevD.88.0440012} {\bibfield
  {journal} {\bibinfo  {journal} {Phys. Rev. D}\ }\textbf {\bibinfo {volume}
  {88}},\ \bibinfo {eid} {044001} (\bibinfo {year} {2013})},\ \Eprint
  {http://arxiv.org/abs/1305.2184} {arXiv:1305.2184 [gr-qc]} \BibitemShut
  {NoStop}%
\bibitem [{\citenamefont {Tanaka}\ \emph {et~al.}(1996)\citenamefont {Tanaka},
  \citenamefont {Tagoshi},\ and\ \citenamefont {Sasaki}}]{TanaTagoSasa96}%
  \BibitemOpen
  \bibfield  {author} {\bibinfo {author} {\bibfnamefont {T.}~\bibnamefont
  {Tanaka}}, \bibinfo {author} {\bibfnamefont {H.}~\bibnamefont {Tagoshi}}, \
  and\ \bibinfo {author} {\bibfnamefont {M.}~\bibnamefont {Sasaki}},\ }\href
  {\doibase 10.1143/PTP.96.1087} {\bibfield  {journal} {\bibinfo  {journal}
  {Progress of Theoretical Physics}\ }\textbf {\bibinfo {volume} {96}},\
  \bibinfo {pages} {1087} (\bibinfo {year} {1996})}\BibitemShut {NoStop}%
\bibitem [{\citenamefont {{Isoyama}}\ \emph
  {et~al.}(2013{\natexlab{a}})\citenamefont {{Isoyama}}, \citenamefont
  {{Fujita}}, \citenamefont {{Sago}}, \citenamefont {{Tagoshi}},\ and\
  \citenamefont {{Tanaka}}}]{IsoyETC13b}%
  \BibitemOpen
  \bibfield  {author} {\bibinfo {author} {\bibfnamefont {S.}~\bibnamefont
  {{Isoyama}}}, \bibinfo {author} {\bibfnamefont {R.}~\bibnamefont {{Fujita}}},
  \bibinfo {author} {\bibfnamefont {N.}~\bibnamefont {{Sago}}}, \bibinfo
  {author} {\bibfnamefont {H.}~\bibnamefont {{Tagoshi}}}, \ and\ \bibinfo
  {author} {\bibfnamefont {T.}~\bibnamefont {{Tanaka}}},\ }\href {\doibase
  10.1103/PhysRevD.87.024010} {\bibfield  {journal} {\bibinfo  {journal} {Phys.
  Rev. D}\ }\textbf {\bibinfo {volume} {87}},\ \bibinfo {eid} {024010}
  (\bibinfo {year} {2013}{\natexlab{a}})},\ \Eprint
  {http://arxiv.org/abs/1210.2569} {arXiv:1210.2569 [gr-qc]} \BibitemShut
  {NoStop}%
\bibitem [{\citenamefont {Damour}\ and\ \citenamefont
  {Nagar}(2007)}]{DamoNaga07}%
  \BibitemOpen
  \bibfield  {author} {\bibinfo {author} {\bibfnamefont {T.}~\bibnamefont
  {Damour}}\ and\ \bibinfo {author} {\bibfnamefont {A.}~\bibnamefont {Nagar}},\
  }\href {\doibase 10.1103/PhysRevD.76.064028} {\bibfield  {journal} {\bibinfo
  {journal} {Phys. Rev. D}\ }\textbf {\bibinfo {volume} {76}},\ \bibinfo {eid}
  {064028} (\bibinfo {year} {2007})},\ \Eprint {http://arxiv.org/abs/0705.2519}
  {arXiv:0705.2519 [gr-qc]} \BibitemShut {NoStop}%
\bibitem [{\citenamefont {Damour}\ \emph {et~al.}(2009)\citenamefont {Damour},
  \citenamefont {Iyer},\ and\ \citenamefont {Nagar}}]{DamoIyerNaga09}%
  \BibitemOpen
  \bibfield  {author} {\bibinfo {author} {\bibfnamefont {T.}~\bibnamefont
  {Damour}}, \bibinfo {author} {\bibfnamefont {B.}~\bibnamefont {Iyer}}, \ and\
  \bibinfo {author} {\bibfnamefont {A.}~\bibnamefont {Nagar}},\ }\href
  {\doibase 10.1103/PhysRevD.79.064004} {\bibfield  {journal} {\bibinfo
  {journal} {Phys. Rev. D}\ }\textbf {\bibinfo {volume} {79}},\ \bibinfo {eid}
  {064004} (\bibinfo {year} {2009})},\ \Eprint {http://arxiv.org/abs/0811.2069}
  {arXiv:0811.2069 [gr-qc]} \BibitemShut {NoStop}%
\bibitem [{\citenamefont {{Pan}}\ \emph {et~al.}(2011)\citenamefont {{Pan}},
  \citenamefont {{Buonanno}}, \citenamefont {{Fujita}}, \citenamefont
  {{Racine}},\ and\ \citenamefont {{Tagoshi}}}]{PanETC11}%
  \BibitemOpen
  \bibfield  {author} {\bibinfo {author} {\bibfnamefont {Y.}~\bibnamefont
  {{Pan}}}, \bibinfo {author} {\bibfnamefont {A.}~\bibnamefont {{Buonanno}}},
  \bibinfo {author} {\bibfnamefont {R.}~\bibnamefont {{Fujita}}}, \bibinfo
  {author} {\bibfnamefont {E.}~\bibnamefont {{Racine}}}, \ and\ \bibinfo
  {author} {\bibfnamefont {H.}~\bibnamefont {{Tagoshi}}},\ }\href {\doibase
  10.1103/PhysRevD.83.064003} {\bibfield  {journal} {\bibinfo  {journal} {Phys.
  Rev. D}\ }\textbf {\bibinfo {volume} {83}},\ \bibinfo {eid} {064003}
  (\bibinfo {year} {2011})},\ \Eprint {http://arxiv.org/abs/1006.0431}
  {arXiv:1006.0431 [gr-qc]} \BibitemShut {NoStop}%
\bibitem [{\citenamefont {{Messina}}\ and\ \citenamefont
  {{Nagar}}(2017)}]{MessNaga17}%
  \BibitemOpen
  \bibfield  {author} {\bibinfo {author} {\bibfnamefont {F.}~\bibnamefont
  {{Messina}}}\ and\ \bibinfo {author} {\bibfnamefont {A.}~\bibnamefont
  {{Nagar}}},\ }\href {\doibase 10.1103/PhysRevD.95.124001} {\bibfield
  {journal} {\bibinfo  {journal} {Phys. Rev. D}\ }\textbf {\bibinfo {volume}
  {95}},\ \bibinfo {eid} {124001} (\bibinfo {year} {2017})},\ \Eprint
  {http://arxiv.org/abs/1703.08107} {arXiv:1703.08107 [gr-qc]} \BibitemShut
  {NoStop}%
\bibitem [{\citenamefont {{Messina}}\ \emph {et~al.}(2018)\citenamefont
  {{Messina}}, \citenamefont {{Maldarella}},\ and\ \citenamefont
  {{Nagar}}}]{MessMaldNaga18}%
  \BibitemOpen
  \bibfield  {author} {\bibinfo {author} {\bibfnamefont {F.}~\bibnamefont
  {{Messina}}}, \bibinfo {author} {\bibfnamefont {A.}~\bibnamefont
  {{Maldarella}}}, \ and\ \bibinfo {author} {\bibfnamefont {A.}~\bibnamefont
  {{Nagar}}},\ }\href {\doibase 10.1103/PhysRevD.97.084016} {\bibfield
  {journal} {\bibinfo  {journal} {Phys. Rev. D}\ }\textbf {\bibinfo {volume}
  {97}},\ \bibinfo {eid} {084016} (\bibinfo {year} {2018})},\ \Eprint
  {http://arxiv.org/abs/1801.02366} {arXiv:1801.02366 [gr-qc]} \BibitemShut
  {NoStop}%
\bibitem [{\citenamefont {{Nagar}}\ \emph {et~al.}(2020)\citenamefont
  {{Nagar}}, \citenamefont {{Pratten}}, \citenamefont {{Riemenschneider}},\
  and\ \citenamefont {{Gamba}}}]{NagaETC20}%
  \BibitemOpen
  \bibfield  {author} {\bibinfo {author} {\bibfnamefont {A.}~\bibnamefont
  {{Nagar}}}, \bibinfo {author} {\bibfnamefont {G.}~\bibnamefont {{Pratten}}},
  \bibinfo {author} {\bibfnamefont {G.}~\bibnamefont {{Riemenschneider}}}, \
  and\ \bibinfo {author} {\bibfnamefont {R.}~\bibnamefont {{Gamba}}},\ }\href
  {\doibase 10.1103/PhysRevD.101.024041} {\bibfield  {journal} {\bibinfo
  {journal} {Phys. Rev. D}\ }\textbf {\bibinfo {volume} {101}},\ \bibinfo {eid}
  {024041} (\bibinfo {year} {2020})},\ \Eprint
  {http://arxiv.org/abs/1904.09550} {arXiv:1904.09550 [gr-qc]} \BibitemShut
  {NoStop}%
\bibitem [{\citenamefont {{Chiaramello}}\ and\ \citenamefont
  {{Nagar}}(2020)}]{ChiaNaga20}%
  \BibitemOpen
  \bibfield  {author} {\bibinfo {author} {\bibfnamefont {D.}~\bibnamefont
  {{Chiaramello}}}\ and\ \bibinfo {author} {\bibfnamefont {A.}~\bibnamefont
  {{Nagar}}},\ }\href {\doibase 10.1103/PhysRevD.101.101501} {\bibfield
  {journal} {\bibinfo  {journal} {Phys. Rev. D}\ }\textbf {\bibinfo {volume}
  {101}},\ \bibinfo {eid} {101501} (\bibinfo {year} {2020})},\ \Eprint
  {http://arxiv.org/abs/2001.11736} {arXiv:2001.11736 [gr-qc]} \BibitemShut
  {NoStop}%
\bibitem [{\citenamefont {Fujita}\ \emph {et~al.}(2018)\citenamefont {Fujita},
  \citenamefont {Sago},\ and\ \citenamefont {Nakano}}]{FujiSagoNaka18}%
  \BibitemOpen
  \bibfield  {author} {\bibinfo {author} {\bibfnamefont {R.}~\bibnamefont
  {Fujita}}, \bibinfo {author} {\bibfnamefont {N.}~\bibnamefont {Sago}}, \ and\
  \bibinfo {author} {\bibfnamefont {H.}~\bibnamefont {Nakano}},\ }\href
  {\doibase 10.1088/1361-6382/aa9ad5} {\bibfield  {journal} {\bibinfo
  {journal} {Class. Quantum Grav..}\ }\textbf {\bibinfo {volume} {35}},\
  \bibinfo {eid} {027001} (\bibinfo {year} {2018})},\ \Eprint
  {http://arxiv.org/abs/1707.09309} {arXiv:1707.09309 [gr-qc]} \BibitemShut
  {NoStop}%
\bibitem [{\citenamefont {{Akcay}}\ \emph {et~al.}(2020)\citenamefont
  {{Akcay}}, \citenamefont {{Dolan}}, \citenamefont {{Kavanagh}}, \citenamefont
  {{Moxon}}, \citenamefont {{Warburton}},\ and\ \citenamefont
  {{Wardell}}}]{AkcaETC20}%
  \BibitemOpen
  \bibfield  {author} {\bibinfo {author} {\bibfnamefont {S.}~\bibnamefont
  {{Akcay}}}, \bibinfo {author} {\bibfnamefont {S.}~\bibnamefont {{Dolan}}},
  \bibinfo {author} {\bibfnamefont {C.}~\bibnamefont {{Kavanagh}}}, \bibinfo
  {author} {\bibfnamefont {J.}~\bibnamefont {{Moxon}}}, \bibinfo {author}
  {\bibfnamefont {N.}~\bibnamefont {{Warburton}}}, \ and\ \bibinfo {author}
  {\bibfnamefont {B.}~\bibnamefont {{Wardell}}},\ }\href@noop {} {\  (\bibinfo
  {year} {2020})},\ \Eprint {http://arxiv.org/abs/1912.09461} {arXiv:1912.09461
  [gr-qc]} \BibitemShut {NoStop}%
\bibitem [{\citenamefont {{Casals}}\ and\ \citenamefont
  {{Ottewill}}(2015)}]{CasaOtte15}%
  \BibitemOpen
  \bibfield  {author} {\bibinfo {author} {\bibfnamefont {M.}~\bibnamefont
  {{Casals}}}\ and\ \bibinfo {author} {\bibfnamefont {A.~C.}\ \bibnamefont
  {{Ottewill}}},\ }\href {\doibase 10.1103/PhysRevD.92.124055} {\bibfield
  {journal} {\bibinfo  {journal} {Phys. Rev. D}\ }\textbf {\bibinfo {volume}
  {92}},\ \bibinfo {eid} {124055} (\bibinfo {year} {2015})},\ \Eprint
  {http://arxiv.org/abs/1509.04702} {arXiv:1509.04702 [gr-qc]} \BibitemShut
  {NoStop}%
\bibitem [{\citenamefont {{Darwin}}(1959)}]{Darw59}%
  \BibitemOpen
  \bibfield  {author} {\bibinfo {author} {\bibfnamefont {C.}~\bibnamefont
  {{Darwin}}},\ }\href {\doibase 10.1098/rspa.1959.0015} {\bibfield  {journal}
  {\bibinfo  {journal} {Proc. R. Soc. Lond. A}\ }\textbf {\bibinfo {volume}
  {249}},\ \bibinfo {pages} {180} (\bibinfo {year} {1959})}\BibitemShut
  {NoStop}%
\bibitem [{\citenamefont {Cutler}\ \emph {et~al.}(1994)\citenamefont {Cutler},
  \citenamefont {Kennefick},\ and\ \citenamefont {Poisson}}]{CutlKennPois94}%
  \BibitemOpen
  \bibfield  {author} {\bibinfo {author} {\bibfnamefont {C.}~\bibnamefont
  {Cutler}}, \bibinfo {author} {\bibfnamefont {D.}~\bibnamefont {Kennefick}}, \
  and\ \bibinfo {author} {\bibfnamefont {E.}~\bibnamefont {Poisson}},\ }\href
  {\doibase 10.1103/PhysRevD.50.3816} {\bibfield  {journal} {\bibinfo
  {journal} {Phys. Rev. D}\ }\textbf {\bibinfo {volume} {50}},\ \bibinfo
  {pages} {3816} (\bibinfo {year} {1994})}\BibitemShut {NoStop}%
\bibitem [{\citenamefont {{Barack}}\ and\ \citenamefont
  {{Sago}}(2010)}]{BaraSago10}%
  \BibitemOpen
  \bibfield  {author} {\bibinfo {author} {\bibfnamefont {L.}~\bibnamefont
  {{Barack}}}\ and\ \bibinfo {author} {\bibfnamefont {N.}~\bibnamefont
  {{Sago}}},\ }\href {\doibase 10.1103/PhysRevD.81.084021} {\bibfield
  {journal} {\bibinfo  {journal} {Phys. Rev. D}\ }\textbf {\bibinfo {volume}
  {81}},\ \bibinfo {eid} {084021} (\bibinfo {year} {2010})},\ \Eprint
  {http://arxiv.org/abs/1002.2386} {arXiv:1002.2386 [gr-qc]} \BibitemShut
  {NoStop}%
\bibitem [{\citenamefont {{Gradshteyn}}\ \emph {et~al.}(2007)\citenamefont
  {{Gradshteyn}}, \citenamefont {{Ryzhik}}, \citenamefont {{Jeffrey}},\ and\
  \citenamefont {{Zwillinger}}}]{GradETC07}%
  \BibitemOpen
  \bibfield  {author} {\bibinfo {author} {\bibfnamefont {I.~S.}\ \bibnamefont
  {{Gradshteyn}}}, \bibinfo {author} {\bibfnamefont {I.~M.}\ \bibnamefont
  {{Ryzhik}}}, \bibinfo {author} {\bibfnamefont {A.}~\bibnamefont {{Jeffrey}}},
  \ and\ \bibinfo {author} {\bibfnamefont {D.}~\bibnamefont {{Zwillinger}}},\
  }\href@noop {} {\emph {\bibinfo {title} {Table of Integrals, Series, and
  Products, Seventh Edition~Elsevier Academic Press, 2007.~ISBN
  012-373637-4}}}\ (\bibinfo {year} {2007})\BibitemShut {NoStop}%
\bibitem [{\citenamefont {Munna}\ and\ \citenamefont
  {Evans}({\natexlab{b}})}]{MunnEvan20b}%
  \BibitemOpen
  \bibfield  {author} {\bibinfo {author} {\bibfnamefont {C.}~\bibnamefont
  {Munna}}\ and\ \bibinfo {author} {\bibfnamefont {C.~R.}\ \bibnamefont
  {Evans}},\ }\href@noop {} {\bibfield  {journal} {\bibinfo  {journal} {to be
  submitted to Phys. Rev. D}\ } ({\natexlab{b}})}\BibitemShut {NoStop}%
\bibitem [{\citenamefont {{Peters}}\ and\ \citenamefont
  {{Mathews}}(1963)}]{PeteMath63}%
  \BibitemOpen
  \bibfield  {author} {\bibinfo {author} {\bibfnamefont {P.~C.}\ \bibnamefont
  {{Peters}}}\ and\ \bibinfo {author} {\bibfnamefont {J.}~\bibnamefont
  {{Mathews}}},\ }\href {\doibase 10.1103/PhysRev.131.435} {\bibfield
  {journal} {\bibinfo  {journal} {Physical Review}\ }\textbf {\bibinfo {volume}
  {131}},\ \bibinfo {pages} {435} (\bibinfo {year} {1963})}\BibitemShut
  {NoStop}%
\bibitem [{\citenamefont {{Wagoner}}\ and\ \citenamefont
  {{Will}}(1976)}]{WagoWill76}%
  \BibitemOpen
  \bibfield  {author} {\bibinfo {author} {\bibfnamefont {R.~V.}\ \bibnamefont
  {{Wagoner}}}\ and\ \bibinfo {author} {\bibfnamefont {C.~M.}\ \bibnamefont
  {{Will}}},\ }\href {\doibase 10.1086/154886} {\bibfield  {journal} {\bibinfo
  {journal} {The Astrophysical Journal}\ }\textbf {\bibinfo {volume} {210}},\
  \bibinfo {pages} {764} (\bibinfo {year} {1976})}\BibitemShut {NoStop}%
\bibitem [{\citenamefont {{Arun}}\ \emph {et~al.}(2009)\citenamefont {{Arun}},
  \citenamefont {{Blanchet}}, \citenamefont {{Iyer}},\ and\ \citenamefont
  {{Sinha}}}]{ArunETC09a}%
  \BibitemOpen
  \bibfield  {author} {\bibinfo {author} {\bibfnamefont {K.~G.}\ \bibnamefont
  {{Arun}}}, \bibinfo {author} {\bibfnamefont {L.}~\bibnamefont {{Blanchet}}},
  \bibinfo {author} {\bibfnamefont {B.~R.}\ \bibnamefont {{Iyer}}}, \ and\
  \bibinfo {author} {\bibfnamefont {S.}~\bibnamefont {{Sinha}}},\ }\href
  {\doibase 10.1103/PhysRevD.80.124018} {\bibfield  {journal} {\bibinfo
  {journal} {Phys. Rev. D}\ }\textbf {\bibinfo {volume} {80}},\ \bibinfo {eid}
  {124018} (\bibinfo {year} {2009})},\ \Eprint {http://arxiv.org/abs/0908.3854}
  {arXiv:0908.3854 [gr-qc]} \BibitemShut {NoStop}%
\bibitem [{\citenamefont {{Peters}}(1964)}]{Pete64}%
  \BibitemOpen
  \bibfield  {author} {\bibinfo {author} {\bibfnamefont {P.~C.}\ \bibnamefont
  {{Peters}}},\ }\href {\doibase 10.1103/PhysRev.136.B1224} {\bibfield
  {journal} {\bibinfo  {journal} {Physical Review}\ }\textbf {\bibinfo {volume}
  {136}},\ \bibinfo {pages} {B1224} (\bibinfo {year} {1964})}\BibitemShut
  {NoStop}%
\bibitem [{\citenamefont {{Blanchet}}\ and\ \citenamefont
  {{Sch{\"a}fer}}(1993)}]{BlanScha93}%
  \BibitemOpen
  \bibfield  {author} {\bibinfo {author} {\bibfnamefont {L.}~\bibnamefont
  {{Blanchet}}}\ and\ \bibinfo {author} {\bibfnamefont {G.}~\bibnamefont
  {{Sch{\"a}fer}}},\ }\href {\doibase 10.1088/0264-9381/10/12/026} {\bibfield
  {journal} {\bibinfo  {journal} {Classical and Quantum Gravity}\ }\textbf
  {\bibinfo {volume} {10}},\ \bibinfo {pages} {2699} (\bibinfo {year}
  {1993})}\BibitemShut {NoStop}%
\bibitem [{\citenamefont {Blanchet}\ \emph {et~al.}(1995)\citenamefont
  {Blanchet}, \citenamefont {Damour},\ and\ \citenamefont
  {Iyer}}]{BlanDamoIyer95}%
  \BibitemOpen
  \bibfield  {author} {\bibinfo {author} {\bibfnamefont {L.}~\bibnamefont
  {Blanchet}}, \bibinfo {author} {\bibfnamefont {T.}~\bibnamefont {Damour}}, \
  and\ \bibinfo {author} {\bibfnamefont {B.~R.}\ \bibnamefont {Iyer}},\ }\href
  {\doibase 10.1103/PhysRevD.51.5360} {\bibfield  {journal} {\bibinfo
  {journal} {Phys. Rev. D}\ }\textbf {\bibinfo {volume} {51}},\ \bibinfo
  {pages} {5360} (\bibinfo {year} {1995})}\BibitemShut {NoStop}%
\bibitem [{\citenamefont {Blanchet}(1996)}]{Blan96}%
  \BibitemOpen
  \bibfield  {author} {\bibinfo {author} {\bibfnamefont {L.}~\bibnamefont
  {Blanchet}},\ }\href {\doibase 10.1103/PhysRevD.54.1417} {\bibfield
  {journal} {\bibinfo  {journal} {Phys. Rev. D}\ }\textbf {\bibinfo {volume}
  {54}},\ \bibinfo {pages} {1417} (\bibinfo {year} {1996})}\BibitemShut
  {NoStop}%
\bibitem [{\citenamefont {{Forseth}}(2016)}]{Fors16}%
  \BibitemOpen
  \bibfield  {author} {\bibinfo {author} {\bibfnamefont {E.~R.}\ \bibnamefont
  {{Forseth}}},\ }\emph {\bibinfo {title} {{High-precision extreme-mass-ratio
  inspirals in black hole perturbation theory and post-Newtonian theory}}},\
  \href@noop {} {Ph.D. thesis},\ \bibinfo  {school} {The University of North
  Carolina at Chapel Hill} (\bibinfo {year} {2016})\BibitemShut {NoStop}%
\bibitem [{\citenamefont {Goldberger}\ and\ \citenamefont
  {Ross}(2010)}]{GoldRoss10}%
  \BibitemOpen
  \bibfield  {author} {\bibinfo {author} {\bibfnamefont {W.}~\bibnamefont
  {Goldberger}}\ and\ \bibinfo {author} {\bibfnamefont {A.}~\bibnamefont
  {Ross}},\ }\href {\doibase 10.1103/PhysRevD.81.124015} {\bibfield  {journal}
  {\bibinfo  {journal} {Phys. Rev. D}\ }\textbf {\bibinfo {volume} {81}},\
  \bibinfo {pages} {124015} (\bibinfo {year} {2010})}\BibitemShut {NoStop}%
\bibitem [{\citenamefont {{Isoyama}}\ \emph
  {et~al.}(2013{\natexlab{b}})\citenamefont {{Isoyama}}, \citenamefont
  {{Fujita}}, \citenamefont {{Nakano}}, \citenamefont {{Sago}},\ and\
  \citenamefont {{Tanaka}}}]{IsoyETC13}%
  \BibitemOpen
  \bibfield  {author} {\bibinfo {author} {\bibfnamefont {S.}~\bibnamefont
  {{Isoyama}}}, \bibinfo {author} {\bibfnamefont {R.}~\bibnamefont {{Fujita}}},
  \bibinfo {author} {\bibfnamefont {H.}~\bibnamefont {{Nakano}}}, \bibinfo
  {author} {\bibfnamefont {N.}~\bibnamefont {{Sago}}}, \ and\ \bibinfo {author}
  {\bibfnamefont {T.}~\bibnamefont {{Tanaka}}},\ }\href {\doibase
  10.1093/ptep/ptt034} {\bibfield  {journal} {\bibinfo  {journal} {Progress of
  Theoretical and Experimental Physics}\ }\textbf {\bibinfo {volume} {2013}},\
  \bibinfo {eid} {063E01} (\bibinfo {year} {2013}{\natexlab{b}})},\ \Eprint
  {http://arxiv.org/abs/1302.4035} {arXiv:1302.4035 [gr-qc]} \BibitemShut
  {NoStop}%
\bibitem [{\citenamefont {Damour}\ and\ \citenamefont
  {Deruelle}(1985)}]{DamoDeru85}%
  \BibitemOpen
  \bibfield  {author} {\bibinfo {author} {\bibfnamefont {T.}~\bibnamefont
  {Damour}}\ and\ \bibinfo {author} {\bibfnamefont {N.}~\bibnamefont
  {Deruelle}},\ }\href {http://eudml.org/doc/76291} {\bibfield  {journal}
  {\bibinfo  {journal} {Annales de l'institut Henri Poincaré (A) Physique
  théorique}\ }\textbf {\bibinfo {volume} {43}},\ \bibinfo {pages} {107}
  (\bibinfo {year} {1985})}\BibitemShut {NoStop}%
\bibitem [{\citenamefont {Damour}\ and\ \citenamefont
  {Sch{\"a}fer}(1988)}]{DamoScha88}%
  \BibitemOpen
  \bibfield  {author} {\bibinfo {author} {\bibfnamefont {T.}~\bibnamefont
  {Damour}}\ and\ \bibinfo {author} {\bibfnamefont {G.}~\bibnamefont
  {Sch{\"a}fer}},\ }\href {\doibase 10.1007/BF02828697} {\bibfield  {journal}
  {\bibinfo  {journal} {Nuovo Cimento B}\ }\bibinfo {series} {11},\ \textbf
  {\bibinfo {volume} {101B}},\ \bibinfo {pages} {127} (\bibinfo {year}
  {1988})}\BibitemShut {NoStop}%
\bibitem [{\citenamefont {Sch{\"a}fer}\ and\ \citenamefont
  {Wex}(1993)}]{SchaWex93}%
  \BibitemOpen
  \bibfield  {author} {\bibinfo {author} {\bibfnamefont {G.}~\bibnamefont
  {Sch{\"a}fer}}\ and\ \bibinfo {author} {\bibfnamefont {N.}~\bibnamefont
  {Wex}},\ }\href {\doibase 10.1016/0375-9601(93)90758-R} {\bibfield  {journal}
  {\bibinfo  {journal} {Phys. Lett.}\ }\textbf {\bibinfo {volume} {174}},\
  \bibinfo {pages} {196} (\bibinfo {year} {1993})}\BibitemShut {NoStop}%
\bibitem [{\citenamefont {Fromholtz}\ \emph {et~al.}(2014)\citenamefont
  {Fromholtz}, \citenamefont {Poisson},\ and\ \citenamefont
  {Will}}]{FromPoisWill14}%
  \BibitemOpen
  \bibfield  {author} {\bibinfo {author} {\bibfnamefont {P.}~\bibnamefont
  {Fromholtz}}, \bibinfo {author} {\bibfnamefont {E.}~\bibnamefont {Poisson}},
  \ and\ \bibinfo {author} {\bibfnamefont {C.}~\bibnamefont {Will}},\ }\href
  {\doibase 10.1119/1.4850396} {\bibfield  {journal} {\bibinfo  {journal} {Am.
  J. Phys.}\ }\textbf {\bibinfo {volume} {82}} (\bibinfo {year} {2014}),\
  10.1119/1.4850396},\ \Eprint {http://arxiv.org/abs/1308.0394}
  {arXiv:1308.0394 [gr-qc]} \BibitemShut {NoStop}%
\bibitem [{\citenamefont {{Barack}}\ \emph {et~al.}(2010)\citenamefont
  {{Barack}}, \citenamefont {{Damour}},\ and\ \citenamefont
  {{Sago}}}]{BaraDamoSago10}%
  \BibitemOpen
  \bibfield  {author} {\bibinfo {author} {\bibfnamefont {L.}~\bibnamefont
  {{Barack}}}, \bibinfo {author} {\bibfnamefont {T.}~\bibnamefont {{Damour}}},
  \ and\ \bibinfo {author} {\bibfnamefont {N.}~\bibnamefont {{Sago}}},\ }\href
  {\doibase 10.1103/PhysRevD.82.084036} {\bibfield  {journal} {\bibinfo
  {journal} {Phys. Rev. D}\ }\textbf {\bibinfo {volume} {82}},\ \bibinfo {eid}
  {084036} (\bibinfo {year} {2010})},\ \Eprint {http://arxiv.org/abs/1008.0935}
  {arXiv:1008.0935 [gr-qc]} \BibitemShut {NoStop}%
\bibitem [{\citenamefont {{Bini}}\ \emph {et~al.}(2019)\citenamefont {{Bini}},
  \citenamefont {{Damour}},\ and\ \citenamefont {{Geralico}}}]{BiniDamoGera19}%
  \BibitemOpen
  \bibfield  {author} {\bibinfo {author} {\bibfnamefont {D.}~\bibnamefont
  {{Bini}}}, \bibinfo {author} {\bibfnamefont {T.}~\bibnamefont {{Damour}}}, \
  and\ \bibinfo {author} {\bibfnamefont {A.}~\bibnamefont {{Geralico}}},\
  }\href {\doibase 10.1103/PhysRevLett.123.231104} {\bibfield  {journal}
  {\bibinfo  {journal} {Phys. Rev. Lett.}\ }\textbf {\bibinfo {volume} {123}},\
  \bibinfo {eid} {231104} (\bibinfo {year} {2019})},\ \Eprint
  {http://arxiv.org/abs/1909.02375} {arXiv:1909.02375 [gr-qc]} \BibitemShut
  {NoStop}%
\bibitem [{\citenamefont {Munna}\ and\ \citenamefont
  {Evans}({\natexlab{c}})}]{MunnEvan20c}%
  \BibitemOpen
  \bibfield  {author} {\bibinfo {author} {\bibfnamefont {C.}~\bibnamefont
  {Munna}}\ and\ \bibinfo {author} {\bibfnamefont {C.~R.}\ \bibnamefont
  {Evans}},\ }\href@noop {} {\bibfield  {journal} {\bibinfo  {journal} {to be
  submitted to Phys. Rev. D}\ } ({\natexlab{c}})}\BibitemShut {NoStop}%
\bibitem [{\citenamefont {Deser}(2014)}]{Dese14}%
  \BibitemOpen
  \bibfield  {author} {\bibinfo {author} {\bibfnamefont {S.}~\bibnamefont
  {Deser}},\ }\href {\doibase 10.1007/s10714-013-1615-9} {\bibfield  {journal}
  {\bibinfo  {journal} {Gen. Rel. Grav}\ }\textbf {\bibinfo {volume} {46}},\
  \bibinfo {pages} {1615} (\bibinfo {year} {2014})},\ \Eprint
  {http://arxiv.org/abs/1307.4776} {arXiv:1307.4776 [gr-qc]} \BibitemShut
  {NoStop}%
\bibitem [{\citenamefont {{Barack}}\ and\ \citenamefont
  {{Sago}}(2011)}]{BaraSago11}%
  \BibitemOpen
  \bibfield  {author} {\bibinfo {author} {\bibfnamefont {L.}~\bibnamefont
  {{Barack}}}\ and\ \bibinfo {author} {\bibfnamefont {N.}~\bibnamefont
  {{Sago}}},\ }\href {\doibase 10.1103/PhysRevD.83.084023} {\bibfield
  {journal} {\bibinfo  {journal} {Phys. Rev. D}\ }\textbf {\bibinfo {volume}
  {83}},\ \bibinfo {pages} {084023} (\bibinfo {year} {2011})},\ \Eprint
  {http://arxiv.org/abs/1101.3331} {arXiv:1101.3331 [gr-qc]} \BibitemShut
  {NoStop}%
\bibitem [{\citenamefont {{Pound}}\ \emph {et~al.}(2013)\citenamefont
  {{Pound}}, \citenamefont {{Merlin}},\ and\ \citenamefont
  {{Barack}}}]{PounMerlBara13}%
  \BibitemOpen
  \bibfield  {author} {\bibinfo {author} {\bibfnamefont {A.}~\bibnamefont
  {{Pound}}}, \bibinfo {author} {\bibfnamefont {C.}~\bibnamefont {{Merlin}}}, \
  and\ \bibinfo {author} {\bibfnamefont {L.}~\bibnamefont {{Barack}}},\
  }\href@noop {} {\bibfield  {journal} {\bibinfo  {journal} {ArXiv e-prints}\ }
  (\bibinfo {year} {2013})},\ \Eprint {http://arxiv.org/abs/1310.1513}
  {arXiv:1310.1513 [gr-qc]} \BibitemShut {NoStop}%
\bibitem [{\citenamefont {{Thompson}}\ \emph {et~al.}(2019)\citenamefont
  {{Thompson}}, \citenamefont {{Wardell}},\ and\ \citenamefont
  {{Whiting}}}]{ThomWardWhit19}%
  \BibitemOpen
  \bibfield  {author} {\bibinfo {author} {\bibfnamefont {J.}~\bibnamefont
  {{Thompson}}}, \bibinfo {author} {\bibfnamefont {B.}~\bibnamefont
  {{Wardell}}}, \ and\ \bibinfo {author} {\bibfnamefont {B.}~\bibnamefont
  {{Whiting}}},\ }\href {\doibase 10.1103/PhysRevD.99.124046} {\bibfield
  {journal} {\bibinfo  {journal} {Phys. Rev. D}\ }\textbf {\bibinfo {volume}
  {99}},\ \bibinfo {eid} {124046} (\bibinfo {year} {2019}),\
  10.1103/PhysRevD.99.124046},\ \Eprint {http://arxiv.org/abs/1811.04432}
  {arXiv:1811.04432 [{gr-qc}]} \BibitemShut {NoStop}%
\end{thebibliography}%

\end{document}